\documentclass[11pt,aps,prd,nofootinbib,superscriptaddress,preprintnumbers]{revtex4}
\usepackage{graphicx,epsfig}
\usepackage{amsmath}
\usepackage{color}
\newcommand{\bea}{\begin{eqnarray}}
\newcommand{\beq}{\begin{equation}}
\newcommand{\eea}{\end{eqnarray}}
\newcommand{\eeq}{\end{equation}}

\newcommand{\nn}{\nonumber}
\newcommand{\Frac}[2]{\frac{\displaystyle{#1}}{\displaystyle{#2}}}
\newcommand{\lsim}{\raise0.3ex\hbox{$\;<$\kern-0.75em\raise-1.1ex\hbox{$\sim\;$}}}
\newcommand{\gsim}{\raise0.3ex\hbox{$\;>$\kern-0.75em\raise-1.1ex\hbox{$\sim\;$}}}
\newcommand{\eq}[1]{Eq.~(\ref{#1})}
\newcommand{\unity}{{\hbox{1\kern-.8mm l}}}
\newcommand{\LL}{{\mbox{\scriptsize LL}}}

\begin{document}

\title{FCNC and CP Violation Observables in a $SU(3)$--flavoured MSSM}
\author{L.~Calibbi}
\affiliation{SISSA/ISAS and INFN, I-34013, Trieste, Italy.}
\author{J.~Jones~P\'erez}
\affiliation{Departament de F\'{\i}sica Te\`orica and IFIC, Universitat de 
Val\`encia-CSIC, E-46100, Burjassot, Spain.}
\author{A.~Masiero}
\affiliation{INFN, Sezione di Padova, via F Marzolo 8, I--35131, Padova, Italy}
\affiliation{Univ. of Padova, Physics Dept. "G. Galieli", Padova, Italy}
\author{Jae-hyeon~Park}
\affiliation{INFN, Sezione di Padova, via F Marzolo 8, I--35131, Padova, Italy}
\author{W.~Porod}
\affiliation{ Institut f\"ur Theoretische Physik und Astrophysik,
 Universit\"at W\"urzburg, D-97074 W\"urzburg, Germany.}
\author{O. Vives}
\affiliation{Departament de F\'{\i}sica Te\`orica and IFIC, Universitat de 
Val\`encia-CSIC, E-46100, Burjassot, Spain.}

\preprint{FTUV-09-0723, IFIC-09-34, SISSA-43/2009/EP}

\begin{abstract}
  A non-Abelian flavour symmetry in a minimal supersymmetric standard model can
  explain the flavour structures in the Yukawa couplings and simultaneously
  solve the SUSY flavour problem. Similarly the SUSY CP problem can be solved
  if CP is spontaneously broken in the flavour sector. In this work, we
  present an explicit example of these statements with a $SU(3)$ flavour
  symmetry and spontaneous CP violation.  In addition, we show that it is
  still possible to find some significant deviation from the SM
  expectations as far as  FCNC and CP violation are concerned. We find that 
  large contributions can be expected in lepton flavour violating decays, as
  $\mu \to e \gamma$ and $\tau \to \mu \gamma$, electric dipole moments, $d_e$
  and $d_n$ and kaon CP violating processes as $\epsilon_K$. We also show that
  without further modifications, it is unlikely for these models to solve the $\Phi_{B_s}$ anomaly at low-moderate $\tan\beta$.
  Thus, these
  flavoured MSSM realizations are 
  phenomenologically sensitive to the experimental searches in the realm of
  flavor and CP violation physics.
\end{abstract}

\maketitle

\section{Introduction}
In the next few years, after a long impasse in the phenomenological searches
for new physics in the energy frontier, all the high-energy particle physics 
community will
be focused on the results from the LHC experiments. Its
first goal will be the study of the physics of electroweak symmetry breaking,
but besides this, we also expect some kind of new physics around the
electroweak scale. Supersymmetry (SUSY) is perhaps the new physics option that
is best motivated. The LHC should find some SUSY particles if SUSY is indeed the
solution to the hierarchy problem and provides a candidate for the dark matter
observed in the universe.

Supersymmetry has been extensively studied in the last decades, but most of
these studies have been done in the framework of the so-called Constrained
Minimal Supersymmetric Standard Model (CMSSM). The CMSSM is one of the simplest
supersymmetric extensions of the Standard Model as it assumes universality of
the supersymmetric soft breaking terms and is completely determined by four
parameters ($M_{1/2}, m_0, A_0, \tan\beta$) plus a sign (sg($\mu$)). This
simplified model is very useful to explore the main features of the SUSY
spectrum in collider experiments, however, nobody really believes
that the realization nature has chosen of supersymmetry is exactly the CMSSM,
specially concerning flavour. In analogy to the known flavour structures
in the Yukawa couplings, we naturally expect non-trivial flavour
structures in the supersymmetry soft-breaking terms.   

Therefore, we should consider other more general flavoured MSSM models as the 
SUSY models we can find when we analyze the experimental results from LHC 
experiments. Nevertheless, it is well-known that the presence of generic
flavour structures in the SUSY soft-breaking terms causes the so-called 
``supersymmetric flavour problem''. Flavour changing neutral currents (FCNC) 
and flavour-dependent CP violation
observables receive too large contributions from loops involving SUSY
particles and can not satisfy the stringent phenomenological bounds on these
processes \cite{Masiero:2001ep,Masiero:2005ua}. Although this statement is
still true, it is important to emphasize that the basis of this 
problem lies clearly on our total ignorance about the origin of the observed 
flavour and CP-violation in our theory, and this includes also the SM Yukawa
couplings. The {\bf real flavour problem} is simply our
inability to understand the complicated structures in the quark and lepton
Yukawa couplings, and likewise for the soft-breaking flavour structures in the
MSSM. It has been recently shown in the literature \cite{Ross:2004qn,Calibbi:2008qt} that an MSSM model with a
non-Abelian flavour symmetry allows a 
simultaneous understanding of the flavour structures in the  Yukawa couplings 
and the SUSY soft-breaking terms, adequately suppressing FCNC and CP violating
phenomena and solving the SUSY flavour problem. 

In this work,  we intend to exhibit a concrete example of such strategy 
considering an MSSM (with the limited number of new parameters present in the 
CMSSM) and showing that the presence of a non-Abelian family (horizontal) 
symmetry can 
simultaneously account for the solution of both flavor problems, namely 
that it is possible to successfully reproduce the correct fermion 
spectrum while adequately suppressing FCNC and CP violating phenomena. 
Regarding the phenomenology of these flavoured MSSM models, we do not expect
large  differences from the expectations in the CMSSM from the point of view
of collider studies, as we know that the departures from flavour universality 
are strongly constrained by the present FCNC and CP 
violation experiments. Much probably the resolution of the new supersymmetric
flavour structures will have to rely on FCNC and CP violation experiments. In
the following we will analyze the FCNC and CP violation phenomenology of our
flavoured MSSM example. We will show that it is still possible to find some 
significant deviation from the the SM expectations as far as the FCNC and CP 
violations are concerned, making these realizations phenomenologically
sensitive to the experimental searches in the realm of flavor physics.

In the next section, we introduce the flavour symmetries in SUSY and study its
effects in the SUSY soft-breaking terms. Then, we illustrate these effects with
an explicit model with a $SU(3)$ flavour symmetry. In section~\ref{FCNCCP}, we
analyze the different FCNC and CP violation observables in this model
including lepton flavour violation, kaon physics, $B$ physics and electric
dipole moments. Section~\ref{sec:combined} is devoted to a combined analysis 
of all the FCNC and CP violation observables and we show the correlations 
between the most interesting observables in this model. Finally, in section
\ref{sec:conclusions} we present our conclusions.

\section{Flavour Symmetries in SUSY}

Flavour symmetries have been used with success in the past to try to extract
some meaning from the complicated structures of fermion masses and mixings in
the Standard Model (SM). Using the Froggat-Nielsen
mechanism~\cite{Froggatt:1978nt}, flavour symmetries explain the structure of
the SM Yukawa couplings as the result of a spontaneously broken symmetry
associated with flavour. In different extensions of the SM, these flavour
symmetries will also constrain the new couplings and masses. For instance, in 
the context of a supersymmetric theory, a flavour
symmetry would apply equally to the fermion and scalar sectors. Therefore,
this implies that in the limit of exact symmetry the soft-breaking scalar
masses and the trilinear couplings must be invariant under the flavour
symmetry and the structures in the soft-breaking terms will be generated after
the spontaneous breaking of the flavour symmetry. Both the flavour structures
in the Yukawa couplings and in the soft-breaking terms are generated by the
same mechanism and thus we can expect a close relation between them\footnote{The relation between the soft-breaking terms and the Yukawa matrices
is present in gravity mediation mechanisms, but it would not be present if the
mediation mechanism is flavour blind, as in the case of gauge mediation or 
anomaly mediation}. 

The flavour-diagonal scalar masses, i.e. couplings $\phi^\dagger\phi$, 
are clearly invariant under any symmetry and are always allowed by the 
flavour symmetry. However, in general,
this does not guarantee that they are family universal. In the case of an 
Abelian
\cite{Froggatt:1978nt,Leurer:1992wg,Pomarol:1995xc,Binetruy:1996xk,Dudas:1996fe,Dreiner:2003yr,Kane:2005va,Chankowski:2005qp} family symmetry, the symmetry does not relate different 
generations, and, generically, different fields will have different diagonal soft masses. On the other hand, a non-Abelian family symmetry  groups two or
three generations in a single multiplet with a common mass in the symmetric
limit, thus helping to solve, in principle, the FCNC problem. 
This was one of the main motivations for the construction of the
first $SU(2)$ flavour models~\cite{Barbieri:1995uv,Barbieri:1996ww}, where the 
first two generation sfermions, facing the strongest constraints, share a
common mass. In the case of an $SU(3)$ flavour symmetry \cite{King:2001uz,King:2003rf,Ross:2004qn,deMedeirosVarzielas:2005ax}, all three generations have the
same mass in the unbroken family symmetry limit. On the contrary, trilinear 
couplings are
completely equivalent to the Yukawa couplings from the point of view of the
symmetry because they involve exactly the same fields (scalar or fermionic
components). Thus, they are generated after
symmetry-breaking as a function of small vevs. Therefore, to solve the ``SUSY
flavour problem'', we will consider in this work non-Abelian family 
symmetries, and more exactly $SU(3)$
theories or discrete versions of this symmetry.

Apart from these renormalizable mass operators in the Lagrangian, we can
construct non-renormalizable operators neutral under the flavour symmetry
inserting an appropriate number of flavon fields. The flavon fields, charged
under the symmetry, are responsible for the spontaneous symmetry breaking
once they acquire a vev. Then, higher dimensional operators involving two SM
fermions and a Higgs field, with several flavon vevs suppressed by a large
mediator mass, generate the observed Yukawa couplings. In the same way, these
flavon fields will couple to the scalar fields in all possible ways allowed by
the symmetry and, after spontaneous symmetry breaking, they will generate a
non-trivial flavour structure in the soft-breaking parameters. Therefore, by
being generated by insertions of the same flavon vevs, we can expect the
structures in the soft-breaking matrices and the Yukawa couplings to be
related.

The structure in the Yukawa couplings is not completely determined by the
observed values of fermionic masses and mixing angles.  To solve this problem
and  fix the Yukawa
couplings, we accept that the smallness of CKM mixing angles is due to the
smallness of the off-diagonal elements in the Yukawa matrices with respect to
the corresponding diagonal elements, and we make the additional simplifying
assumption of choosing the matrices to be symmetric. With these two
theoretical assumptions,  and using the ratio of masses at the GUT scale to
define the expansion parameters in the up and down sector as $\bar
\varepsilon= \sqrt{m_s/m_b}$ and $\varepsilon= \sqrt{m_c/m_t}$, we can fix the
Yukawa textures in the quark sector to be:
\begin{eqnarray}  
\label{fit}
Y_d\propto\left( 
\begin{array}{ccc}
0 & x^d_{12}\,\bar\varepsilon^3 &x^d_{13}\,\bar\varepsilon^3 \\ 
x^d_{12}\,\bar\varepsilon^3 & \bar\varepsilon^2 & 
x^d_{23}\,\bar\varepsilon^{2} \\ 
x^d_{13}\,\bar\varepsilon^3 & x^d_{23}\,\bar\varepsilon^2 & 1%
\end{array}%
\right),~~~~~~ Y_u\propto \left( 
\begin{array}{ccc}
0 & x^u_{12}\,\varepsilon^3 & x^u_{13}\,\varepsilon^3 \\ 
 x^u_{12}\,\varepsilon^3 & \varepsilon^2 & 
x^u_{23}\,\varepsilon^{2} \\ 
 x^u_{13}\,\varepsilon^{3} & x^u_{23}\,\varepsilon^{2} & 1%
\end{array}
\right ) \, ,
\end{eqnarray}
where  $\bar \varepsilon\simeq 0.15$, $\varepsilon \simeq 0.05$ and the $x^a_{ij}$ 
are $O(1)$ coefficients fixed by the observed values of fermion masses and 
mixings. In the Appendix we show the full structure of the Yukawas, and find the best fit values $x^d_{12}\simeq 1.7$, 
$x^d_{13}\simeq 0.4$, $x^d_{23}\simeq 1.8$, $x^u_{12}\simeq 1.4$, $x^u_{13}\simeq 2$, $x^u_{23}\simeq 2$.
In the leptonic sector we will follow the same strategy as in
Ref.~\cite{Calibbi:2008qt} and require unification of charged lepton and down-quark
flavour matrices, i.e. we embed our model in a grand unified framework, for
instance $SO(10)$,   and we try to explain simultaneously quark and lepton
Yukawas. The detailed structure of the leptonic Yukawa matrix is also shown in the Appendix.

Taking this Yukawa structure as our starting point, we will generate the 
flavour structure of the soft-breaking terms in generic non-Abelian $SU(3)$
flavour  symmetries or discrete versions,  like
$\Delta(27)$ or $\Delta(54)$ \cite{deMedeirosVarzielas:2005qg,deMedeirosVarzielas:2006fc,Kobayashi:2006wq,Ishimori:2008uc,Ishimori:2009ew}. Under these symmetries, the three 
generations of SM fields, both $SU(2)_L$-doublets and singlets, 
are triplets ${\bf 3}$ and the Higgs fields are singlets. 
Therefore Yukawa couplings and trilinear terms are not allowed by the 
symmetry. In these models, we add several flavon fields. For instance in 
\cite{Ross:2004qn} we have $\theta_3$, $\theta_{23}$
(anti-triplets ${\bf \bar 3}$), $\bar \theta_3$ and $\bar \theta_{23}$
(triplets ${\bf 3}$), while in \cite{deMedeirosVarzielas:2005ax} we have also $\theta_{123}$ and
$\bar \theta_{123}$. The symmetry is broken in several steps, first $SU(3)$ is broken into $SU(2)$
by the vev of $\theta_3$ and $\bar\theta_3$, with
$\langle\theta_3\rangle=(0,0,a)$ and $a$ being of the same order as the
  mediator mass $M_f$, i.e. $a/M_f \simeq O(1)$.
Subsequently 
$\theta_{23}$ and $\bar \theta_{23}$ get a smaller vev $\propto (0,b,b)$, with
$b/M_d\simeq \bar \varepsilon $ and  $b/M_u\simeq\varepsilon$. Notice
that in principle, we have three different mediator masses, $M_f =M_L, M_u, M_d$, because 
the flavour symmetry must commute with the SM symmetry and therefore the 
vector-like mediator fields must have the SM quantum numbers of the usual 
particles\footnote{For simplicity, we take $M_d$ and $M_u=M_L$ in all our numerical calculations.}.
In some models we also have $\theta_{123}$ and $\bar\theta_{123}$ getting a 
lower vev $\propto (c,c,c)$ with  $c/M_d\simeq \bar \varepsilon^2$.

Notice that the mentioned vevs require a vacuum alignment mechanism. In this work we do not specify a particular mechanism, but we refer the interested reader to the examples in~\cite{Ross:2004qn, Altarelli:2005yp, deMedeirosVarzielas:2005qg, Kobayashi:2008ih, Seidl:2008yf}.

The basic structure of the Yukawa superpotential (for quarks and leptons) is 
then given by
\begin{eqnarray}
\label{thetaW}
W_{\rm Y} &=& H\psi _{i}\psi _{j}^{c} \left[\theta _{3}^{i}  \theta
_{3}^{j}+\theta _{23}^{i} \theta _{23}^{j}
\left(\theta_3\overline{\theta}_3\right)+ 
\epsilon ^{ikl} \overline{ \theta}_{23,k} {\overline{ \theta }_{3,l}} \theta _{23}^{j}\left(
 \theta _{23} {\overline{\theta} _{3}}\right) +\dots \right],
\end{eqnarray} 
in the absence of $\theta_{123}$ fields, or,
\begin{eqnarray}
\label{theta123W}
W_{\rm Y} &=& H\psi _{i}\psi _{j}^{c} \left[\theta _{3}^{i}  \theta
_{3}^{j}+\theta _{23}^{i} \theta _{23}^{j}  + 
\theta_{23}^i \theta_{123}^j + \theta_{123}^i \theta_{23}^j \dots \right],
\end{eqnarray} 
in the models with $\theta_{123}$.
All flavon fields in these equations should be understood  as $\theta_i/M_f$. 
Note, however, that the $SU(3)$ symmetry is not enough by itself to determine 
the required structure in the superpotential. Generically, we have to
introduce additional global symmetries to forbid unwanted terms, like a mixed
term $\theta_3 \theta_{23}$, that would spoil the Yukawa structure. 
Nevertheless, the structure in Eqs.~(\ref{thetaW}) and (\ref{theta123W}) 
is quite general for the different $SU(3)$ models we can build,
and for additional details we refer to
\cite{King:2001uz,King:2003rf,Ross:2004qn}. 

In the same way,  the scalar soft masses deviate from exact universality after $SU(3)$ breaking. As explained above, $\phi _{i}^{\dagger }\phi _{i}$ is
completely neutral under gauge and global symmetries and gives rise to a
common contribution for the family triplet. However, after $SU(3)$ breaking,
terms with additional flavon fields give rise to important corrections
\cite{Ross:2002mr,Ross:2004qn,Antusch:2007re,Olive:2008vv}. Any invariant
combination of flavon fields can also contribute to the sfermion masses (at
least with Planck scale suppression). In
this case, it is easy to see that the following terms will always contribute
to the sfermion mass matrices (the presence of the $\theta_{123}$ field  will depend on the model):
\begin{align}
\label{eq:minimal}
(M^2_{\tilde f})_{i}^j = m_0^2 \bigg(\delta_{i}^j &
  +\frac{\displaystyle{1}}{\displaystyle{M_f^{2}}}\left[{\theta _{3}^{\dagger}}_i
\theta_3^j  + \overline{ \theta} _{3,i} {\overline{ \theta} _{3}^{\dagger}}^j + 
{\theta _{23}^{\dagger }}_i\theta_{23}^j + \overline{ \theta} _{23,i} 
{\overline{ \theta}_{23}^{\dagger}}^j 
+{\theta _{123}^{\dagger}}_i \theta_{123}^{j} + \overline{ \theta} _{123,i} 
{\overline{ \theta}_{123}^{\dagger}}^j \right] \nonumber \\&+  \Frac{1}{M_f^4}(\epsilon_{ikl}{\overline{\theta }_{3}^\dagger}^k
{\overline{\theta }_{23}^{\dagger }}^l) (\epsilon^{jmn}
\overline{\theta }_{3,m}\overline{\theta }_{23,n})+\quad\ldots\quad\bigg),
\end{align}
where $f$ represents the $SU(2)$ quark and lepton doublets or the up
(neutrino) and down (charged-lepton) singlets, so that $M_f=M_L, M_u, M_d$.
From here we see
that taking  $\langle \theta_{123} \rangle/M_d = \langle \bar \theta_{123}
\rangle /M_d = \bar \varepsilon^2$, both models with and without
$\theta_{123}$ have the same structure up to corrections $O(\varepsilon^4)$
that will be always subdominant in the relevant soft mass matrices in the
basis of diagonal Yukawa matrices.

Notice that in \eq{eq:minimal} only the fields that enter the 
superpotential have their vevs and associated mediators masses fixed by the 
$\varepsilon$ or $\bar \varepsilon$ parameters. For instance in the discrete 
model of Ref.\cite{deMedeirosVarzielas:2006fc}, $\bar \theta_{123}$ (the $\bf 3$-field that in this 
reference is written as $\theta_{123}$) does not enter the superpotential. 
Therefore, its contributions to the soft masses can be suppressed even having 
a large vev if the associated mediator mass is high enough\footnote{We thank
  G.~G.~Ross and I.~de Medeiros Varzielas for clarifying this point.}.
However, in the continuous version of this model~\cite{deMedeirosVarzielas:2005ax},
 the vevs of $\bar \theta_{123}/M_f$ and $\bar
\theta_{123}/M_f$ are constrained to be both $O(\varepsilon^2)$  by D-flatness 
and thus they are not dangerous in the soft-mass matrices.
Moreover, we have to remember that these deviations from 
universality in the soft-mass matrices
proportional to flavour symmetry breaking come always
through corrections in the K\"ahler potential and these effects will be 
important only in gravity-mediation SUSY models. 

In the case of the trilinear couplings we  have to emphasize that from the
point of view of the flavour symmetry these couplings are completely
equivalent to the corresponding Yukawa coupling. This means that they
necessarily involve the same combination of flavon vevs, although order one
coefficients are generically different because they require at least an
additional coupling to a field mediating SUSY breaking (in general coupled in
different ways in the various contributions). Therefore, from our point of view, we expect that the trilinear couplings have the same structure as the Yukawa matrices in the flavour basis. However in general they are {\bf not} proportional to the Yukawas, because of different $O(1)$ coefficients in the different elements. Thus, we can expect that going to the SCKM basis does not diagonalize the trilinear matrices. In fact, the trilinear matrices maintain the same structure as in the flavour basis and only the $O(1)$ coefficients are modified.

Therefore, we can see that in these $SU(3)$-like models with the three
generations unified in a single field, we have basically the same ``leading
order'' structures in the soft mass matrices and the trilinear couplings
directly related to the structures in the Yukawa couplings. In the following
we concentrate in the $SU(3)$ model of~\cite{Ross:2004qn} because it has
a complete phase structure in the flavon vevs consistent with the CKM
phase. Remember however, that the main features of the soft terms are similar 
in other models and in principle the phase structure could be also adapted in
these models.

\subsection{SU(3) Flavour Model}
\label{models}

Let us now specify more explicitely the $SU(3)$ flavour model that we use as
our main example.  The full superpotential is determined by $SU(3)$ and
several global symmetries are used to forbid unwanted terms that would spoil the
observed structure of the Yukawa couplings. Using the charges presented in
Table~\ref{Table:RVV1}, the leading terms in the superpotential are, 
\begin{eqnarray}
W_{\rm Y} &=& H\psi _{i}\psi _{j}^{c} \left[\theta _{3}^{i}  \theta
_{3}^{j}+\theta _{23}^{i} \theta _{23}^{j} \Sigma + \left(
\epsilon ^{ikl} \overline{ \theta}_{23,k} {\overline{ \theta }_{3,l}} \theta
_{23}^{j} +  \epsilon ^{jkl} \overline{ \theta}_{23,k} {\overline{ \theta
  }_{3,l}} \theta _{23}^{i} \right) \left(
 \theta _{23} {\overline{\theta} _{3}}\right) + \right. \nn \\
&& \left.  \epsilon ^{ijl} \overline{ \theta}_{23,l} \left(
 \theta _{23} {\overline{\theta} _{3}}\right)^2 +  \epsilon ^{ijl} \overline{
 \theta}_{3,l} \left(\theta _{23} {\overline{\theta} _{3}}\right)\left(\theta
 _{23} {\overline{\theta} _{23}}\right) + \dots \right],
\end{eqnarray} 
where to simplify the notation, the flavon fields have been normalized to the
corresponding mediator mass, which means that all the flavon fields in this 
equation
should be understood  as $\theta_i/M_f$. The field $\Sigma$ is a
Georgi-Jarlskog field that gets a vev in the $B - L$ direction, distinguishing
leptons and quarks. Furthermore, as said above this model is embeded in a 
$SO(10)$ grand unified structure at high scales, which allow us to relate
quark and lepton (including neutrino) Yukawa couplings. However, the $SU(2)_R$
subgroup of $SO(10)$ must be broken as we need different mediator masses for
the up and down sector and, in fact $\theta_3$ and
$\overline{\theta}_3$ are $\bf{3}\oplus\bf{1}$ representations of
$SU(2)_R$ which is broken by their vevs 
\cite{Ross:2002fb,King:2003rf,Ross:2004qn}.

After spontaneous symmetry breaking the flavon fields get the following vevs: 
\begin{align}
\label{vevs}
\langle\theta_3\rangle=\left( 
\begin{array}{c}
0 \\ 
0 \\ 
1\end{array}
\right)&\otimes \left(
\begin{array}{cc}
a_3^u & 0 \\ 
0 & a_3^d~ e^{i \chi}
\end{array}
\right);& \langle\bar{\theta}_3\rangle=\left( 
\begin{array}{c}
0 \\ 
0 \\ 
1\end{array}
\right)&\otimes \left(
\begin{array}{cc}
a_3^u~ e^{i \alpha_u} & 0 \\ 
0 & a_3^d~ e^{i \alpha_d}
\end{array}
\right);  \nonumber \\
\langle\theta_{23}\rangle=&\left( 
\begin{array}{c}
0 \\ 
b_{23} \\ 
b_{23}~e^{i\beta_3}
\end{array}
\right);& \langle\bar\theta_{23}\rangle=&\left( 
\begin{array}{c}
0 \\ 
b_{23}~e^{i\beta^{\prime}_2} \\ 
b_{23}~e^{i(\beta^{\prime}_2 - \beta_3)}%
\end{array}
\right);
\end{align}
where we require the following relations:
\begin{align}
\label{expansion}
\left(\frac{\displaystyle{a_3^u}}{\displaystyle{M_u}}\right)^2& = y_t,&
\left(\frac{\displaystyle{a_3^d}}{\displaystyle{M_d}}\right)^2& = y_b, \nonumber \\
\frac{\displaystyle{b_{23}}}{\displaystyle{M_u}}& = \varepsilon,& \frac{\displaystyle{b_{23}}}{\displaystyle{M_d}}& = \bar \varepsilon.
\end{align}
These relations are valid at the flavour breaking scale, that  we take as the
GUT scale in the numerical evaluation.

\begin{table}
\begin{center}
\begin{tabular}{|c||c | c | c | c|}
\hline
Flavon Phase & $\alpha_u-\alpha_d$ & $\chi$ & $\beta_3$ & $\beta'_2$ \\
\hline
Allowed Values & $(-40\pm2)^\circ$ & $(20\pm10)^\circ+180^\circ n$ & $-(20\pm10)^\circ+180^\circ n$ & Unconstrained \\
\hline
\end{tabular}
\end{center}
\caption{\label{Table:Phases} Flavon phases after imposing CKM constraints, with $n\in N$. Notice that the approximate errors we quote in $\chi$ and $\beta_3$ are highly correlated, as shown in the Appendix.}
\end{table}

It is straight-forward to see that this superpotential reproduces correctly
the required Yukawa structure in \eq{fit}. For completeness, we also list in Table~\ref{Table:Phases} the values that the flavon phases can take, given the constraints imposed by the CKM matrix. The analysis leading to such constraints is found in the Appendix.

We can now turn to the soft
breaking terms.  As mentioned in the previous section, a universal, flavour 
diagonal mass term will
always be allowed. Moreover, in a SUSY theory, the same messenger
fields as in the Yukawas will couple the flavons to the scalar fields in the
soft terms. Thus, the
$\varepsilon$ and $\bar\varepsilon$ parameters still act as expansion
parameters, and represent important corrections to the soft terms.

Clearly any coupling involving a flavon field and its hermitian conjugate
(i.e. $\theta_3^{i\dagger}\theta_3^j$) is invariant under the flavour
symmetry.  From this we can deduce that the soft mass terms get a minimum structure determined uniquely by the flavon content of the model and on their vevs. This minimum structure is obtained from the following effective terms:
\begin{align}
\label{eq:softminimal}
(M^2_{\tilde f})_i^{j} = m_0^2 \bigg(\delta_i^j &
  +\left[{\theta _3^{\dagger}}_i\theta _3^j+\overline{\theta}_{3,i}{\overline{\theta}_{3}^\dagger}^j + 
{\theta _{23}^{\dagger }}_i\theta_{23}^j+\overline{\theta}_{23,i} 
{\overline{\theta}_{23}^\dagger}^j\right] \nonumber \\
& +(\epsilon_{ikl}\theta_3^k\theta_{23}^l)(\epsilon^{jmn}{\theta_3^\dagger}_m{\theta_{23}^\dagger}_n) 
+ (\epsilon_{ikl}{\overline{\theta }_3^\dagger}^k{\overline{\theta }_{23}^\dagger}^l)(\epsilon^{jmn}
\overline{\theta }_{3,m}\overline{\theta }_{23,n})+\quad\ldots\quad\bigg).
\end{align}
In Table~\ref{Table:RVV1} we show a choice of global charges that reproduces
the correct Yukawa structure and does not allow other terms at leading order
in the K\"ahler potential (soft-masses). 

\begin{table}
\begin{center}
\begin{tabular*}{1.00\textwidth}{@{\extracolsep{\fill}}|c|c c c c c c c c|}
\hline
${\bf Field}$ & $\psi$ & $\psi^c$ & $H$& $\Sigma$ & $\theta_3$ & $\theta_{23}$ & $\bar\theta_3$ &
$\bar\theta_{23}$~~~ \\
\hline
${\bf SU(3)}$ & 3 & 3 & 1 & 1 & $\bar3$ & $\bar3$ & 3 & 3~~~ \\
${\bf U(1)}$ & 0 & 0 & 0 & 1 & 0 & -1 & 1 & 0~~~ \\
${\bf U'(1)}$ & -1 & -1 & 0 & 2 & 1 & 0 & -1 & 4~~~ \\
${\bf U''(1)}$ & 1 & 1 & 0 & -3 & -1 & 1 & 0 & -4~~~ \\
\hline
\end{tabular*}
\end{center}
\caption{\label{Table:RVV1} Charges required to build the minimal RVV1 Model.}
\end{table}

In the squark sector, after rephasing the fields such that the CKM matrix
elements $V_{ud}$, $V_{us}$, $V_{cb}$ and $V_{tb}$ are real, the soft masses in the SCKM basis are:
\begin{subequations}
\label{sckm1}
\begin{eqnarray}
\left(M_{\tilde u_R^c}^2\right)^T & = &\left(\begin{array}{ccc}
1+\varepsilon^2\,y_t & -\varepsilon^3\,e^{i\omega'} & -\varepsilon^3\,e^{i(\omega'-2\chi)} \\
-\varepsilon^3\,e^{-i\omega'} & 1+\varepsilon^2 & \varepsilon^2\,e^{-2i\chi} \\
-\varepsilon^3\,e^{-i(\omega'-2\chi)} & \varepsilon^2\,e^{2i\chi} & 1+y_t
\end{array}\right)m_0^2\\
\left(M_{\tilde d_R^c}^2\right)^T & = & \left(\begin{array}{ccc}
1+\bar\varepsilon^2\,y_b & -\bar\varepsilon^3\,e^{i\omega_{us}} & -\bar\varepsilon^3\,e^{i\omega_{us}} \\
-\bar\varepsilon^3\,e^{-i\omega_{us}} & 1+\bar\varepsilon^2 & \bar\varepsilon^2 \\
-\bar\varepsilon^3\,e^{-i\omega_{us}} & \bar\varepsilon^2 & 1+y_b
\end{array}\right)m_0^2\\
M_{\tilde Q}^2 & = & \left(\begin{array}{ccc}
1+\varepsilon^2\,y_t & -\varepsilon^2\bar\varepsilon\,e^{i\omega_{us}} & -\bar\varepsilon^3\,y_t e^{i\omega_{us}} \\
-\varepsilon^2\bar\varepsilon\,e^{-i\omega_{us}} & 1+\varepsilon^2 & \bar\varepsilon^2\,y_t \\
-\bar\varepsilon^3\,y_t e^{-i\omega_{us}} & \bar\varepsilon^2\,y_t & 1+y_t
\end{array}\right)m_0^2 \\
\left(M_{\tilde e_R^c}^2\right)^T & = & \left(\begin{array}{ccc}
1+\bar\varepsilon^2\,y_b & -\bar\varepsilon^3 & -\bar\varepsilon^3\,e^{i(\chi-\beta_3)} \\
-\bar\varepsilon^3 & 1+\bar\varepsilon^2 & \bar\varepsilon^2\,e^{i(\chi-\beta_3)} \\
-\bar\varepsilon^3\,e^{-i(\chi-\beta_3)} & \bar\varepsilon^2\,e^{-i(\chi-\beta_3)} & 1+y_b
\end{array}\right)m_0^2 \\
M_{\tilde L}^2 & = & \left(\begin{array}{ccc}
1+\varepsilon^2\,y_t & -\varepsilon^2\bar\varepsilon & -\bar\varepsilon^3\,y_t\,e^{i(\chi-\beta_3)} \\
-\varepsilon^2\bar\varepsilon & 1+\varepsilon^2 & \bar\varepsilon^2\,y_t\,e^{i(\chi-\beta_3)} \\
-\bar\varepsilon^3\,y_t e^{-i(\chi-\beta_3)} & \bar\varepsilon^2\,y_t\,e^{-i(\chi-\beta_3)} & 1+y_t
\end{array}\right)m_0^2
\end{eqnarray}
\end{subequations}
where $M_{\tilde Q}^2$ ($M_{\tilde L}^2$) is in the basis where $Y_d$ ($Y_e$) is diagonal, $\omega_{us}$ is related to the CKM phase,
and $\omega'=\omega_{us}-(\delta_u-\delta_d)$.  The $\omega_{us}$ and $\delta_i$ phases can be found in the Appendix.
The structure of $M_{\tilde Q}^2$ in
the basis where $Y_u$ is diagonal is similar to $M_{\tilde u_R^c}^2$. 
Notice that, although the structure in terms of $\varepsilon$, $\bar\varepsilon$ in $M_{\tilde e_R^c}^2$ and $M_{\tilde L}^2$ is the same as that of $M_{\tilde d_R^c}^2$ and $M_{\tilde Q}^2$, respectively, the coefficients from the SCKM rotation and RGE evolution are different due to the Georgi-Jarlskog field $\Sigma$. This can be seen in the Appendix, and is summarized in Table~\ref{tab:deltas}, in Section~\ref{lfv}.
The phase structure of the slepton matrices is different to the one of squarks, since the latter have been rephased in order for the CKM matrix to follow the standard phase convention.\footnote{As physical observables are independent of phase conventions, we could also rephase the slepton superfields and use the same phase structure as squarks.}

In Eq.~(\ref{sckm1}), we have written only the leading contribution in $\varepsilon$ and $\bar\varepsilon$
to each element with a leading phase, omitting effective complex $O(1)$ coefficients. This sets the size of the modulus of the mass-insertion. The $O(1)$ coefficients are the remaining part of the full element, after factorizing the terms explicitly written in Eq.~(\ref{sckm1}). There are many contributions to these coefficients: first, we have the real $O(1)$ constants in front of each term in the K\"ahler potential, at $M_{GUT}$. We also have contributions from the real $O(1)$ constants in the Yukawa matrices, coming from the rotation into the SCKM basis. A third contribution comes from the RGE evolution to the EW scale. And finally, it is possible to have further contributions from subleading terms in the K\"ahler potential, still at $M_{GUT}$.

All of these contributions can involve the flavon phases, so the effective $O(1)$ can be complex. As we are factorizing the leading phases, the phase in each coefficient shall only appear in subleading terms. If the leading phases cancel for a particular observable, these subleading phases shall be important. Such a cancellation can happen in EDMs. For
example, in mass-insertion notation~\cite{Gabbiani:1996hi, Hagelin:1992tc},
the corresponding effective $O(1)$ coefficients of $(\delta^d_{13})_{RR}$ and $(\delta^d_{13})_{LL}$ have the following structure:
\begin{subequations}
\begin{eqnarray}
 D_{13}^{RR} & \sim & 1 - y_b\,e^{-2i(\chi-\beta_3)} \\
 Q_{13}^{LL} & \sim & 1 - (\varepsilon^2/\bar\varepsilon^2)\,e^{-2i(\chi-\beta_3)}.
\end{eqnarray}
\end{subequations}

Although \eq{eq:softminimal} is the minimal structure present for all possible
models, it is possible, for particular choices of the global symmetries and
charges, to build other symmetry-dependent soft-mass structures. In fact, the
observed structure in the Yukawa couplings does not fix completely the
introduced global charges and it is possible to add new invariant combinations
of flavon fields to the K\"ahler potential without modifying the Yukawas.

\begin{table}
\begin{center}
\begin{tabular*}{1.00\textwidth}{@{\extracolsep{\fill}}|c|c c c c c c c c|}
\hline
${\bf Field}$ & $\psi$ & $\psi^c$ & $H$& $\Sigma$ & $\theta_3$ & $\theta_{23}$ & $\bar\theta_3$ &
$\bar\theta_{23}$~~~ \\
\hline
${\bf SU(3)}$ & 3 & 3 & 1 & 1 & $\bar3$ & $\bar3$ & 3 & 3~~~ \\
${\bf U(1)}$ & -2 & -2 & 0 & -4 & 2 & 3 & 0 & -2~~~ \\
${\bf U'(1)}$ & 0 & 0 & 0 & 1 & 0 & -1 & 1 & 0~~~ \\
\hline
\end{tabular*}
\end{center}
\caption{\label{Table:RVV2} Charges for RVV2 Model.}
\end{table}
The first example of these new combinations of flavon fields in the K\"ahler
is achieved by allowing a $\theta_3^i\bar\theta_{23}^j$ term (RVV2). The
required charges are shown in Table~\ref{Table:RVV2}. It is easy to check
that, with these charges, the structure of the Yukawa couplings in the
superpotential remains unchanged. This is due to the fact that the 
superpotential is a holomorphic function of the fields while the K\"ahler is
only a real function.  When rotated to the SCKM basis, the soft-mass matrices 
become:
\begin{subequations}
\label{sckm2}
\begin{eqnarray}
\left(M_{\tilde u_R^c}^2\right)^T & = &\left(\begin{array}{ccc}
1+\varepsilon^2\,y_t & -\varepsilon^3\,e^{i\omega'} & -\varepsilon^2\,y_t^{0.5} e^{i(\omega'-2\chi+\beta_3-\beta'_2)} \\
-\varepsilon^3\,e^{-i\omega'} & 1+\varepsilon^2 & \varepsilon\,y_t^{0.5} e^{-i(2\chi-\beta_3+\beta'_2)} \\
-\varepsilon^2\,y_t^{0.5} e^{-i(\omega'-2\chi+\beta_3-\beta'_2)} & \varepsilon\,y_t^{0.5} e^{i(2\chi-\beta_3+\beta'_2)} & 1+y_t
\end{array}\right)m_0^2\\
\left(M_{\tilde d_R^c}^2\right)^T & = & \left(\begin{array}{ccc}
1+\bar\varepsilon^2\,y_b & -\bar\varepsilon^3\,e^{i\omega_{us}} & -\bar\varepsilon^2\,y_b^{0.5} e^{i(\omega_{us}-\chi+\beta_3-\beta'_2)} \\
-\bar\varepsilon^3\,e^{-i\omega_{us}} & 1+\bar\varepsilon^2 & \bar\varepsilon\,y_b^{0.5} e^{-i(\chi-\beta_3+\beta'_2)} \\
-\bar\varepsilon^2\,y_b^{0.5} e^{-i(\omega_{us}-\chi+\beta_3-\beta'_2)} & \bar\varepsilon\,y_b^{0.5} e^{i(\chi-\beta_3+\beta'_2)} & 1+y_b
\end{array}\right)m_0^2\\
M_{\tilde Q}^2 & = & \left(\begin{array}{ccc}
1+\varepsilon^2\,y_t & -\varepsilon^2\bar\varepsilon\,e^{i\omega_{us}} & \varepsilon\bar\varepsilon\,y_t^{0.5} e^{i(\omega_{us}-2\chi+\beta_3+\beta'_2)} \\
-\varepsilon^2\bar\varepsilon\,e^{-i\omega_{us}} & 1+\varepsilon^2 & \varepsilon\,y_t^{0.5} e^{-i(2\chi-\beta_3-\beta'_2)} \\
\varepsilon\bar\varepsilon\,y_t^{0.5} e^{-i(\omega_{us}-2\chi+\beta_3+\beta'_2)} & \varepsilon\,y_t^{0.5} e^{i(2\chi-\beta_3-\beta'_2)} & 1+y_t
\end{array}\right)m_0^2 \\
\left(M_{\tilde e_R^c}^2\right)^T & = & \left(\begin{array}{ccc}
1+\bar\varepsilon^2\,y_b & -\bar\varepsilon^3 & -\bar\varepsilon^2\,y_b^{0.5} e^{-i\beta'_2} \\
-\bar\varepsilon^3 & 1+\bar\varepsilon^2 & \bar\varepsilon\,y_b^{0.5} e^{-i\beta'_2} \\
-\bar\varepsilon^2\,y_b^{0.5} e^{i\beta'_2} & \bar\varepsilon\,y_b^{0.5} e^{i\beta'_2} & 1+y_b
\end{array}\right)m_0^2 \\
M_{\tilde L}^2 & = & \left(\begin{array}{ccc}
1+\varepsilon^2\,y_t & -\varepsilon^2\bar\varepsilon & \varepsilon\bar\varepsilon\,y_t^{0.5}\,e^{-i(\chi-\beta'_2)} \\
-\varepsilon^2\bar\varepsilon & 1+\varepsilon^2 & \varepsilon\,y_t^{0.5}\,e^{-i(\chi-\beta'_2)} \\
\varepsilon\bar\varepsilon\,y_t^{0.5}\,e^{i(\chi-\beta'_2)} & \varepsilon\,y_t^{0.5}\,e^{i(\chi-\beta'_2)} & 1+y_t
\end{array}\right)m_0^2
\end{eqnarray}
\end{subequations}
One can see that the effect of this term in $m_{\tilde d_R^c}^2$ is to exchange one power of $\bar\varepsilon$ by a $y_b^{0.5}$ suppression in $(\delta^d_{13})_{RR}$ and $(\delta^d_{23})_{RR}$. In $m_{\tilde Q}^2$, the same terms change an $\bar\varepsilon^2$ by an $\varepsilon\,y_t^{0.5}$. However, for $\tan\beta=10$, and considering that $\varepsilon\approx\bar\varepsilon^2$, such replacements leave the structure of the mass matrices very similar numerically to the original one (notice that $y_t$ and $y_b$ are taken at $M_{GUT}$). Nonetheless, it must be remarked that the phase structure of the whole mass matrix is modified.

As in RVV1, it is crucial to take into account that relative phases exist
within the effective $O(1)$ coefficients. For instance, although the global
phase of $(\delta^d_{12})_{AA}$ is still $\omega_{us}$, the $O(1)$ structure
of $(\delta^d_{12})_{LL}$ in RVV2 is now:
\begin{equation}
\label{rvv2-12}
(Q_{12}^{LL})^{RVV2} \sim 1-\left(\bar\varepsilon^2/\varepsilon\right) (1+e^{i(2\chi-\beta_3-\beta'_2)}).
\end{equation}
Notice that the factor $(\bar\varepsilon^2/\varepsilon)=0.45$ does not really provide any suppression at all. This means that the imaginary part of $(\delta^d_{12})_{LL}$ is larger than just $\varepsilon^2\bar\varepsilon\sin\omega_{us}$. We label this new, larger, effective phase as $\omega'_{us}$.

\begin{table}
\begin{center}
\begin{tabular*}{1.00\textwidth}{@{\extracolsep{\fill}}|c|c c c c c c c c|}
\hline
${\bf Field}$ & $\psi$ & $\psi^c$ & $H$& $\Sigma$ & $\theta_3$ & $\theta_{23}$ & $\bar\theta_3$ &
$\bar\theta_{23}$~~~ \\
\hline
${\bf SU(3)}$ & 3 & 3 & 1 & 1 & $\bar3$ & $\bar3$ & 3 & 3~~~ \\
${\bf U(1)}$ & -1 & -1 & 0 & 5 & 1 & -2 & 0 & 6~~~ \\
${\bf U'(1)}$ & 0 & 0 & 0 & -1 & 0 & 0 & 1 & -2~~~ \\
\hline
\end{tabular*}
\end{center}
\caption{\label{Table:RVV3} Charges for RVV3 Model.}
\end{table}

A second possibility is to allow a
$\left(\epsilon^{ikl}\theta_3^k\theta_{23}^l\right)\theta_3^j$ term (RVV3) in
the K\"ahler, with the charges being shown in Table~\ref{Table:RVV3}. The soft matrices, when rotated into the SCKM basis, have the following structure:
\begin{subequations}
\label{sckm3}
\begin{eqnarray}
\left(M_{\tilde u_R^c}^2\right)^T & = &\left(\begin{array}{ccc}
1+\varepsilon^2\,y_t & -\varepsilon^3\,e^{i\omega'} & \varepsilon\,y_t e^{i(\omega'-2\chi+\beta_3-\delta_u)} \\
-\varepsilon^3\,e^{-i\omega'} & 1+\varepsilon^2 & \varepsilon^2\,e^{-2i\chi} \\
\varepsilon\,y_t e^{-i(\omega'-2\chi+\beta_3-\delta_u)} & \varepsilon^2\,e^{2i\chi} & 1+y_t
\end{array}\right)m_0^2\\
\left(M_{\tilde d_R^c}^2\right)^T & = & \left(\begin{array}{ccc}
1+\bar\varepsilon^2\,y_b & -\bar\varepsilon^3\,e^{i\omega_{us}} & \bar\varepsilon\,y_b e^{i(\omega_{us}+\beta_3-\delta_d)} \\
-\bar\varepsilon^3\,e^{-i\omega_{us}} & 1+\bar\varepsilon^2 & \bar\varepsilon^2 \\
\bar\varepsilon\,y_b e^{-i(\omega_{us}+\beta_3-\delta_d)} & \bar\varepsilon^2 & 1+y_b
\end{array}\right)m_0^2\\
M_{\tilde Q}^2 & = & \left(\begin{array}{ccc}
1+\varepsilon^2\,y_t & -\varepsilon\bar\varepsilon^2\,y_t e^{i(\omega_{us}-2\chi+\beta_3+\delta_d)} & \varepsilon\,y_t e^{i(\omega_{us}-2\chi+\beta_3+\delta_d)} \\
-\varepsilon\bar\varepsilon^2\,y_t e^{-i(\omega_{us}-2\chi+\beta_3+\delta_d)} & 1+\varepsilon^2 & \bar\varepsilon^2\,y_t \\
\varepsilon\,y_t e^{-i(\omega_{us}-2\chi+\beta_3+\delta_d)} & \bar\varepsilon^2\,y_t & 1+y_t
\end{array}\right)m_0^2 \nonumber \\ \\
\left(M_{\tilde e_R^c}^2\right)^T & = & \left(\begin{array}{ccc}
1+\bar\varepsilon^2\,y_b & -\bar\varepsilon^3 & \bar\varepsilon\,y_b e^{i(\chi-\delta_d)} \\
-\bar\varepsilon^3 & 1+\bar\varepsilon^2 & \bar\varepsilon^2\,e^{i(\chi-\beta_3)} \\
\bar\varepsilon\,y_b e^{-i(\chi-\delta_d)} & \bar\varepsilon^2\,e^{-i(\chi-\beta_3)} & 1+y_b
\end{array}\right)m_0^2 \\
M_{\tilde L}^2 & = & \left(\begin{array}{ccc}
1+\varepsilon^2\,y_t & -\varepsilon\bar\varepsilon^2\,y_t\,e^{-i(2\chi-\beta_3-\delta_d)} & \varepsilon\,y_t\,e^{-i(\chi-\delta_d)} \\
-\varepsilon\bar\varepsilon^2\,y_t\,e^{i(2\chi-\beta_3-\delta_d)} & 1+\varepsilon^2 & \bar\varepsilon^2\,y_t\,e^{i(\chi-\beta_3)} \\
\varepsilon\,y_t\,e^{i(\chi-\delta_d)} & \bar\varepsilon^2\,y_t\,e^{-i(\chi-\beta_3)} & 1+y_t
\end{array}\right)m_0^2
\end{eqnarray}
\end{subequations}
with $\delta_i$ defined in the Appendix.

This model shows larger deviations from RVV1 in the $LL$ sector. It is
important to notice that $(\delta^d_{12})_{LL}$ is now of order
$\varepsilon\bar\varepsilon^2$ instead of $\varepsilon^2\bar\varepsilon$,
which will have considerable consequences in processes such as $\mu\to
e\gamma$. Likewise, $(\delta^d_{13})_{LL}$ is of order $\varepsilon\,y_t$
instead of $\bar\varepsilon^3$, so an enhancement in $\tau\to e\gamma$ should
be expected. Regarding the $RR$ sector, for $\tan\beta=10$, the $y_b$
suppression at $M_{GUT}$ has roughly the same size as an $\bar\varepsilon$
suppression, so once again the structure of $m_{\tilde d_R^c}^2$ is
numerically similar to RVV1 for this value of $\tan \beta$.

The trilinear couplings, on the other hand, follow the same symmetries as the Yukawas. Thus, they have the same flavon structure in RVV1, RVV2 and RVV3. Nonetheless, although they have the same structure, they do not have the same $O(1)$ constants, which means that the rotation into the SCKM basis does not diagonalize them. In the SCKM basis, after rephasing the fields, the trilinears have the following structure:
\begin{subequations}
 \label{aterms}
\begin{eqnarray}
A_u=\left(\begin{array}{ccc}
0 & \varepsilon^3\,e^{-i\omega'} & \varepsilon^3\,e^{i(2\chi-\omega')} \\
\varepsilon^3\,e^{i\omega'} & \varepsilon^2\,\Sigma_u & \varepsilon^2\,\Sigma_u\,e^{2i\chi} \\
\varepsilon^3\,e^{i(\omega'+2\beta_3-2\chi)} & \varepsilon^2\,\Sigma_u\,e^{2i(\beta_3-\chi)} & 1
\end{array}\right)A_0\,y_t \\
A_d=\left(\begin{array}{ccc}
0 & \bar\varepsilon^3\,e^{-i\omega_{us}} & \bar\varepsilon^3\,e^{-i\omega_{us}} \\
\bar\varepsilon^3\,e^{-i\omega_{us}} & \bar\varepsilon^2\,\Sigma_d & \bar\varepsilon^2\,\Sigma_d \\
\bar\varepsilon^3\,e^{i(\omega_{us}+2\beta_3-2\chi)} & \bar\varepsilon^2\,\Sigma_d\,e^{2i(\beta_3-\chi)} & 1
\end{array}\right)A_0\,y_b
\end{eqnarray}
\end{subequations}
where we have neglected the $O(1)$ coefficients. One can get $A_e$ by taking $A_d$ with $\left<\Sigma_e\right>=3\left<\Sigma_d\right>$.

The soft mass matrices in Eqs.~(\ref{sckm1},\ref{sckm2},\ref{sckm3},\ref{aterms})
are given at the large scale $\sim M_{GUT}$. Then, we have to include also effects
coming from the running from $M_{GUT}$ to $ M_{W}$. Although these effects
produce further non-universal contributions, they are usually smaller than the 
terms presented here. In the quark sector, the misalignment of the $Y_u$ and
$Y_d$ matrices gives sizeable contributions to the $LL$ and $LR$ sectors, analogous to the MFV contributions of CMSSM models. In the lepton sector with RH neutrinos, the same happens due to the misalignment of $Y_\nu$ and $Y_e$~\cite{Borzumati:1986qx,Casas:2001sr,Masiero:2002jn}. Both of these contributions are unavoidable, albeit the $Y_\nu$ contribution is highly model dependent.

Moreover, in the present case there are new contributions to the running given
by the intrinsic non-universality of the soft mass matrices. Although these
effects contribute even in the $RR$ sector, it turns out that the magnitude of
the generated off-diagonal terms are, at most, of the same order as those at
$M_{GUT}$. This means that the addition of running effects will only change
the already unknown $O(1)$ constants, such that the low-energy phenomenology
can still be understood by analyzing Eqs.~(\ref{sckm1}), (\ref{sckm2}),
(\ref{sckm3}) and (\ref{aterms}).

Finally, we have to emphasize once more that all the soft matrices presented
here have unknown $O(1)$ coefficients in all non-universal terms. The flavour
symmetry allows us to fix the order in $\varepsilon$ and $\bar \varepsilon$ of
the different entries but the final values could easily vary up or down by
factors of two and this has to be taken into account when analyzing our
numerical results. 

\section{FCNC and CP Violation Observables}
\label{FCNCCP}
In the following, we shall study flavour and CP violating observables,
presenting
analytical estimates in mass insertion approximation (MIA). Moreover,
we will perform a full
numerical analysis in the SUGRA parameter space through a scan in
$m_0$ and
$M_{1/2}$ for fixed values of $\tan\beta$ and $a_0 = A_0/m_0$.
The numerical analysis is done defining the Yukawa, trilinear
and soft mass matrices at $M_{flav}=2\times10^{16}$ GeV,
for the three different versions of the model, as explained in
Section~\ref{models}.
Then the different flavour matrices are evolved to the
electro-weak scale, solving 1-loop RGEs with SPheno~\cite{Porod:2003um}.
$O(1)$ coefficients in the Yukawa matrices are determined by
requiring a good fit on the fermion masses
and quark mixings at $M_Z$~\cite{Roberts:2001zy}. The result of such fit
is presented in the Appendix.

Regarding the unknown $O(1)$ constants in both the superpotential and K\"ahler potential, as we do not have a full high-scale model, we shall fix each $O(1)$s at a random value, between $0.5$ and 2. Notice they can be of either sign. Thus, a model is characterized by both the choice of symmetries involved, and by the particular $O(1)$s we have in front of each effective term. In Section~\ref{sec:combined} we shall take into account their variation.

After running the resulting matrices down to the $M_Z$ scale, we
diagonalize
the Yukawas in order to obtain the left and right mixing matrices and
rotate
the soft matrices into the SCKM basis. 
At the $M_Z$ scale, for each point of the SUSY parameter space,
we compute the SUSY spectrum and check that the electroweak
symmetry breaking does take place and no tachyonic particles arise.
Moreover, to be
conservative, we require that the Lightest Supersymmetric Particle
(LSP) is the lightest
neutralino.

In this work we do not include the non-holomorphic corrections to the Yukawa couplings, so our results are valid for low and moderate values of $\tan\beta$. In this regime, the effects of these non-holomorphic corrections are usually small. The only exception is found in the case for the neutron EDM, where new imaginary parts induced by these corrections can give sizeable contributions. We give details on how we introduce such corrections to this observable in Section~\ref{sec:edms}.

Finally, we apply the following constraints: 

\begin{itemize}
\item Bounds on the sparticle masses from direct searches at LEP and
Tevatron~\cite{Amsler:2008zzb}.
\item Lower bound on Higgs masses from LEP~\cite{Barate:2003sz}. Although the SM bound on the Higgs mass is of 114 GeV, in SUSY the modification of the lightest Higgs coupling could reduce this constraint. In order to take this effect into account, for each point we calculate the Higgs mixing angle $\alpha$ and use Eq.~(17) of Ref.~\cite{Djouadi:2001yk} to estimate the correct Higgs mass bound. We then use SPheno to calculate the two loop Higgs mass, and take into account a theoretical uncertainty of 3 GeV~\cite{Allanach:2004rh,Heinemeyer:2004xw}.
\item  $b\to s\gamma$. The experimental world average from the
    CLEO~\cite{Chen:2001fja}, Belle~\cite{bellebsg1}
and BaBar~\cite{Aubert:2005cua} collaborations is given
by~\cite{hfag-web}:
\begin{eqnarray}
\label{bsgamma}
{\rm BR} (B\to X_s \gamma)_{E_\gamma > 1.6 {\rm GeV}}
=
( 3.52 \pm 0.23\pm 0.9) \times 10^{-4}\; .
\end{eqnarray}
We have to compare these values with the MSSM predictions. In our numerical calculation, we use the expression presented in
Ref.~\cite{Hurth:2003dk} in which the branching ratio is explicitly given in terms of arbitrary complex Wilson
coefficients $C_7$ and $C_8$. At present, the SM contribution to this decay is already available at NNLO while the SUSY contribution in a general MSSM is partially known at NLO. In this work, we include the NLO SM contribution with a modified low value of the scale for the charm mass to reproduce the NNLO SM contribution ($\textrm{BR}(B\to X_s\gamma)^{SM}=3.15\times10^{-4}$)~\cite{Degrassi:2007kj,Becher:2006pu,Lunghi:2006hc}. We add the supersymmetric contributions at one loop, and require that the total result does not deviate from the experimental value of Eq.~(\ref{bsgamma}) in more than 2-sigma.

\item Muon anomalos magnetic moment, $a_\mu=(g-2)_\mu/2$.
At present, the experimental result for this
observable is given by~\cite{Bennett:2006fi},
\beq
a_\mu^{\rm exp} =  11\, 659\, 2080\, (63) \times 10^{-11},
\eeq
while, computing the hadronic contribution by means of
the hadronic $e^+e^-$ annihilation data, the SM theoretical expectation is \cite{Miller:2007kk,Passera:2008jk,Jegerlehner:2009ry},
\beq
a_{\mu}^{\rm SM} = 11\, 659\, 181\, (8) \times 10^{-10}.
\eeq
The resulting discrepancy is:
\beq
\Delta a_{\mu} = a_\mu^{\rm exp}-a_{\mu}^{\rm SM} = +302(88) \times 10^{-11}.
\label{eq:delta_amu}
\eeq
It is well-known that in the MSSM $a_\mu$ receives contributions from
${\tilde\chi}^0$--$\tilde \mu$ and ${\tilde \chi}^\pm$--$\tilde \nu$
loops~\cite{Moroi:1995yh}. Such contributions are approximately given by the
following expression:
\beq
\frac{a^{\rm MSSM}_\mu}{ 1 \times 10^{-9}} \approx 1.5\left(\frac{\tan\beta }{10} \right)
\left( \frac{300~\rm GeV}{m_{\tilde \nu}} \right)^2
\left(\frac{\mu M_2}{m^{2}_{\tilde \nu}} \right)~.
\label{eq:g_2}
\eeq
A comparison with Eq.~(\ref{eq:delta_amu}) implies that the present discrepancy
strongly favours the $\mu>0$ region of the SUSY parameter space. In the case of the theoretical
prediction based on $\tau$ decay \cite{Davier:2002dy}, the difference of Eq.~(\ref{eq:delta_amu})
is reduced to $\sim 1\sigma$ \cite{Passera:2008jk} but it still requires a positive
correction and disfavours strongly a sizable negative contribution.
\end{itemize} 

\subsection{Lepton Flavour Violation}
\label{lfv}

As pointed out in \cite{Calibbi:2008qt}, flavour models based on SU(3) give
rise to potentially large rates of LFV processes, such that positive signals
of LFV can be found in the currently running or near-future experiments, at least for SUSY masses within the reach of the LHC. 
The arising of large mixing among flavours relies on the features of the SU(3) model discussed in the previous sections:
the presence of nonuniversal scalar masses already at the scale where the SUSY breaking terms appear, and the fact that the trilinear
$A_f$ matrices are in general not aligned with the corresponding Yukawa matrices.
Let's start considering the case $A_0=0$, where the latter effect is strongly reduced so that, in terms of mass insertions, BR($l_i \to l_j \gamma$) mainly depends on $|(\delta^e_{LL})_{ij}|^2$ and $|(\delta^e_{RR})_{ij}|^2$.
Looking at the structure of the slepton soft mass matrices in the three versions of the model (Table~\ref{tab:deltas}),
we see that RVV1 and RVV2 are expected to give similar predictions for ${\rm
  BR}(\mu\to e \gamma)$ and ${\rm BR}(\tau\to \mu \gamma)$, with possibly 
sizeable
contributions coming from both the LL and the RR sector.  In the case of RVV3, the prediction for ${\rm BR}(\tau\to \mu \gamma)$ will be also similar to the previous two cases, while we expect ${\rm BR}(\mu\to e \gamma)$ to be strongly enhanced. In fact, for RVV3, the LL mass insertion is larger
by a factor $9\,y_t\,\bar\varepsilon/\varepsilon = \mathcal{O}(10)$ with respect to RVV1 and RVV2, and the ${\rm BR}(\mu\to e \gamma)$ is consequently increased by two orders of magnitude. 

\begin{table}
\begin{center}
\begin{tabular}{|c|ccc|ccc|}
\hline
& $|(\delta^e_{LL})_{12}|$  & $|(\delta^e_{LL})_{13}|$& $|(\delta^e_{LL})_{23}|$ &
$|(\delta^e_{RR})_{12}|$ & $|(\delta^e_{RR})_{13}|$ & $|(\delta^e_{RR})_{23}|$\\
\hline \hline
RVV1 & $ \frac{1}{3}\varepsilon^2\bar\varepsilon $  &
$ y_t \bar\varepsilon^3 $ &  $ 3 y_t \bar\varepsilon^2 $ & $ \frac{1}{3}\bar\varepsilon^3 $  & $\frac{1}{3}\bar\varepsilon^3$ & $\bar\varepsilon^2$  \\
RVV2 & $ \frac{1}{3}\varepsilon^2\bar\varepsilon $  &
$  \frac{1}{3} \sqrt{y_t} \varepsilon \bar\varepsilon $ & $ \sqrt{y_t}\varepsilon $ & $ \frac{1}{3}\bar\varepsilon^3 $  & $\frac{1}{3} \sqrt{y_b}\bar\varepsilon^2$ & $\sqrt{y_b} \bar\varepsilon$  \\
RVV3 & $ 3 y_t \varepsilon\bar\varepsilon^2 $  &
$  y_t \varepsilon$ & $ 3 y_t \bar\varepsilon^2 $ & $ \frac{1}{3} \bar\varepsilon^3 $  & $ y_b \bar\varepsilon $ & $ \bar\varepsilon^2$  \\
\hline
\end{tabular}
\end{center}
\caption{\label{tab:deltas} Order of magnitude of LFV mass-insertions, for the three models.}
\end{table} 

\begin{table}
\begin{center}
\begin{tabular}{|c|c|c|}
\hline
& Present Bound & Future Sensitivity  \\
\hline \hline
$BR(\mu\to e\gamma)$ & $ 1.2\times10^{-11}$ \cite{Ahmed:2001eh} & $\mathcal{O}(10^{-13})$ \cite{meg} \\
$BR(\tau\to\mu\gamma)$ & $ 1.6\times10^{-8}$ \cite{Banerjee:2007rj, Lusiani:2007cb} & $\mathcal{O}(10^{-9})$ \cite{Bona:2007qt} \\
$BR(\tau\to e\gamma)$ & $ 1.1\times10^{-7}$ \cite{Aubert:2005wa} &  $\mathcal{O}(10^{-8})$  \cite{Akeroyd:2004mj}  \\
\hline
\end{tabular}
\end{center}
\caption{\label{tab:lfv} Present bounds and future experimental sensitivities of lepton flavour violating processes.}
\end{table}

To summarize, let's compare the expections for the different LFV processes. In the case $A_0 = 0$, considering for simplicity only the contribution from $\delta^e_{LL}$, we have:
\begin{eqnarray}
\label{brs1}
{BR(\tau\to e \,\gamma) \over BR(\mu\to e\,\gamma) } &  \approx &
\left({m_\tau \over m_\mu}\right)^5  {\Gamma_\mu \over \Gamma_\tau} 
{(\delta^e_{\LL})^2_{13} \over  (\delta^e_{\LL})^2_{12}} \approx ~
{\mathcal O}(1) ~(\rm RVV1,\,RVV2,\,RVV3) \\
\label{brs2}
{BR(\tau\to\mu\,\gamma) \over BR(\mu\to e\,\gamma) } & \approx &
\left({m_\tau \over m_\mu}\right)^5  {\Gamma_\mu \over \Gamma_\tau} 
{(\delta^e_{\LL})^2_{23} \over  (\delta^e_{\LL})^2_{12}} \approx ~
{\mathcal O}(10^3) ~(\rm RVV1,\,RVV2), ~  {\mathcal O}(10)~ (\rm RVV3)
\end{eqnarray}
where $\Gamma_\mu$ ($\Gamma_\tau$) is the $\mu$ ($\tau$) full width. 
Given the present limits and future sensitivities of LFV processes shown in Tab. \ref{tab:lfv}, we see that ${\rm BR}(\tau\to e \gamma)$ is not able to constrain the parameter space better than ${\rm BR}(\mu\to e \gamma)$ in none of the three models. 
On the other hand, we expect from Eq. (\ref{brs2}) that the present
constraints given by $\mu \to e \gamma$ and $\tau \to \mu \gamma$, that
differ by three orders of magnitude, are comparable for RVV1 and RVV2, while $\mu \to e \gamma$ should give the strongest constraint in the case of RVV3.

In the case $A_0 \neq 0$, generally large $\delta^e_{LR}$ insertions arise as a consequence of the misalignment between $A_f$ and the corresponding Yukawa matrix $Y_f$. In this case, the neutralino contribution to ${\rm BR}(\mu\to e \gamma)$ gets strongly enhanced \cite{Calibbi:2008qt} and the present (or future) bound requires heavier SUSY masses to be fulfilled, specially in the region where the  gaugino mass is much larger than the common sfermion mass. Nevertheless, we expect this effect to be visible only in the case of RVV1 and RVV2, while for RVV3 the very large $(\delta^e_{LL})_{12}$ should still give the dominant contribution. 
\begin{figure}
\includegraphics[width=0.32\textwidth]{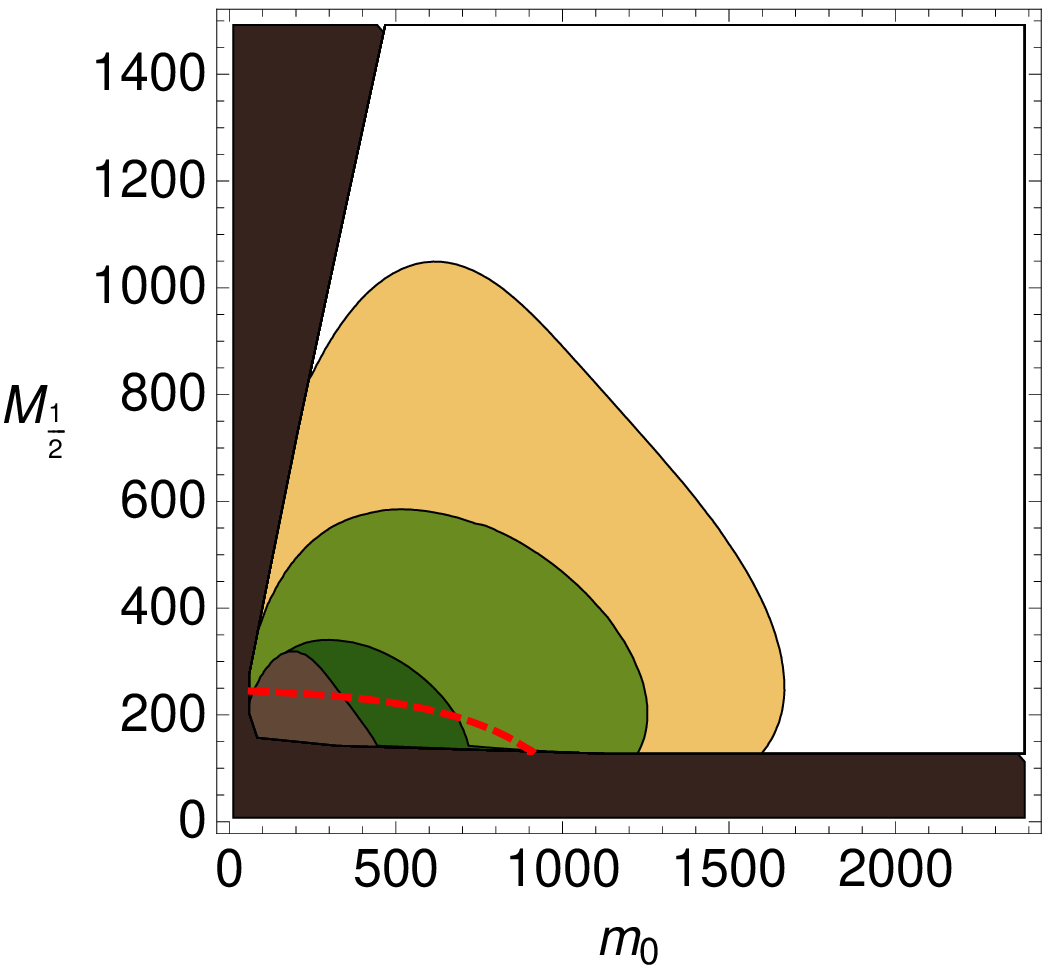}
\includegraphics[width=0.32\textwidth]{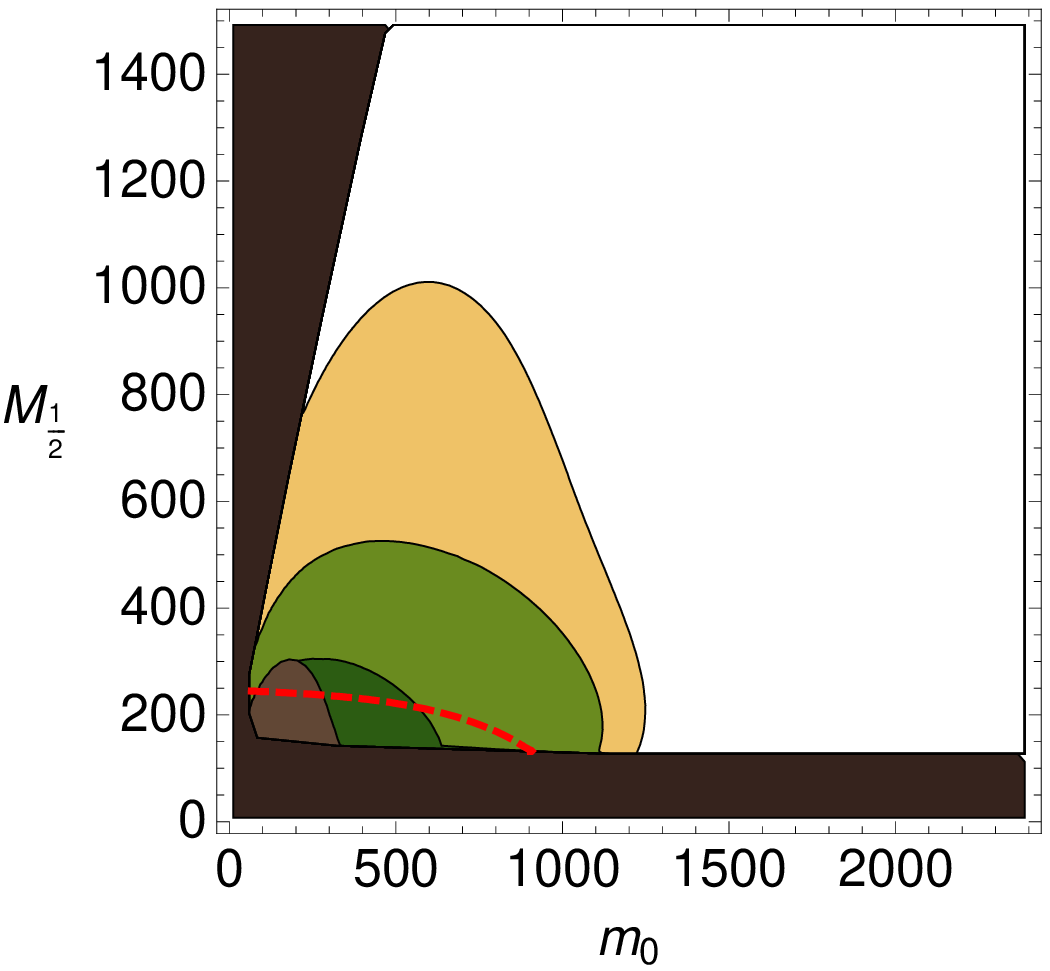} 
\includegraphics[width=0.32\textwidth]{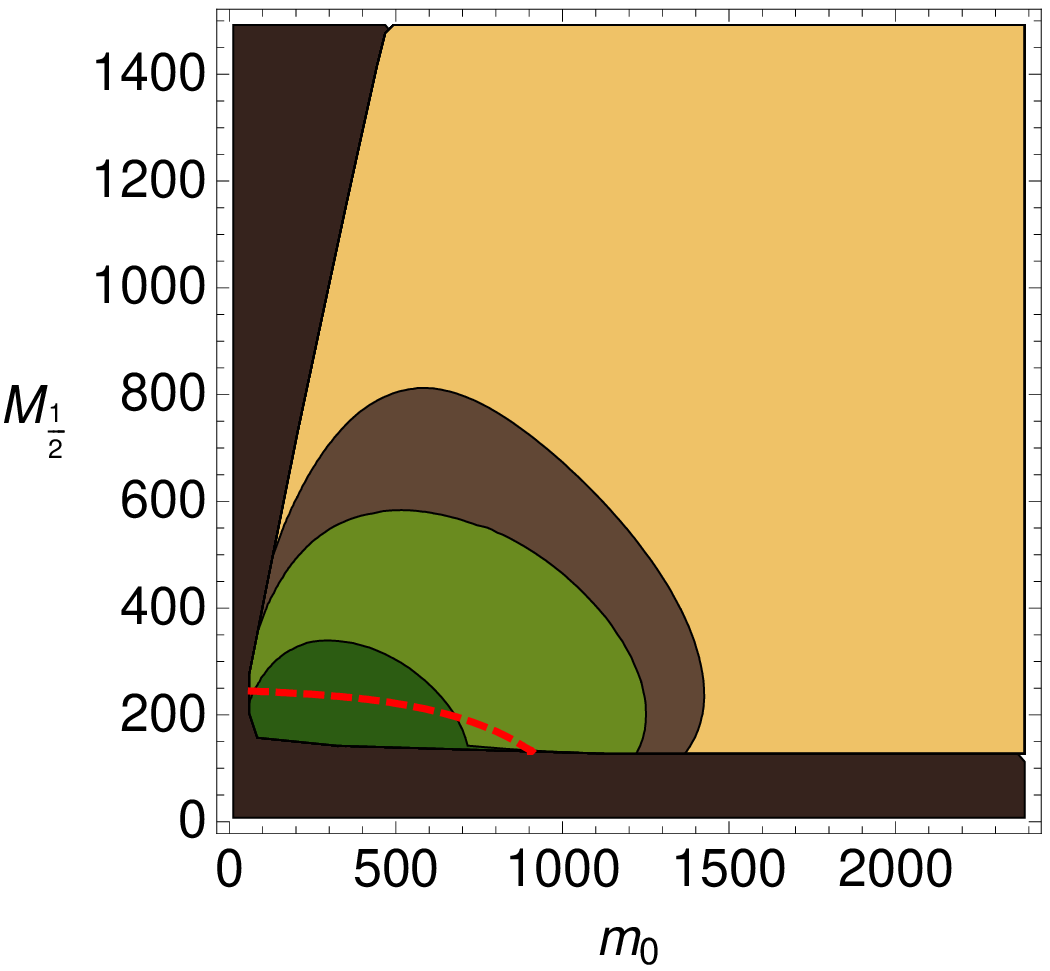} 
\includegraphics[width=0.32\textwidth]{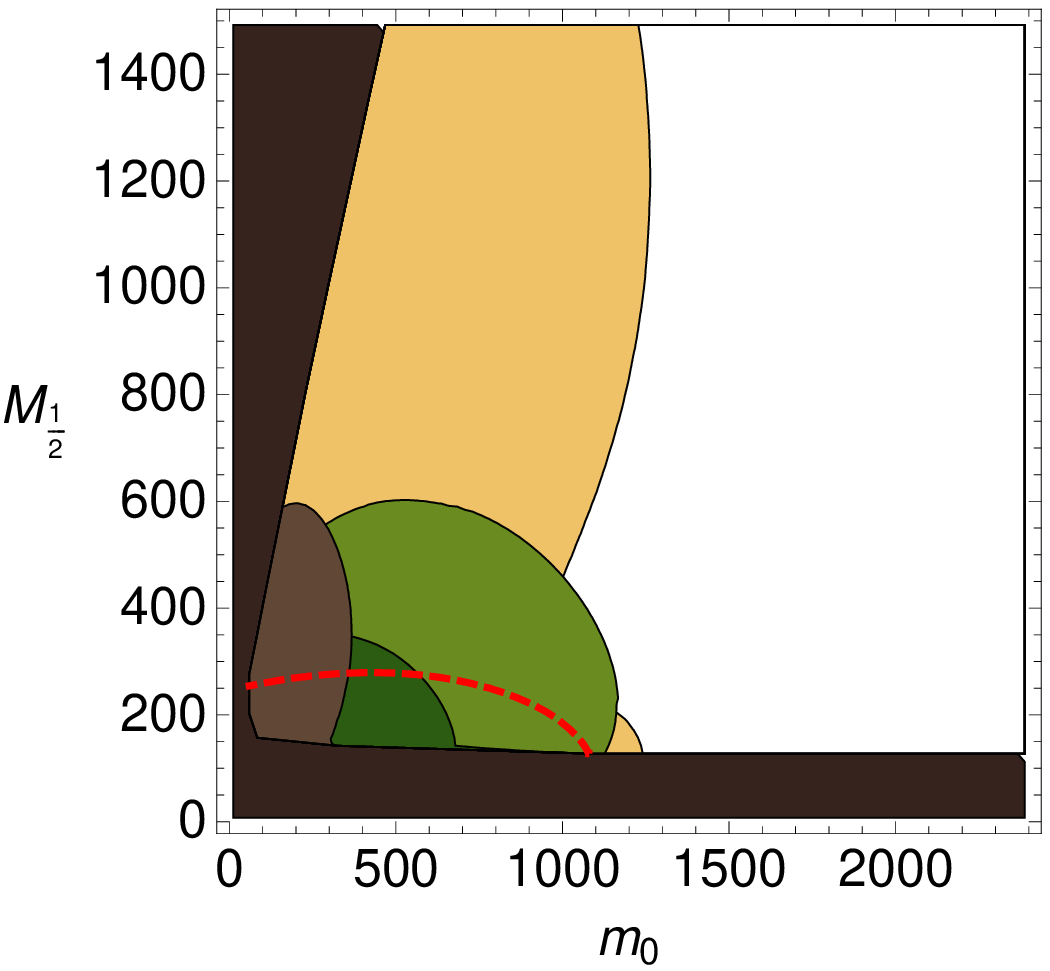}
\includegraphics[width=0.32\textwidth]{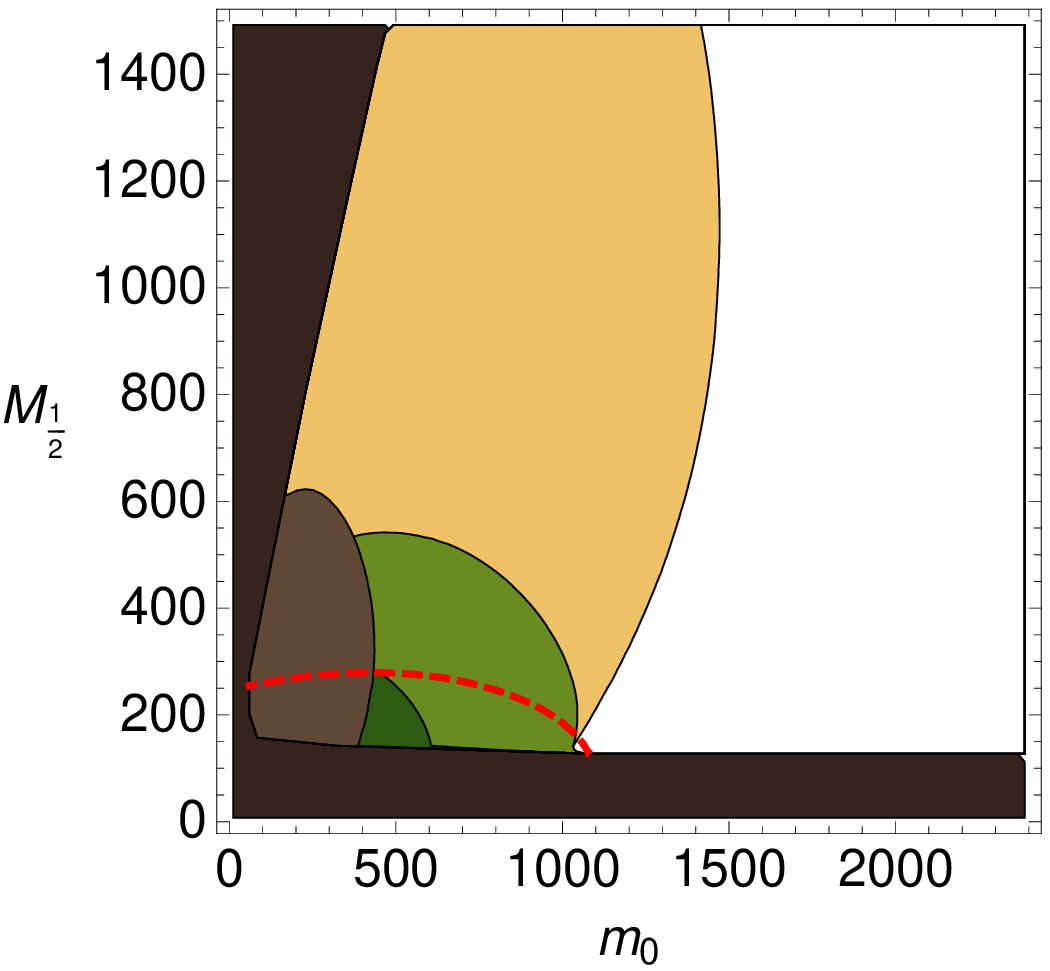} 
\includegraphics[width=0.32\textwidth]{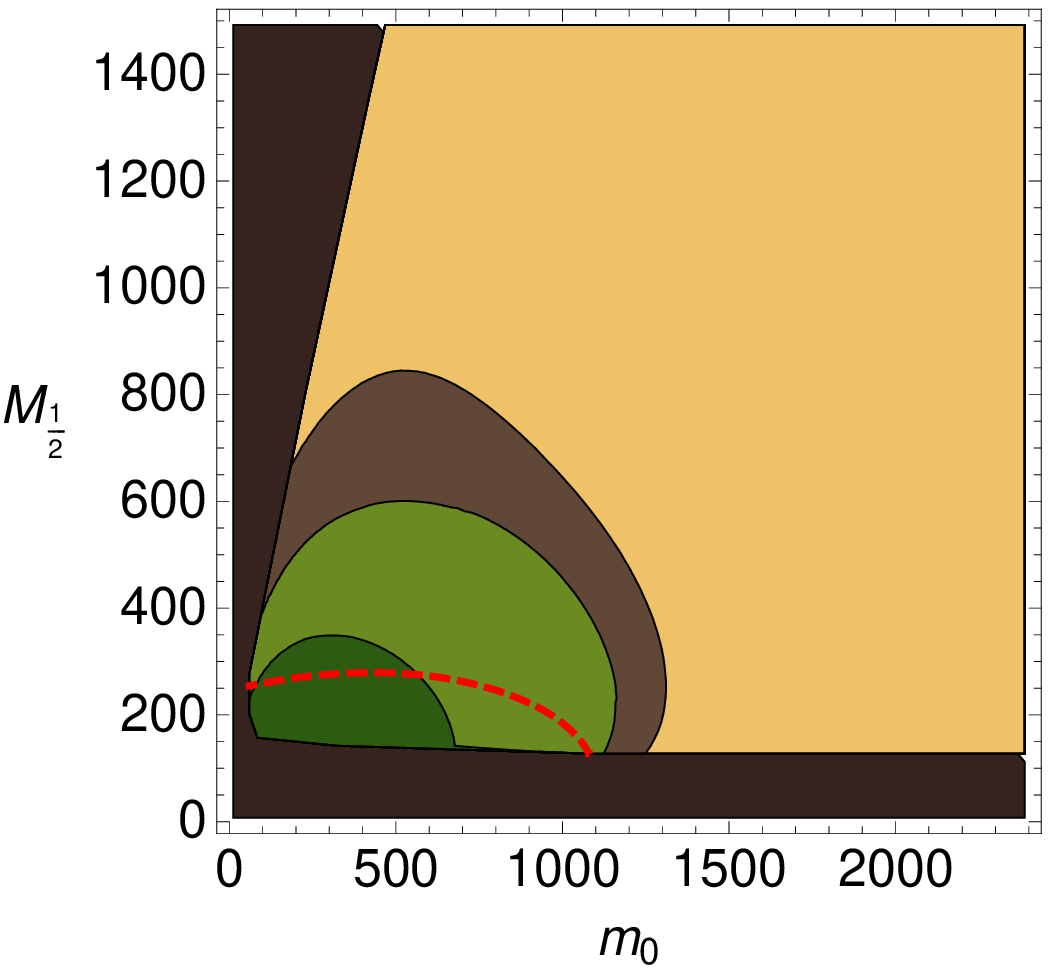} 
\caption{Current LFV constraints in the $m_0$-$M_{1/2}$ plane, for RVV1 (left), RVV2 (center) and RVV3 (right) for $\tan\beta=10$ and 
$A_0=0$ (first row), $A_0=m_0$ (second row). The brown region corresponds to the present limit ${\rm BR}(\mu\to e \gamma) \lesssim 1.2 \cdot 10^{-11}$, the light brown region represents the sensitivity of the MEG experiment ($10^{-13}$). The present BaBar+Belle combined limit ${\rm BR}(\tau\to \mu \gamma) \lesssim 1.6 \cdot 10^{-8}$ is shown in dark green, and the light green region corresponds to the sensitivity of a Super Flavour Factory ($10^{-9}$). The dark brown region show areas excluded by having a charged LSP or by LEP, excepting the Higgs mass bound, which is shown in thick dashed red lines.}
\label{fig:meg}
\end{figure}

Let us now consider the results of the numerical analysis for the LFV decays. After fixing the unknown $O(1)$ parameters to random values, we present in Fig.~\ref{fig:meg} the current bounds provided by $\mu\to e\gamma$ and $\tau\to \mu\gamma$ in the ($m_0$, $M_{1/2}$) plane, and also the final reach of the MEG experiment. 
In the first row, the $A_0=0$ case is displayed (for $\tan\beta=10$). We see
that, as expected, RVV1 and RVV2 give similar results both for ${\rm
  BR}(\mu\to e \gamma)$ and ${\rm BR}(\tau\to \mu \gamma)$. The regions of the
parameter space excluded by the present bounds ${\rm BR}(\mu\to e \gamma)
\lesssim 1.2 \cdot 10^{-11}$, ${\rm BR}(\tau\to \mu \gamma) \lesssim 1.6 \cdot
10^{-8}$ are comparable, as obtained by the naive estimate in
Eq.~(\ref{brs2}), although $\tau\to \mu \gamma$ turns out to be more
constraining for $m_0 > M_{1/2}$. The presently allowed region is approximately $(m_0,~ M_{1/2})\gtrsim
(700,~300)~ {\rm GeV}$. In the case of RVV3, $\mu\to e \gamma$ already gives a
strong constraint, $(m_0,~ M_{1/2})\gtrsim (1400,~800)~ {\rm GeV}$, which is
much more stringent than the one provided by $\tau\to \mu \gamma$. Although
the precise values may vary with different $O(1)$ parameters, these plots give
an idea of the reach of these LFV experiments.  As a consequence, for SUSY masses lying within the LHC reach, RVV3 results already rather disfavoured, while RVV1 and RVV2 are not strongly constrained. Considering the sensitivity expected at the MEG experiment for ${\rm BR}(\mu\to e \gamma)$, $\mathcal{O}(10^{-13})$, we see that also RVV1 and RVV2 will be tested in most of the parameter space accessible to the LHC, while RVV3 will be completely probed well beyond the LHC reach. Moreover, in case of larger values of $\tan\beta$ (e.g. $\tan\beta=30$), since ${\rm BR}(\mu\to e \gamma)\propto \tan^2\beta$, MEG will be able to test all the displayed parameter space also for RVV1 and RVV2 (cfr. \cite{Calibbi:2008qt}, where a similar model has been studied). 

In the second row of Fig.~\ref{fig:meg}, the three versions of the model are displayed for the case $A_0=m_0$, in order to show the potentially large flavour mixing induced by the A-terms. As expected, we see that, for RVV1 and RVV2, the $\mu\to e \gamma$ bound is increased with respect to the $A_0=0$ case, specially for $M_{1/2} \gtrsim m_0$, as a consequence of the large $(\delta^e_{LR})_{12}$ insertion contributing in diagrams with pure $\tilde{B}$ exchange. Also in this case, MEG has a very high capability of testing the parameter space. Indeed, for moderate slepton masses the neutralino contribution is so large that the models will be fully probed. Only in the case of rather heavy sleptons, the ${\rm BR}(\mu\to e \gamma)$ can be suppressed enough to escape the reach of MEG. In the case of RVV3, the contribution from LL insertion remains dominant and no substantial changes are observed with respect to the case of vanishing trilinear terms.

As RVV3 is heavily constrained by LFV, in the following we shall exclude it from our analysis, and concentrate exclusively on RVV1 and RVV2.

\subsection{Kaon physics}

\label{epsK}
Supersymmetric contributions to $K^0$--$\overline K^0$ mixing are mainly determined by the mixing between the first two generations of down type squarks.
In terms of mass insertions, a major hadronic constraint on 
$\Im m\{(\delta^d_{12})_{AB}\}$ comes from $\epsilon_K$.
The kaon mass difference, $\Delta M_K$, does constrain $|(\delta^d_{12})_{AB}|$. However, this observable is less sensitive to small MIs  since the
long-distance effects are difficult to estimate.

On the experimental side,
current data \cite{Amsler:2008zzb} shows that
$|\epsilon_K^\text{exp}| = (2.229\pm 0.012)\times 10^{-3}$
and the phase of $\epsilon_K$ is $\phi_\epsilon = (43.51\pm 0.05)^\circ$.
On the theoretical side, many improvements are being made
in the determination of input parameters needed
to calculate the SM prediction of $\epsilon_K$.
We have good information on the CKM matrix elements thanks to the
vast amount of data from $B$ physics.
Also available are new lattice estimates of the $\widehat{B}_K$ parameter
which enters the $\Delta S = 2$ hadronic matrix element.
However, these developments have lead to a puzzle in understanding the data.
The authors of Refs.~\cite{Buras:2008nn,Buras:2009pj}
realized that within the SM, the theoretical prediction of $\epsilon_K$
might not be enough to account for the measured value if
one accepts the unitarity triangle (UT) fit using $B$ physics observables.
One can summarize their result in the numerical form~\cite{Buras:2008nn,Buras:2009pj},
\begin{equation}
  \label{eq:epsKSM}
  \begin{aligned}
|\epsilon_K^\text{SM}| =
1.78\cdot 10^{-3} \times &
\left(\frac{\kappa_\epsilon}{0.92} \right)
\left(\frac{\widehat{B}_K}{0.72} \right)
\left(\frac{|V_{cb}|}{0.0412}\right)^2
\left(\frac{R_t}{0.914}\right) 
\times
\\
\Biggl\{&
0.74\times
\left(\frac{|V_{cb}|}{0.0412}\right)^2
\left(\frac{R_t}{0.914}\right)
\left(\frac{\sin 2 \beta}{0.675}\right) 
\left(\frac{\eta_{tt}}{0.5765}\right) 
\left(\frac{S_0(x_t)}{2.30}\right) + 
\\
\phantom{\Biggl\{}&
\left(\frac{\sin\beta}{0.362}\right) 
\left[
0.40\times
\left(\frac{\eta_{ct}}{0.47}\right) 
\left(\frac{S_0(x_c,x_t)}{0.00221}\right) - 
0.14\times
\left(\frac{\eta_{cc}}{1.43}\right) 
\left(\frac{x_c}{0.000250}\right) 
\right]
\Biggr\} .
  \end{aligned}
\end{equation}
The definition of each symbol and its value
can be found in the above references. Notice that the $\sin\beta$, $\sin2\beta$ appearing in this equation refer to the UT angle $\beta$, not to the ratio of the Higgs vevs in SUSY.
The denominator of each fraction is the central value
of the input parameter appearing in the numerator, quoted in
Ref.~\cite{Buras:2009pj}.
Major uncertainties are those in
$\widehat{B}_K$, $|V_{cb}|^4$, $R_t^2$, which are
5\%, 11\%, 7\%, respectively. 
One important refinement made in this expression,
which had been overlooked in most of the literature,
is the factor $\kappa_\epsilon$ that parameterizes suppression of
the result due to
$\phi_\epsilon \neq \pi/4$ and the imaginary part of
the 0-isospin amplitude of $K\rightarrow\pi\pi$.
In addition, latest lattice values of $\widehat{B}_K$
are lower than previous determinations.
Here we take $\widehat{B}_K = 0.720(13)(37)$
\cite{Antonio:2007pb,Allton:2008pn} (see also \cite{Aubin:2009jh}).
These two facts cooperate to make $|\epsilon_K^\text{SM}|$ insufficient for the central values,
obtaining $|\epsilon_K^\text{SM}| = 1.78\times 10^{-3}$, which is smaller 
than the observed $|\epsilon_K|$
by 18\%. This deficit should be compensated by some new source;
since our model has extra sources of CP violation, we are lead to invoke these
new sources to resolve the puzzle.

Before we proceed further,
a clarification is in order regarding the $\epsilon_K$ puzzle.
It stems from the tension among different observables
used to determine the UT on the
$(\overline{\rho}, \overline{\eta})$ plane.
The regions preferred by $\epsilon_K$,
$S_{\psi K}$, and $\Delta M_{B_s} / \Delta M_{B_d}$ do not precisely
overlap with one another (see Ref.~\cite{Lunghi:2008aa} for example).
As we said, if one accepts the regions determined by
$S_{\psi K}$ and $\Delta M_{B_s} / \Delta M_{B_d}$, then
$\epsilon_K$ should be modified.
Conversely, one may also conceive a scenario where
the $\epsilon_K$ region is in fact correct whereas
one or more of the $B$ sector observables have been contaminated
by new physics effects.
This could be an equally legitimate solution in a general case.
However, it is not viable within our framework.
The reason is that supersymmetric contributions to the above three
observables are (up to $O(1)$ uncertainties) correlated and that
the $\epsilon_K$ region is modified much more than the other two.
We shall add more quantitative comments on this in the next subsection.

In order to evaluate
the supersymmetric effects on $\epsilon_K$, we should calculate
$ M^\text{SUSY}_{K12} = \langle K|H^\text{SUSY}_\text{eff}|\overline{K}\rangle$.
For this, we follow Ref.~\cite{Ciuchini:1998ix}
which presents the $\Delta S = 2$ Wilson coefficients from
the gluino-squark box graphs (which are the largest SUSY contributions in the
presence of sizeable squark mass insertions) 
and the expressions for evolving those Wilson coefficients
 from the sparticle mass scale down to the hadronic scale.
After that, the SUSY contribution to $\epsilon_K$ is given by the formula,
\begin{equation}
  \epsilon_K^\text{SUSY} = e^{i \phi_\epsilon} \sin\phi_\epsilon
  \ \frac{\mathrm{Im} (M^\text{SUSY}_{K12})}{\Delta M_K}.
\end{equation}
And then we can express the supersymmetric contribution to $\epsilon_K$
in terms of mass insertions as,
\begin{equation}
  \label{eq:epsKSUSY}
  \begin{aligned}
  \frac{\epsilon_K^\text{SUSY}}{\epsilon_K^\text{exp}} =
  \sqrt{2} &\sin\phi_\epsilon\
  {\rm Im} \Biggl\{
  \frac{(\delta^d_{12})_{LL}^2 + (\delta^d_{12})_{RR}^2}{0.0062^2}
  \left(\frac{\widehat{B}_K}{0.72}\right)
  \left[
    2.2 \left(\frac{\tilde{f}_6(x)}{(-1/30)}\right) -
    1.2 \left(\frac{x f_6(x)}{1/20}\right)
  \right]
 -
 \\
\phantom{\Biggl\{}&
  \frac{(\delta^d_{12})_{LL} (\delta^d_{12})_{RR}}{0.00013^2}
  \left(\frac{B_4(\mu)}{0.70}\right)
  \left[
    0.05 \left(\frac{\tilde{f}_6(x)}{(-1/30)}\right) +
    0.95 \left(\frac{x f_6(x)}{1/20}\right)
\right]
  \Biggr\}
  \left(\frac{500\ \mathrm{GeV}}{m_{\tilde{q}}}\right)^2 ,
  \end{aligned}
\end{equation}
where $x \equiv m^2_{\tilde{g}} / m^2_{\tilde{q}}$ and
functions $f_6(x)$ and $\tilde{f}_6(x)$ can be found in
Ref.~\cite{Ciuchini:1998ix}.
We have omitted the term proportional to $B_5(\mu)$ which
happens to contribute only $O(1\%)$.


In the numerical analysis, we set $\widehat{B}_K = 0.72$
\cite{Antonio:2007pb,Allton:2008pn}
and use the central values of $B_{2,\ldots,5}(\mu)$
from Ref.~\cite{Nakamura:2006eq}.

In  our flavour models, RVV1 and RVV2, we have
$|(\delta^d_{12})_{RR}| \sim \bar\varepsilon^3 \simeq 0.003$ and
$|(\delta^d_{12})_{LL}| \sim \varepsilon^2\bar\varepsilon \simeq 0.0004$ with
different phases in the two models, $\omega_{us} \sim O(\bar \varepsilon)$ and
$\omega_{us}^\prime \sim O(1)$ respectively, as can be seen from 
Eq.~(\ref{sckm1}) and the paragraph following Eq.~(\ref{rvv2-12}). 
Therefore, we can expect that the supersymmetric contribution to
$\epsilon_K$ can easily be comparable to $\epsilon_K^\text{exp}$,
in particular from the contribution in the second term in 
Eq.~(\ref{eq:epsKSUSY}).
The additional contribution might well be just what we need
to fill the gap between $\epsilon_K^\text{SM}$ and
$\epsilon_K^\text{exp}$, which could amount to 18\% or more.
\begin{figure}
\includegraphics[scale=.5]{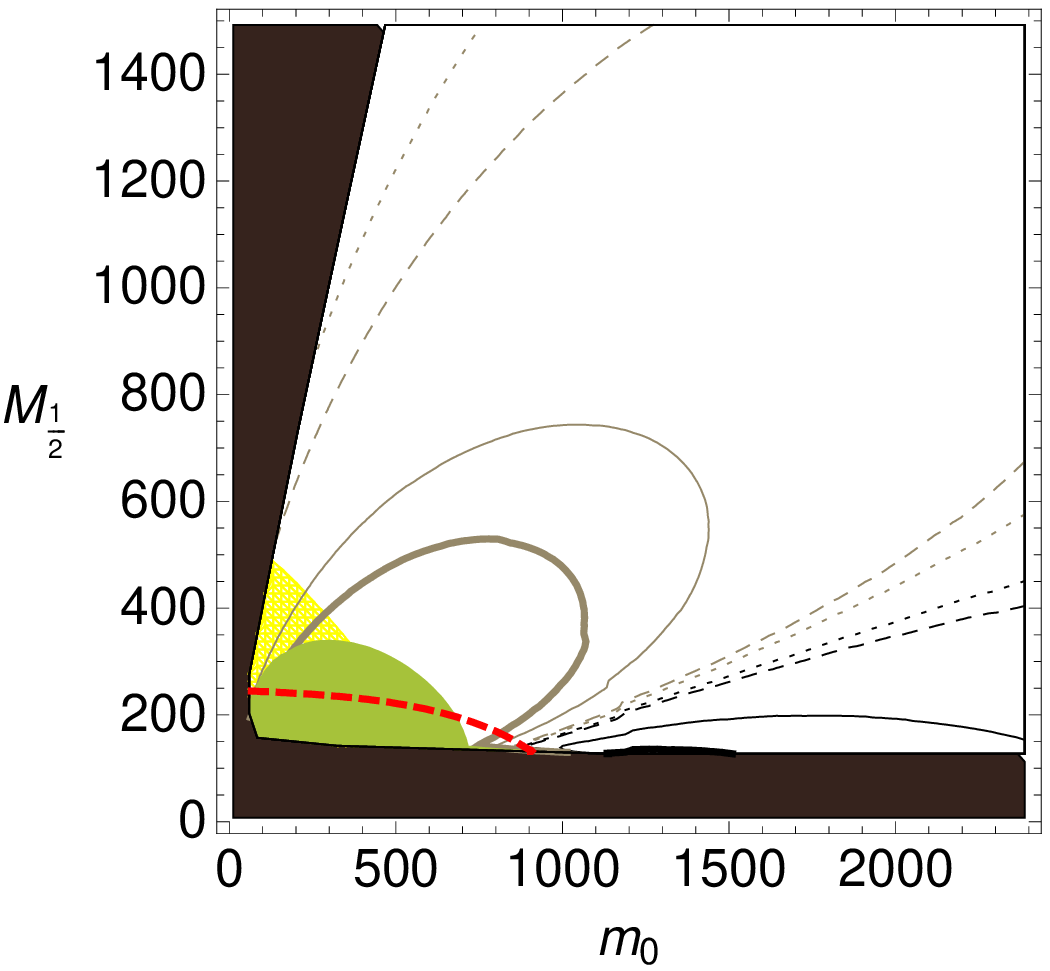}
\includegraphics[scale=.5]{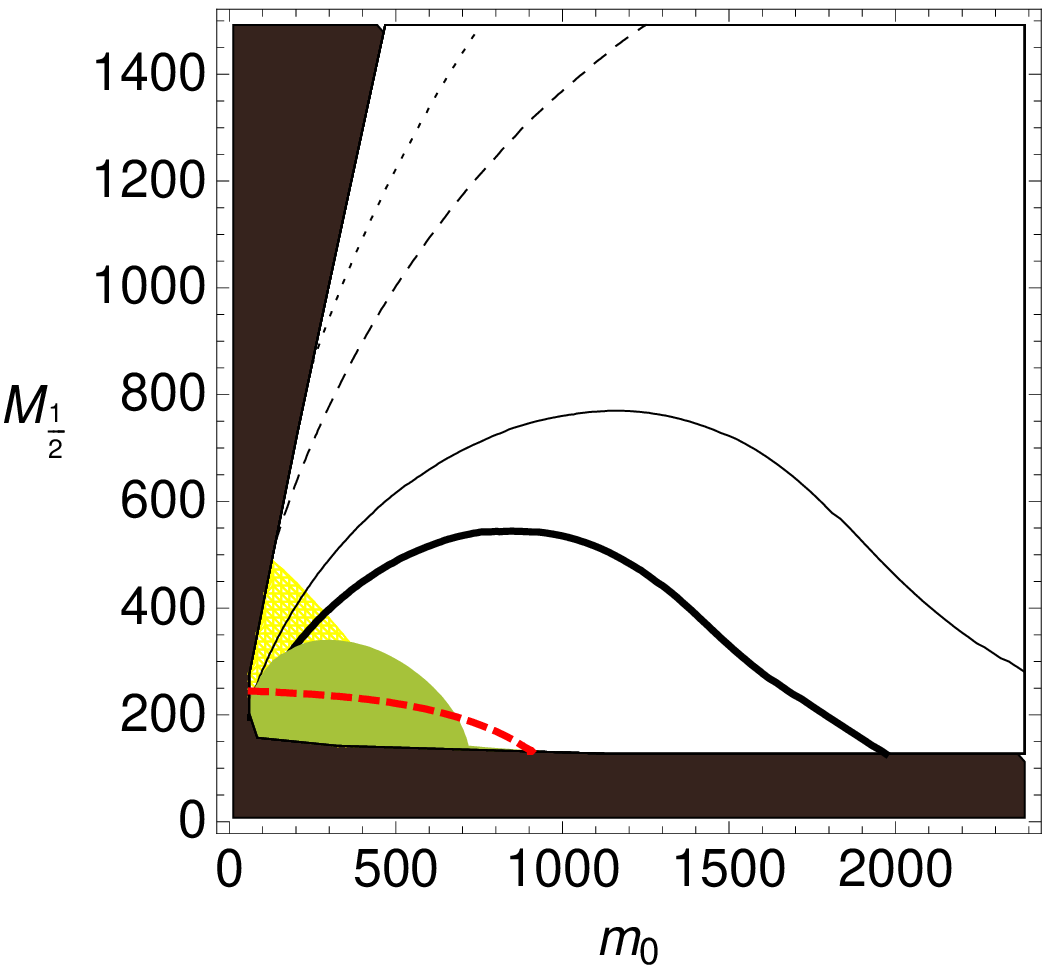}
  \caption{Contours of $|\epsilon_K^\text{SUSY}|$ for RVV1,
in the $m_0$-$M_{1/2}$ plane for $\tan\beta =10$ and $A_0=0$.
The lines are (from thicker to dotted)
$|\epsilon_K^\text{SUSY}| = 10^{-3}, 5\times 10^{-4}, 10^{-4}, 5\times10^{-5}$,
respectively.
 We show predictions for positive (left) and negative (right) O(1)'s
 for $M_{\tilde d_R^c}^2$, giving in some cases $\epsilon_K^{SUSY}$
with the wrong phase, which are shown in grey. Current LFV bounds are also shown in green, and the $(g-2)_\mu$ favoured region is shown hatched in yellow. The dark brown region show areas excluded by having a charged LSP or by LEP, excepting the Higgs mass bound, which is shown in thick dashed red lines.}
  \label{fig:epsK}
\end{figure}
In Figs.~\ref{fig:epsK},
we present contours of $\epsilon_K$ on the $(m_0, M_{1/2})$ plane for RVV1.
The left and the right figures are for positive and negative O(1)'s
for $M_{\tilde d_R^c}^2$, respectively.

The phase of $\epsilon_K^\text{SUSY}$ can be either
$43.51^\circ$ or $-136.49^\circ$ which are respectively marked in
black and gray in Figs.~\ref{fig:epsK}.
Note that the phase of $\epsilon_K^\text{SM}$ is $43.51^\circ$
and that $\epsilon_K^\text{SUSY}$ and $\epsilon_K^\text{SM}$ should
interfere constructively to fit the experimental value.
Therefore, gray contours worsen compatibility between
the theoretical and the experimental values of $\epsilon_K$
by increasing their discrepancy.
The phase of $\epsilon_K^\text{SUSY}$ is determined by the sign
of Eq.~(\ref{eq:epsKSUSY}).
It is often dominated by the term proportional to
$\mathrm{Im}[(\delta^d_{12})_{LL} (\delta^d_{12})_{RR}]$.
In this case,
the phase of $\epsilon_K^\text{SUSY}$ can be shifted by $\pi$
if one multiplies either
$(\delta^d_{12})_{LL}$ or $(\delta^d_{12})_{RR}$ by $-1$ (of course, the modulus $|\epsilon_K^\text{SUSY}|$ does not remain
exactly the same due to the first term).
For example, we can flip the sign of the RR insertion
by changing the signs of the O(1) coefficients in $M_{\tilde d_R^c}^2$.
This is the reason why the region with gray contours
on the left plot turns to black on the right.

For RVV2, the phase in $(\delta^d_{12})_{LL}$ is much larger than that in RVV1, as we can see in Eq.~(\ref{rvv2-12}). This makes the SUSY $(\delta^d_{12})_{RR}(\delta^d_{12})_{LL}$ contribution to $\epsilon_K^{\rm SUSY}$ dominant in all of the evaluated parameter space, so only one of the two possible phases for $(\delta^d_{12})_{RR}$ solves the $\epsilon_K$ tension. Furthermore, each of the contours in Figure~\ref{fig:epsK} would be scaled roughly by a factor $1/\bar\varepsilon\simeq6$.


The requirement that the supersymmetric contribution solves
the $\epsilon_K$ puzzle would define a strip in this plane
which will be shown in all of the following plots, for both RVV1 and RVV2.
One virtue of such a strip is that
the parameter space therein will lead to a definite correlation
between observables which are not apparently related to each other.
We will elaborate on this in Sec.~\ref{sec:combined}.

Another important quantity in kaonic CP violation is
$\epsilon'/\epsilon$.
It is specially sensitive to chirality-flipping mass insertions,
$(\delta^d_{12})_{LR}$ and $(\delta^d_{12})_{RL}$ \cite{Gabbiani:1996hi}.
Using the formulae in Ref.~\cite{Buras:1999da},
one can write
the supersymmetric contribution to $\epsilon'/\epsilon$ in the form,
\begin{equation}
\frac{(\epsilon')^\text{SUSY}}{(\epsilon')^\text{exp}} =
{\rm Im} \left[
\frac{(\delta^d_{12})_{LR}-(\delta^d_{12})_{RL}}{1.6\times 10^{-5}}
\times B_G
\right]
\left(\frac{500\ \mathrm{GeV}}{m_{\tilde{q}}}\right) ,
\end{equation}
which has been derived for $m^2_{\tilde{g}} / m^2_{\tilde{q}} = 1$.
The $O(1)$ constant $B_G$ parameterizes the uncertainty in
the hadronic matrix element of the chromomagnetic operator.
We have suppressed other factors that can
change by $O(1)$ for $m^2_{\tilde{g}} / m^2_{\tilde{q}} \neq 1$,
since the following argument depends only on the order of magnitude
of $\epsilon'/\epsilon$.

In this model, there can be flavor-violating A-terms
arising from mismatch of the $O(1)$ coefficients between
the Yukawas and the A-terms.
To estimate their possible effects,
we define  $A_0 \equiv a_0~ m_0$, with $A_0$ being
the overall dimensionful coefficient of the A-terms.
Using Eq.~(\ref{aterms}b) and (\ref{sckm1}b),
one can get rough relationships among different 1--2 mass insertions,
\begin{equation}
  {(\delta^d_{12})_{LR}} \sim 
  {(\delta^d_{12})_{RL}} \sim 
  a_0~\frac{m_b}{m_0} \times {(\delta^d_{12})_{RR}}
  \sim a_0~
  \frac{\bar\varepsilon^2}{\varepsilon^2} \frac{m_b}{m_0} \times
  {(\delta^d_{12})_{LL}} ,
\end{equation}
where $m_b$ is the $b$-quark mass at the GUT scale.
If we require that the size of the second term in Eq.~(\ref{eq:epsKSUSY})
does not exceed unity, we find that the extra contribution to
$\epsilon'/\epsilon$ has an upper limit like
\begin{equation}
\left|\frac{(\epsilon')^\text{SUSY}}{(\epsilon')^\text{exp}}\right|
\lesssim 8.1\times \frac{\bar\varepsilon}{\varepsilon}~ a_0~ \frac{m_b}{m_0} ,
\end{equation}
for purely imaginary $(\delta^d_{12})_{LR}$ or $(\delta^d_{12})_{RL}$.
In this case, the supersymmetric fraction within $\epsilon'/\epsilon$
should be lower than 40\% for a reasonably high
$m_0 \gtrsim 100\ \mathrm{GeV}$.
In fact, the phases of these flavor and chirality
changing insertions are $\pm\omega_{us}$ which are smaller than maximal,
as can be seen from Eq.~(\ref{omega_us}).
Thus their effects on $\epsilon'/\epsilon$
are smaller than the above estimate.
We have numerically checked that
the gluino-squark loop contribution to $\epsilon'/\epsilon$
is less than 15\%
on the strip compatible with $\epsilon_K$.
Given the large theoretical uncertainty in $\epsilon'/\epsilon$,
this amount of contamination by new physics should be hard to disentangle.

Another process that could, in principle, be interesting is the decay
$K_L\to\pi^0\nu\bar\nu$~\cite{Nir:1997tf,Buras:1997ij,Colangelo:1998pm}. The
dominant contribution to this process requires the presence of CP violating
phases, and therefore we could expect to have a contribution in our model. As,
due to gauge invariance,
this decay depends on $SU(2)_L$-breaking, it requires the presence of two Higgs vevs, either within $\delta_{LR}$ insertions in the squark sector, or within gaugino-higgsino mixing on the neutralino and chargino sectors. At moderate $\tan\beta$, this disfavours the gluino contribution, since $\delta^d_{LR}$ insertions are, at most, proportional to the bottom mass. Of much more interest is the chargino contribution, as the $(\delta^u_{33})_{LR}$ insertions are proportional to the top mass. Nonetheless, the required $(\delta^u_{23})_{LL}$ and $(\delta^u_{31})_{RR}$ insertions depend on powers of $\varepsilon$, instead of $\bar\varepsilon$, which constrains greatly any chargino contribution dependent on the $SU(3)$ structure. The chargino contribution will then be dominated by the flavour-changing of the CKM matrix, and thus cannot present any larger deviations than those predicted by MFV models.

\subsection{ $B$ physics}

Let us now discuss new physics effects on $B_s$ mixing.
For this, it is convenient to adopt the following parametrization~\cite{Bona:2008jn}:
\begin{equation}
  \label{eq:phiBs}
  C_{B_s} e^{2 i \phi_{B_s}} =
  \frac{M^\text{SM}_{s12}+M^\text{SUSY}_{s12}}{M^\text{SM}_{s12}} .
\end{equation}
The $B_s$--$\overline{B_s}$ transition amplitude is divided into
two parts, one arising from the SM loops and the other from
the gluino-squark which are the largest SUSY contributions in our model:
\begin{equation}
  \label{eq:M12s}
  M^\text{SM}_{s12} =
  \langle B_s| H^\text{SM}_\text{eff} |\overline{B_s}\rangle, \quad
  M^\text{SUSY}_{s12} = 
  \langle B_s| H^\text{SUSY}_\text{eff} |\overline{B_s}\rangle .
\end{equation}
Here, we focus on the phase $\phi_{B_s}$ rather than on $C_{B_s}$, since
the hadronic uncertainty in the latter is larger
than the extra contributions that can be expected in this model.

The current data of $B_s$ mixing phase is showing an interesting deviation
 from the SM prediction.
A constrained fit, performed by HFAG, results in
\cite{Barberio:2008fa}:
\begin{equation}
  \label{eq:phiBsHFAG68}
  \phi_{B_s} = -0.36^{+0.19}_{-0.17} \text{ \ or \ }
  \mbox{$-1.17$}^{+0.17}_{-0.19} .
\end{equation}
Recall that non-vanishing $\phi_{B_s}$ is an indication of new physics.
The above fit is away from the SM at the level of $2.4\ \sigma$,
and the 90\% CL range is:
\begin{equation}
  \label{eq:phiBsHFAG90}
  \phi_{B_s} \in [-0.61, -0.045] \cup [-1.48, -0.92] .
\end{equation}
It would be amusing to see whether or not our model could push
this phase close to its best fit value.

In order to estimate maximal size of $\phi_{B_s}$ within this model,
one should consider the ratio of
$M^\text{SUSY}_{s12}$ to $M^\text{SM}_{s12}$ appearing in Eq.~(\ref{eq:phiBs}).
We follow Ref.~\cite{Becirevic:2001jj} to express
the supersymmetric amplitude in terms of mass insertions.
The SM amplitude is available in Ref.~\cite{Buchalla:1995vs} for instance.
For this, we use $V_{ts}$ and $V_{tb}$ from the full fit
by the UTfit collaboration \cite{UTfit}.
The result can be written in the form,
\begin{equation}
  \label{eq:M12Bs-mia}
  \frac{M^\text{SUSY}_{s12}}{M^\text{SM}_{s12}} =
  e^{2 i \beta_s} \left[
  \frac{(\delta^d_{23})_{LL}^2 + (\delta^d_{23})_{RR}^2}{0.71^{2}} -
  \frac{(\delta^d_{23})_{LL} (\delta^d_{23})_{RR}}{0.062^{2}}
  \right]
  \left(\frac{500\ \mathrm{GeV}}{m_{\tilde{q}}}\right)^2 ,
\end{equation}
where we have taken the ratio $m^2_{\tilde{g}} / m^2_{\tilde{q}} = 1$,
since we need only the order of magnitude of the above ratio.

\begin{figure}
\includegraphics[scale=.5]{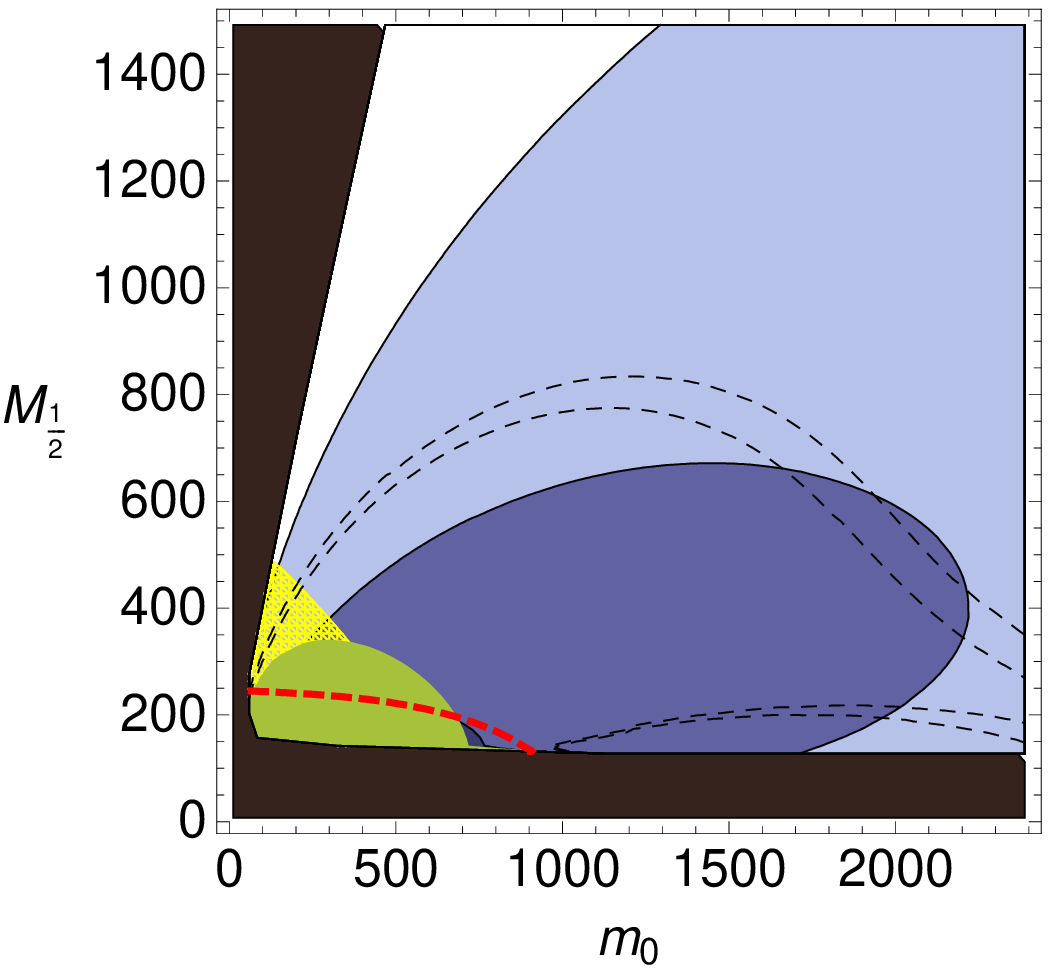} \qquad \includegraphics[scale=.5]{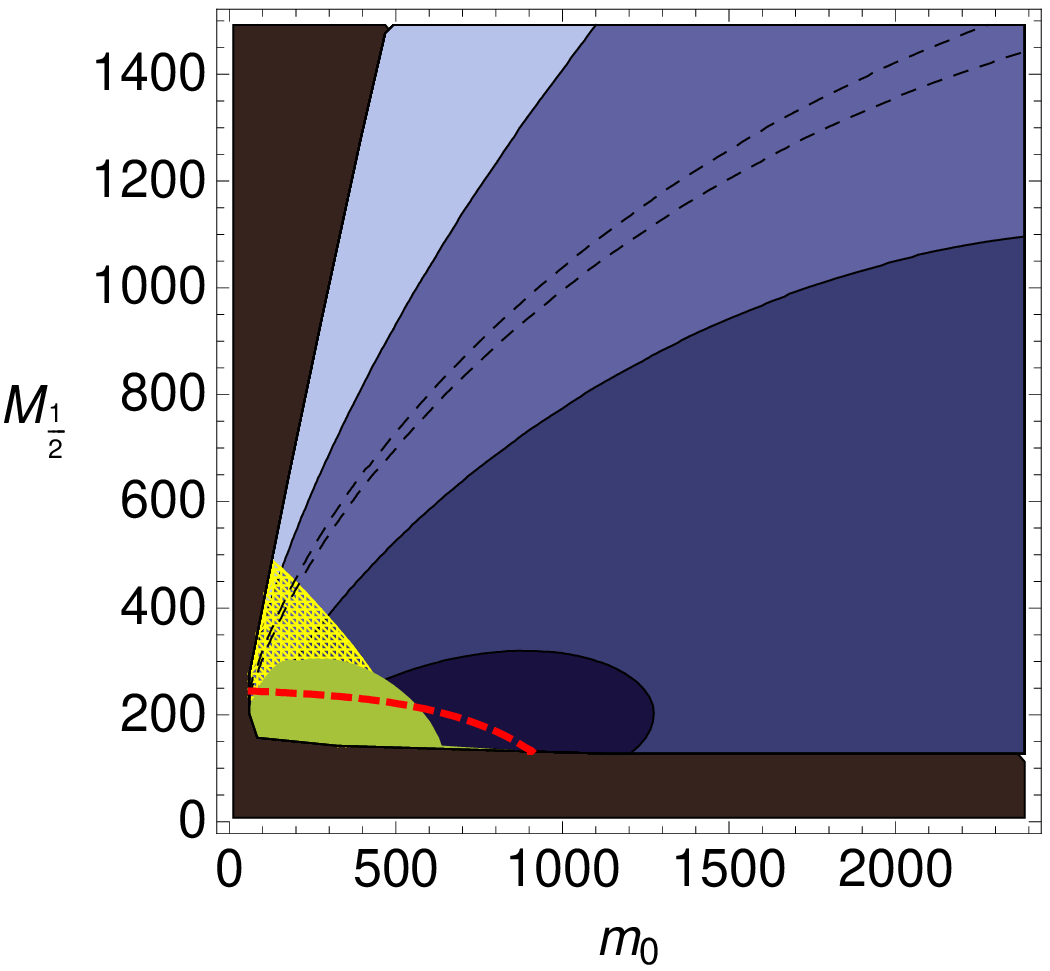} \\
\includegraphics[scale=.5]{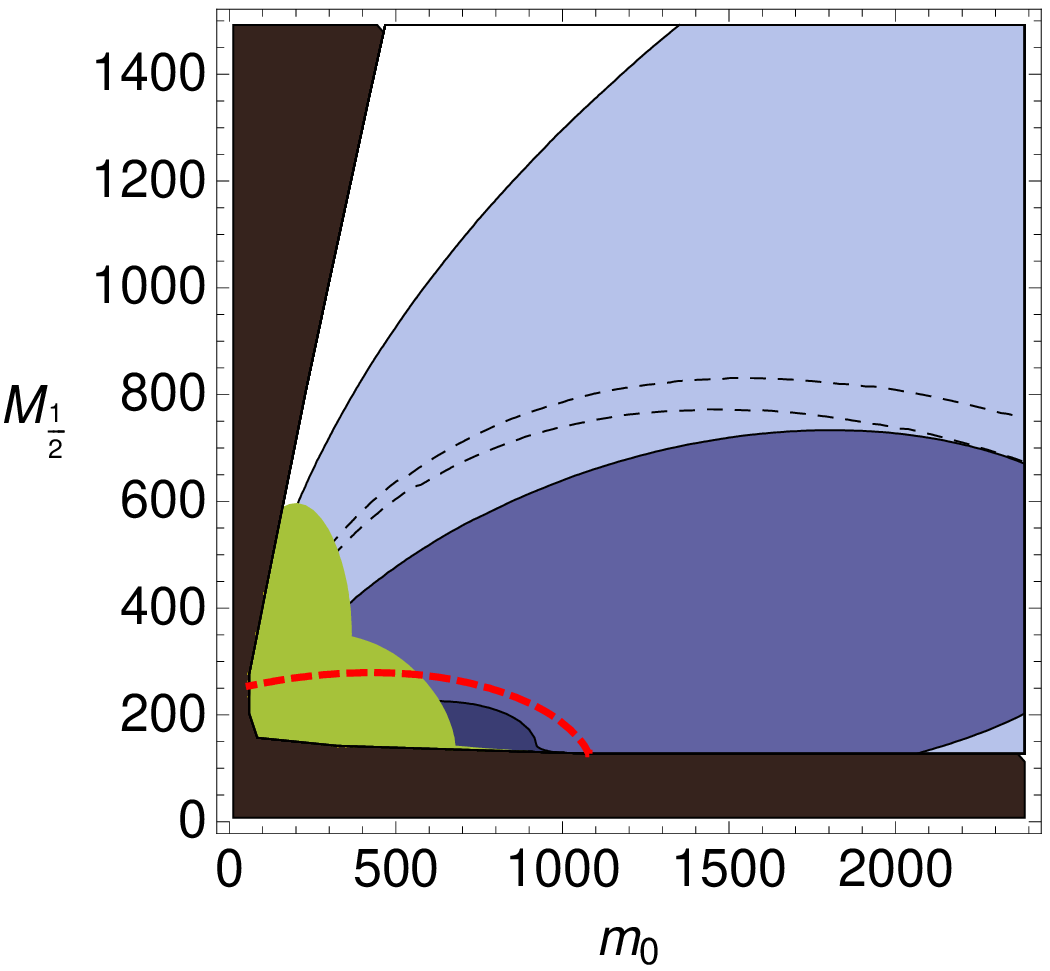} \qquad \includegraphics[scale=.5]{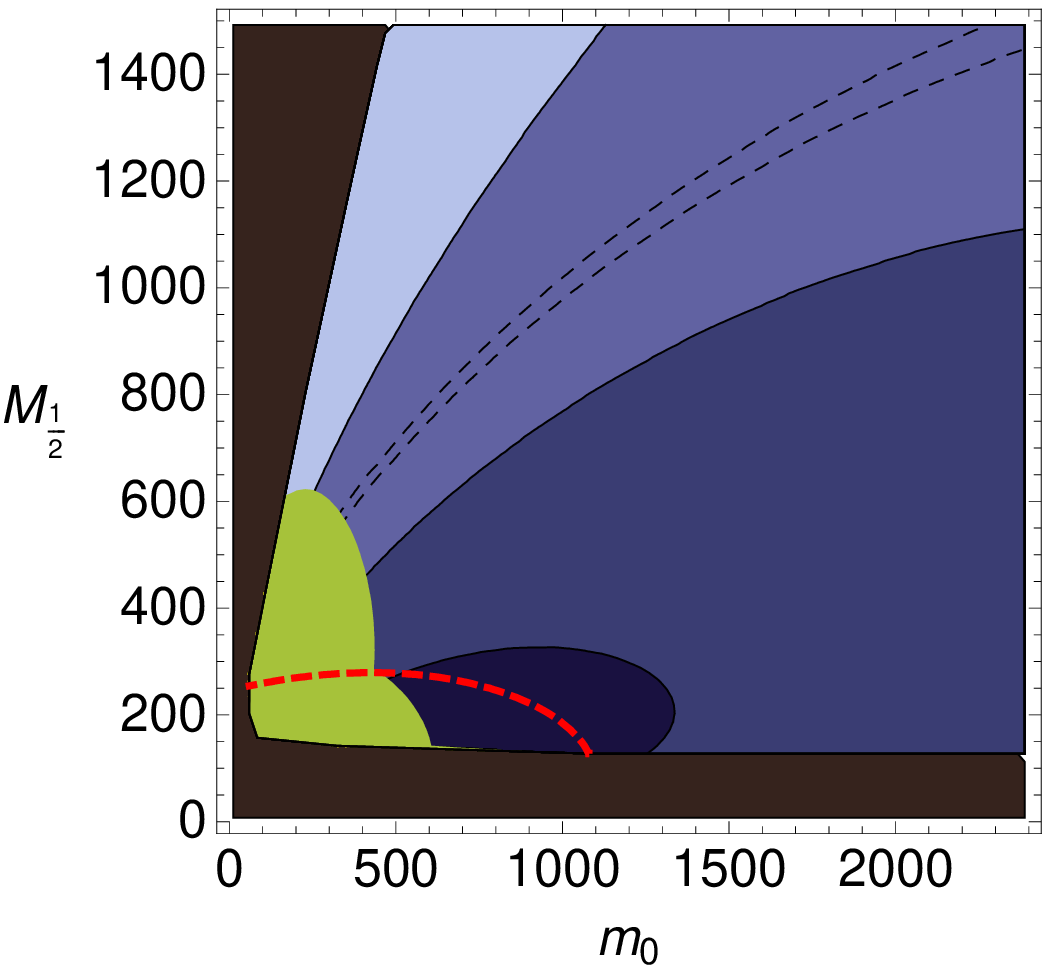} \\
  \caption{Contours of $\phi_{B_s}$ for RVV1 (left) and RVV2 (right)
in the $m_0$-$M_{1/2}$ plane for $\tan\beta =10$ and $A_0=0$ (top) and $A_0=m_0$ (bottom).
The color of an area indicates the size of $\phi_{B_s}$,
which decreases as one moves
in the order of dark to light blue.
Each line between two adjacent areas
is one of $\phi_{B_s}= 10^{-2}, 10^{-3}, 10^{-4}, 10^{-5}$ (the contour
corresponding to $\phi_{B_s} = -0.045$, which would solve the discrepancy, is
within the LFV excluded region in RVV2, and forbidden by direct bounds in
RVV1). As in previous plots, green areas
correspond to the currect LFV constraints and yellow area to the $(g-2)_\mu$
favoured region. The area between the dashed black lines solve the $\epsilon_K$ tension, and the dark brown region show areas excluded by having a charged LSP or by LEP, excepting the Higgs mass bound, which is shown in thick dashed red lines.}
  \label{fig:phiBs}
\end{figure}

The first factor on the right hand side
comes from the phase of $M^\text{SM}_{s12}$
which is equal to $-2 \beta_s = -0.04$.
Now, from  \eq{eq:phiBs}, it is clear that a change in the $B_s$ phase, ${\rm
  Arg} \left[M^\text{SM}_{s12} + M^\text{SUSY}_{s12}\right]$, would
require $|M^\text{SUSY}_{s12}/M^\text{SM}_{s12}| \simeq O(1)$ and a sizeable
phase $\phi_{B_s}^{\rm SUSY} \equiv {\rm Arg}
\left[M^\text{SUSY}_{s12}/M^\text{SM}_{s12}\right]$. 
This fact has strong implications on the size and phases of the mass 
insertions appearing above.
These mass insertions differ in different variations of our model,
for instance, that largest MIs can be expected in RVV2.
In this model, the mass insertions are
$(\delta^d_{23})_{LL} \sim \varepsilon \simeq 0.05$ and
$(\delta^d_{23})_{RR} \sim \bar\varepsilon y_b^{0.5} \simeq
0.015 \sqrt{\sec\!\beta}$,
as one can find in Eqs.~(\ref{sckm2}).
The size of the RR insertion depends on $\tan\beta$ and 
taking $\tan\beta = 10$ for example, we have
$(\delta^d_{23})_{RR} \sim 0.05$.
These values appear to be large enough to make a significant change
in the $B_s$ phase from the second term of Eq.~(\ref{eq:M12Bs-mia}).
Moreover, the product of these two insertions can have an $O(1)$ phase.
Notice that this is possible only in RVV2
where the leading terms in these mass insertions have phases. In 
Fig.~\ref{fig:phiBs}, we show the different contours of $\phi_{B_s}$.

Indeed, in model RVV2 there is a region on the 
$(m_0, M_{1/2})$ plane
where it is possible to shift $\phi_{B_s}$ into the interval in 
Eq.~(\ref{eq:phiBsHFAG90}), at relatively high $m_0$ and low $M_{1/2}$
\cite{Dutta:2008xg,Ko:2008zu}.
However, these regions of large SUSY contributions to $\phi_{B_s}$ are
excluded by LFV and tend to
have a too large contribution to $\epsilon_K$ as well.
If we confine ourselves on a strip allowed by $\epsilon_K$, the
maximal value of $\phi_{B_s}$ is around $10^{-4}$.
From the plots we can see that, in the case of RVV1, we are again in the same
situation and for the strips allowed by $\epsilon_K$ we always have
$\phi_{B_s} \lsim 10^{-4}$. Although these values could in
principle vary due to the unknown $O(1)$ coefficients in the soft terms, we
can not expect variations in the order of magnitude.
Thus, for low and moderate $\tan\beta$, our models are not able to provide a solution to the $\phi_{B_s}$
anomaly, and, in this situation, we would expect this anomaly to 
disappear with the inclusion of further data\footnote{The authors of \cite{Altmannshofer:2009ne} claim to be able to satisfy the $\Phi_{B_s}$ anomaly and the $\epsilon_K$ tension in RVV2. We believe these points correspond to a large value of $\tan\beta\gtrsim50$. Moreover, we checked that the parameters can conspire to give $\Phi_{B_s}$ of $O(0.1)$, even for moderate values of $\tan\beta$.}. If this is the case,
our predicted deviation should also be hard to observe at LHCb
since it is much smaller than
the precision of $\phi_{B_s}$ attainable
after 5 years of run ($10\ \mathrm{fb}^{-1}$), which is estimated to be
$5\times 10^{-3}$ \cite{phisLHCb}.

Similarly, we can repeat a parallel discussion on $B_d$ mixing with the 
following parameterization,
\begin{equation}
  \label{eq:phiBd}
  C_{B_d} e^{2 i \Delta\beta} =
  \frac{M^\text{SM}_{d12}+M^\text{SUSY}_{d12}}{M^\text{SM}_{d12}} ,
\end{equation}
where each $M_{d12}$ is the same as that in Eqs.~(\ref{eq:M12s})
with the replacement $s \rightarrow d$.
The supersymmetric to SM contribution ratio is given by
\begin{equation}
  \label{eq:M12Bd-mia}
  \frac{M^\text{SUSY}_{d12}}{M^\text{SM}_{d12}} =
  e^{-2 i \beta} \left[
  \frac{(\delta^d_{13})_{LL}^2 + (\delta^d_{13})_{RR}^2}{0.15^2} -
  \frac{(\delta^d_{13})_{LL} (\delta^d_{13})_{RR}}{0.013^2}
  \right]
  \left(\frac{500\ \mathrm{GeV}}{m_{\tilde{q}}}\right)^2 .
\end{equation}
One can get this result in the same way as for Eq.~(\ref{eq:M12Bs-mia}).

In this case, each denominator in the square brackets is smaller than
the corresponding one in Eq.~(\ref{eq:M12Bs-mia})
by the factor $|V_{td}/V_{ts}|^2$.
Again, we use the CKM matrix elements from the full fit
by the UTfit collaboration \cite{UTfit}.
In RVV2, we have
$(\delta^d_{13})_{LL} (\delta^d_{13})_{RR} \sim \varepsilon\bar\varepsilon^3\,y_b^{0.5}
\sim 1.7\times 10^{-5} \sqrt{\sec\!\beta}$, and thus we could expect a sizeable effect on
$S_{\psi K}$ if it were not for $\epsilon_K$.
However, $\Delta S_{\psi K}$ is suppressed below $1\times 10^{-3}$
on an $\epsilon_K$ strip in the same way as $\phi_{B_s}$ is.
This is below the sensitivity of a super B factory to $S_{\psi K}$
whose estimate is around $5\times 10^{-3}$ \cite{Bona:2007qt}.
In RVV1, the new physics effect is even smaller, since the phases are suppressed.
In the end, $B_d$--$\overline{B_d}$ mixing
is not very much affected in both models.
This fact, a posteriori, justifies
the way we determine the $O(1)$ coefficients in Yukawas:
we tune the coefficients so that they reproduce the CKM matrix elements
which were obtained under the assumption of no new physics.

Let us come back to the alternative solutions of the $\epsilon_K$ puzzle
that we mentioned in the previous subsection.
If one is to blame the UT fit tension on $S_{\psi K}$,
one would need a change due to new physics of the amount
$\Delta S_{\psi K} \sim 0.07$ \cite{Lunghi:2008aa}.
However, this is two orders of magnitude bigger than
what can be maximally expected in our models when
$\epsilon_K^\text{SUSY}$ is at the level of 18\%,
as we have seen above.
In an alternative solution, $\epsilon_K^\text{SUSY}$ should be
more suppressed, which in turn suppresses
$\Delta S_{\psi K}$ as well, thereby making it
far less sufficient.
The other possibility of invoking modification of
$\Delta M_{B_s} / \Delta M_{B_d}$ does not work for a similar reason.
The authors of Ref.~\cite{Altmannshofer:2009ne}
need a new physics contribution to this ratio
at the level of $-22\%$ in order to get an exact agreement.
In our models however,
the fractional change is about 0.5\% at most,
which is too small compared to what is needed.

Finally, we can look to the expected values in our model for the decays
$b\to s\mu^+\mu^-$ and $b\to s\nu\bar\nu$ that are similar to 
$K_L\to\pi^0\nu\bar\nu$ in the $B$ sector. We find that the values of the branching ratios for these
processes, in both RVV1 and RVV2 models for $A_0 =0$ and $A_0 = m_0$,
do not produce deviations from the SM values larger than the per cent level,
 in spite of the presence of sizeable flavour changing mass insertions.
Therefore, these proceses are not interesting tests of new physics in our
scenario.  This is consistent with the work in~\cite{Yamada:2007me}, where it was found that
the deviation from the SM prediction is small
once the constraints from $b\to s \gamma$ and $B_s \to\mu^+\mu^-$ are taken into account.

\subsection{Electric Dipole Moments}
\label{sec:edms}

The EDMs of fermions, such as the electron and the neutron, provide very
stringent constraints on CP-violation in new physics.
In the SM, the electron EDM from the Kobayashi-Maskawa phase is predicted to
be  $\lsim O(10^{-40})$ $e\textrm{ cm}^{-1}$~\cite{SMeEDM}. The SM expectation
for the neutron EDM is of $O(10^{-31})$ $e\textrm{ cm}^{-1}$~\cite{SMnEDM},
assuming a vanishing $\theta_{QCD}$. We can see that EDMs are highly
suppressed in the SM, and thus they are excellent observables where to look for new physics.

The electron EDM was studied in~\cite{Calibbi:2008qt} within the context of
RVV1. In these models CP is spontaneously broken in the flavour
sector. Therefore, the phases in the $\mu$ parameter and diagonal $A_f$ terms
are very suppressed and can be neglected. In such a case, the imaginary parts
required for EDMs only appear from flavour-changing mass insertions~\cite{Hisano:2008hn}. For the
electron EDM, only neutralino-mediated diagrams contribute. The most important contribution to $d_e$, when $A_0=0$, comes from a bino-mediated diagram proportional to $\Im m\left[(\delta^e_{13})_{LL}(\delta^e_{33})_{LR}(\delta^e_{31})_{RR}\right]$, which is enhanced by $m_\tau\tan\beta$.

\begin{figure}
\includegraphics[scale=.5]{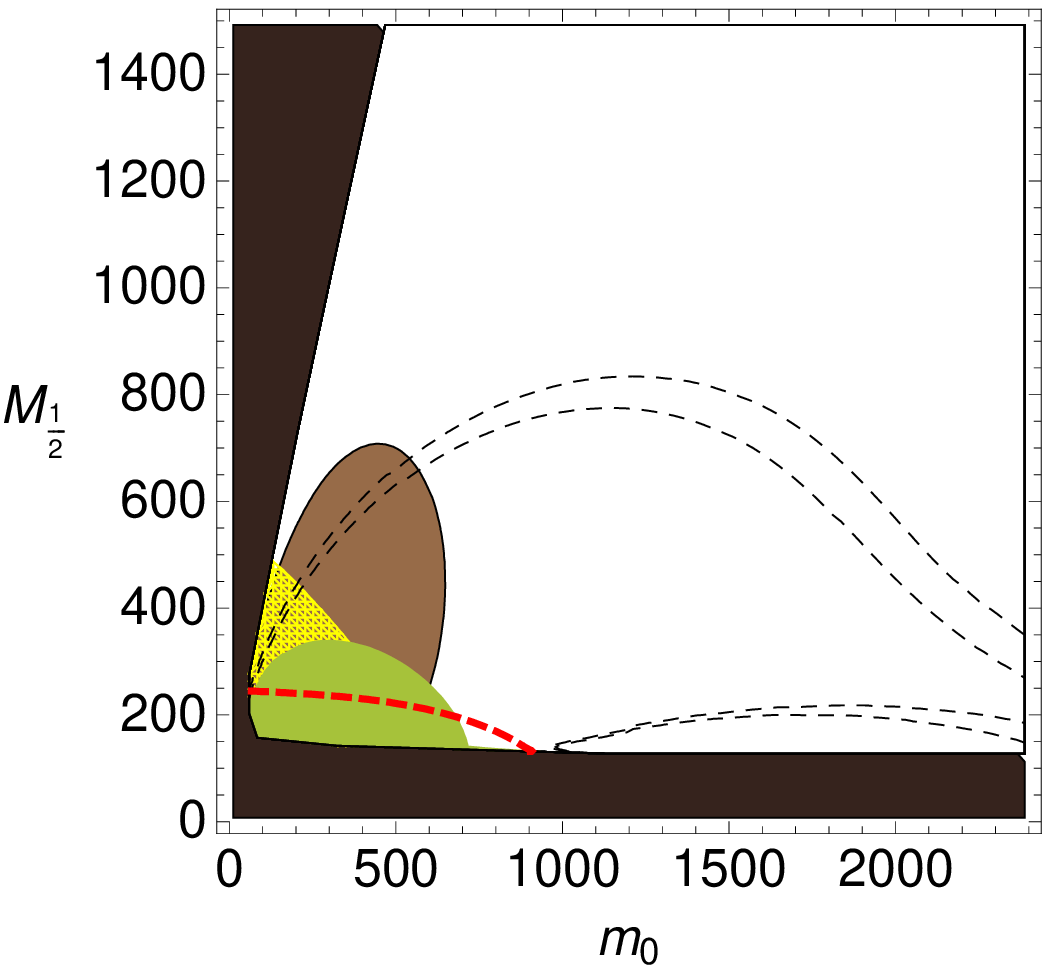} \qquad \includegraphics[scale=.5]{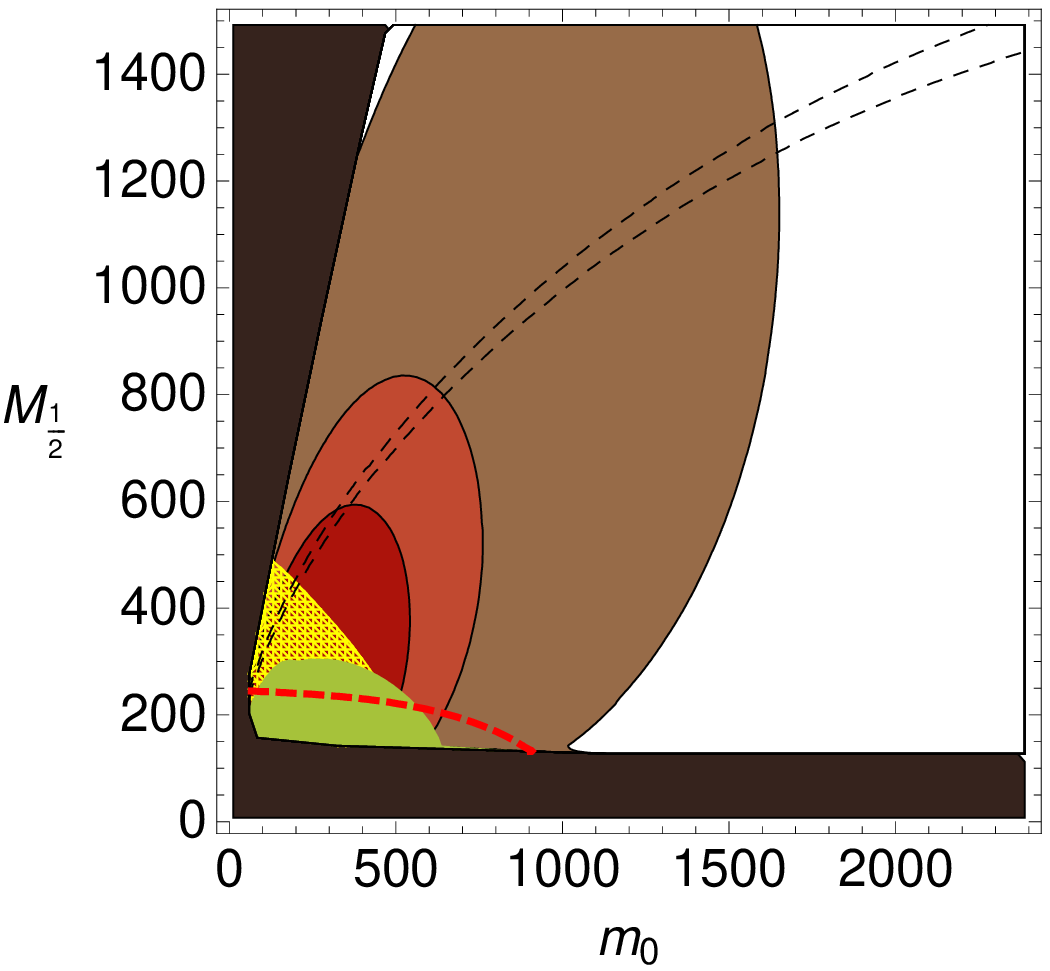} \\ 
\includegraphics[scale=.5]{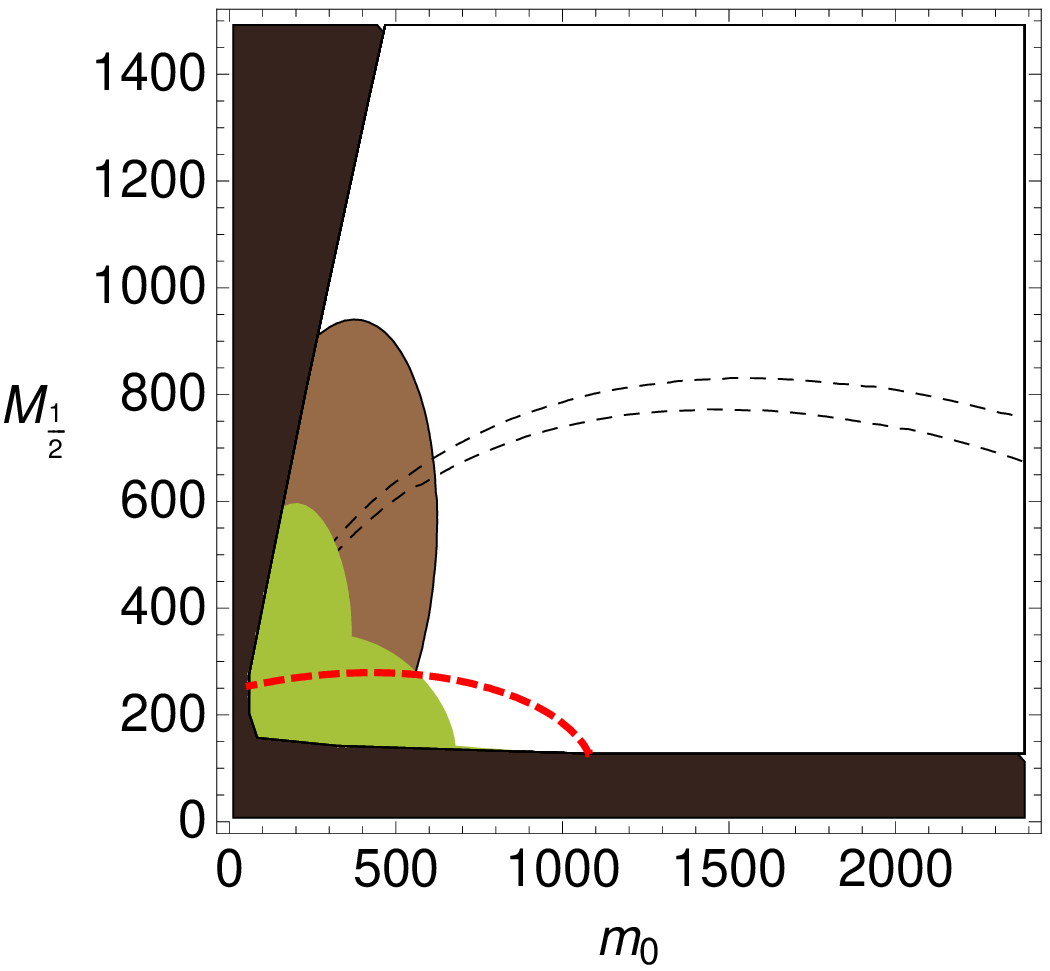} \qquad \includegraphics[scale=.5]{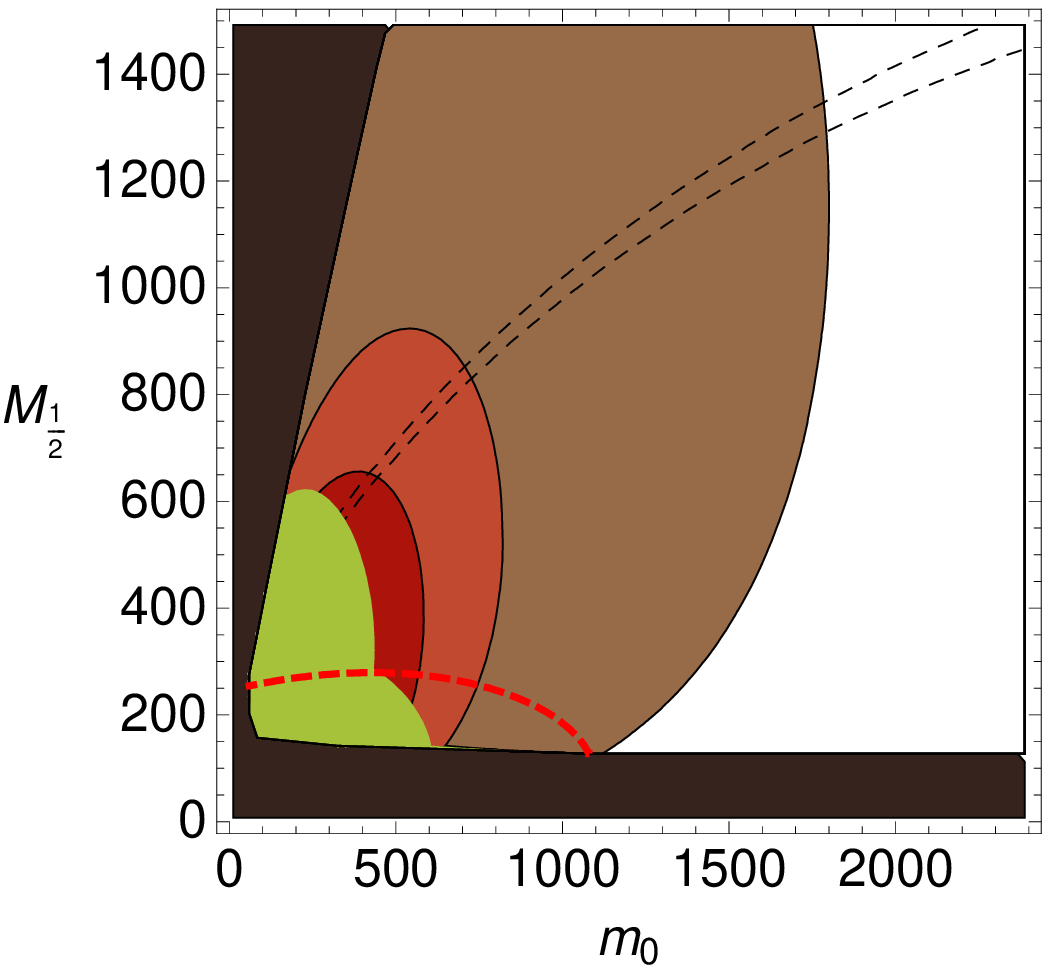} \\ 
\caption{Contours of $|d_e|=1\times10^{-28}$ e cm (dark red), $|d_e|=5\times10^{-29}$ e cm (light red) and $|d_e|=1\times10^{-29}$ e cm (brown) in the $m_0$-$M_{1/2}$ plane for $\tan\beta =10$ and $A_0=0$ (top), $A_0=m_0$ (bottom). We show predictions for RVV1 (left) and RVV2 (right). Current LFV bounds are also shown in green, and the $(g-2)_\mu$ favoured region is shown hatched in yellow. The area between the dashed black lines solve the $\epsilon_K$ tension, and the dark brown region show areas excluded by having a charged LSP or by LEP, excepting the Higgs mass bound, which is shown in thick dashed red lines.}
\label{fig:eedms}
\end{figure}

In this work, we have updated our analysis on the electron EDM, taking into
account the constraints on the $\beta_3$ and $\chi$ phases imposed by the fit
of CKM matrix and the $O(1)$ coefficients in $Y_e$ (see the Appendix for
details). The results are shown for the RVV1 and RVV2 models in
Figure~\ref{fig:eedms}. An inspection of the slepton soft mass matrices
indicates that, for RVV1, the leading phases in the product
$(\delta^e_{13})_{LL}(\delta^e_{33})_{LR}(\delta^e_{31})_{RR}$ cancel and only
subleading phases contribute, while RVV2 
has non-vanishing leading phases. We then have the following predictions:
\begin{eqnarray}
\label{EDMdel}
 (d_e)_{\textnormal{RVV1}} & \sim & \bar\varepsilon^6\,y_b\,y_t\sin(2(\chi-\beta_3)) \\
 (d_e)_{\textnormal{RVV2}} & \sim & \varepsilon\bar\varepsilon^3\frac{(y_b\,y_t)}{9}^{0.5}\sin(\chi-2\beta'_2) 
\end{eqnarray}

In Figure~\ref{fig:eedms} we present the sensitivity of current~\cite{Regan:2002ta}
and future~\cite{Lamoreaux:2001hb,FuteEDM} $d_e$ experiments, for the two models. In
this Figure, current constraints do not impose any restrictions on the
parameter space of either RVV1 or RVV2. Nevertheless, electron EDM predictions
are large enough to be probed at future EDM experiments. For relatively light
SUSY masses we obtain $d_e\sim10^{-29}$ $e\textrm{ cm}^{-1}$ and
$d_e\sim10^{-28}$ $e\textrm{ cm}^{-1}$, for RVV1 and RVV2, respectively. The
latter predicts a value of $d_e$ about one order of magnitude larger than
the former for any particular value of $m_0$ and $M_{1/2}$ due to the larger
$\varepsilon$ suppresion as seen in \eq{EDMdel}. This means that by reaching
$d_e\sim10^{-29}$ $e\textrm{ cm}^{-1}$ one could probe a much larger part of
the evaluated parameter space, with $m_0\lesssim1500$ GeV,
$M_{1/2}\lesssim2000$ GeV. In particular, for RVV2, observation of SUSY at the LHC
and solving the $\epsilon_K$ tension would force $d_e$ to be larger than
$10^{-29}$ $e\textrm{ cm}^{-1}$. However, we have to take into account that
these values will vary by factors $O(1)$ because of the unknown $O(1)$
coefficients to the different MIs.



We now turn to the neutron EDM, $d_n$. In this observable we have an additional difficulty, since its calculation as a function of partonic EDMs is not straightforward. Here we use two different approaches to this problem, the
quark-parton \cite{Ellis:1996dg} and chiral quark \cite{Manohar:1983md} models for $d_n$.

In the quark-parton model~\cite{Ellis:1996dg}, $d_n$ is expressed in terms of the up, down and strange quark EDMs, weighed by the fractional contribution $\Delta_n^q$ of each quark to the spin of the neutron:
\begin{equation}
d_n^{QP}=\eta^E\left(\Delta_n^u d_u + \Delta_n^d d_d + \Delta_n^s d_s\right)
\end{equation}
where $\eta^E=0.61$ is a QCD correction factor~\cite{Degrassi:2005zd} and
typical values for $\Delta_n^q$ are $\Delta_n^u=-0.508$, $\Delta_n^d=0.746$ and $\Delta_n^s=-0.226$.

The chiral quark model~\cite{Manohar:1983md} uses naive dimensional analysis to establish:
\begin{equation}
d_n^{CQ}=\frac{4}{3}\tilde{d}_d-\frac{1}{3}\tilde{d}_u
\end{equation}
where $\tilde{d}_q$ is the Wilson coefficient of the EDM operator at the hadronic scale. $\tilde{d}_q$ differs from $d_q$, which is calculated at the SUSY-breaking scale, since the former receives contributions from the quark chromo-electric dipole moment, $d^c_q$, and Weinberg's gluonic dimension-six operator, $d_G$~\cite{Arnowitt:1990eh, Weinberg:1989dx}:
\begin{equation}
 \tilde{d}_q=\eta^E d_q + \eta^C \frac{e}{4\pi} d^c_q + \eta^G \frac{e\Lambda}{4\pi} d_G
\end{equation}
These contributions arise from the operator mixing ocurring in the running
from the electroweak scale to the hadronic scale. In our estimates, we take the QCD correction factor 
$\eta^C=3.4$~\cite{Arnowitt:1990eh}, and ignore $d_G$, since it can be at most
of the same order of $d_q$ and $d^c_q$ and we do not expect it to make big
changes in our order of magnitude prediction.

The (C)EDM of each quark has contributions from diagrams with gluinos
$d_q^{\tilde{g}}$, charginos $d_q^{\tilde{\chi}^\pm}$ and neutralinos
$d_q^{\tilde{\chi}^0}$. For gluinos and neutralinos, we can use the same
arguments as in~\cite{Calibbi:2008qt} to establish that the main part of each
contribution will come from $\Im
m\left[(\delta^q_{i3})_{LL}(\delta^q_{33})_{LR}(\delta^q_{3i})_{RR}\right]$
insertions. Chargino contributions come from both pure higgsino and wino-higgsino diagrams.
The mixed wino-higgsino diagrams require no LR flip on the squark line. Contrary to the case with leptons, it is possible to have only one $\delta_{LL}$ insertion, provided that the CKM matrix element $V_{ij}$ is not flavour-diagonal. For instance, the largest contributions to the down quark (C)EDM are proportional to $\Im m \left[V_{qd}(\delta^u_{q q'})_{LL} V^*_{q' d}Y_d\right]$.
Pure higgsino diagrams require a LR flip on the squark line. One would expect this contribution to be strongly suppressed, as there are two Yukawa couplings on the vertices. However, it is possible to exchange one of the small Yukawa couplings for a $y_t$ or $y_b$ coupling through a CKM off-diagonal term. As an example, the main contribution to the pure higgsino diagram for $d_d^{\tilde{\chi}^\pm}$ is proportional to $\Im m \left[V_{td} \, y_t (\delta^u_{33})_{RL} (\delta^u_{32})_{LL} V^*_{c d} \, y_d\right]$. Given the flavour structure of the $SU(3)$ models, we find that the pure higgsino part shall be larger than the mixed wino-higgsino part, and will dominate the chargino contribution to $d_q^{(c)}$.

It was noticed in~\cite{Hisano:2006mj} that, in addition to these contributions, it is important to consider beyond-leading-order (BLO) effects. These effects can be represented as new complex effective couplings~\cite{Buras:2002vd}, and although these are noticeable mainly for large values of $\tan\beta$, for EDMs they can also give significant contributions for moderate and small values of $\tan\beta$. In particular, diagrams with no imaginary part at LO can become complex from these effective vertices~\cite{Hisano:2008hn}.

The most important BLO effect for the down quark EDM, for the values of $\tan\beta$ we are using, has been found to be the $H^\pm$ contribution~\cite{Hisano:2008hn}. Although the BLO corrections also affect the gluino, neutralino, chargino and neutral Higgs loops, these contributions represent only small corrections and do not play an important role. Therefore, the only BLO effects we shall include will come from the $H^\pm$ diagrams.

The BLO effects can also enhance the coupling between the $H^\pm$ and fermions. At LO, the \textit{flavour} suppression for an $H^\pm$-mediated dipole operator would be of order $\left[y_dV^*_{td}\,y^2_t\,V_{td}\right]\sim\bar\varepsilon^{10}$. The BLO vertex allows us to change the $y_d\,V^*_{td}$ term for a $(\delta^d_{13})_{RR}\,y_b\,V^*_{tb}$ term, having then a total flavour suppression of order $\left[(\delta^d_{13})_{RR}\,y_b\,V^*_{tb}\,y^2_t\,V_{td}\right]\sim\bar\varepsilon^{6}$. This is comparable to the gluino flavour suppression at LO.

Even though the flavour suppression of the BLO $H^\pm$ contribution is of an order of magnitude comparable to the LO gluino contribution, the question that needs to be answered is how much does the extra loop suppression affect the $H^\pm$ diagram. This calculation has been done in~\cite{Hisano:2006mj}, for equal SUSY masses and moderate $\tan\beta$, giving:
\begin{equation}
\frac{d_d^{H^\pm}}{d_d^{\tilde{g}}}\sim\left[\frac{\alpha_2}{9\pi}\frac{m_t^2}{m_W^2}\right]
\left[\frac{m^2_{\tilde{q}}}{m^2_{H^\pm}}\right]
\left[\frac{\Im m\left[V^*_{td}(\delta^d_{31})_{RR}\right]}{\Im m\left[(\delta^d_{13})_{LL}(\delta^d_{31})_{RR}\right]}\right]
f\left(\frac{m^2_t}{m^2_{H^\pm}}\right)
\end{equation}
The first square bracket gives a factor of $O(0.1)$, which is compensated by the loop function $f(m^2_t/m^2_{H^\pm})$, which is of $O(10)$. The third bracket is the flavour suppression ratio, which is of $O(1)$. Thus, in this approximation, the importance of the $H^\pm$ contribution with respect to the gluino contribution depends mainly on the ratio between the squark and charged Higgs masses squared. Higgs bosons decouple differently from squarks and gauginos, as gluino and squark masses increase faster than the $H^\pm$ mass for increasing $m_0$ and $M_{1/2}$, the BLO contributions shall be more important for large values of these parameters.

We shall include the $H^\pm$ contribution through the mass-insertion formulae of~\cite{Hisano:2008hn}. Such an approach contemplates three types of contributions. Two of them involve the CKM matrix, and are of the type $\Im m \left[V_{3i}^*(\delta^d_{3i})_{LL}\right]$ and $\Im m \left[V_{3i}^*(\delta^d_{3i})_{RR}\right]$. The third one is similar to that for the gluino loops: $\Im m \left[(\delta^d_{i3})_{LL}(\delta^d_{3i})_{RR}\right]$. The latter two shall be the ones that will allow us to avoid the $y_d$ suppression. The second term shall be particularly important for RVV1, where the gluino contribution has a cancellation between $(\delta^d_{i3})_{LL}$ and $(\delta^d_{3i})_{RR}$.

\begin{figure}
\includegraphics[scale=.5]{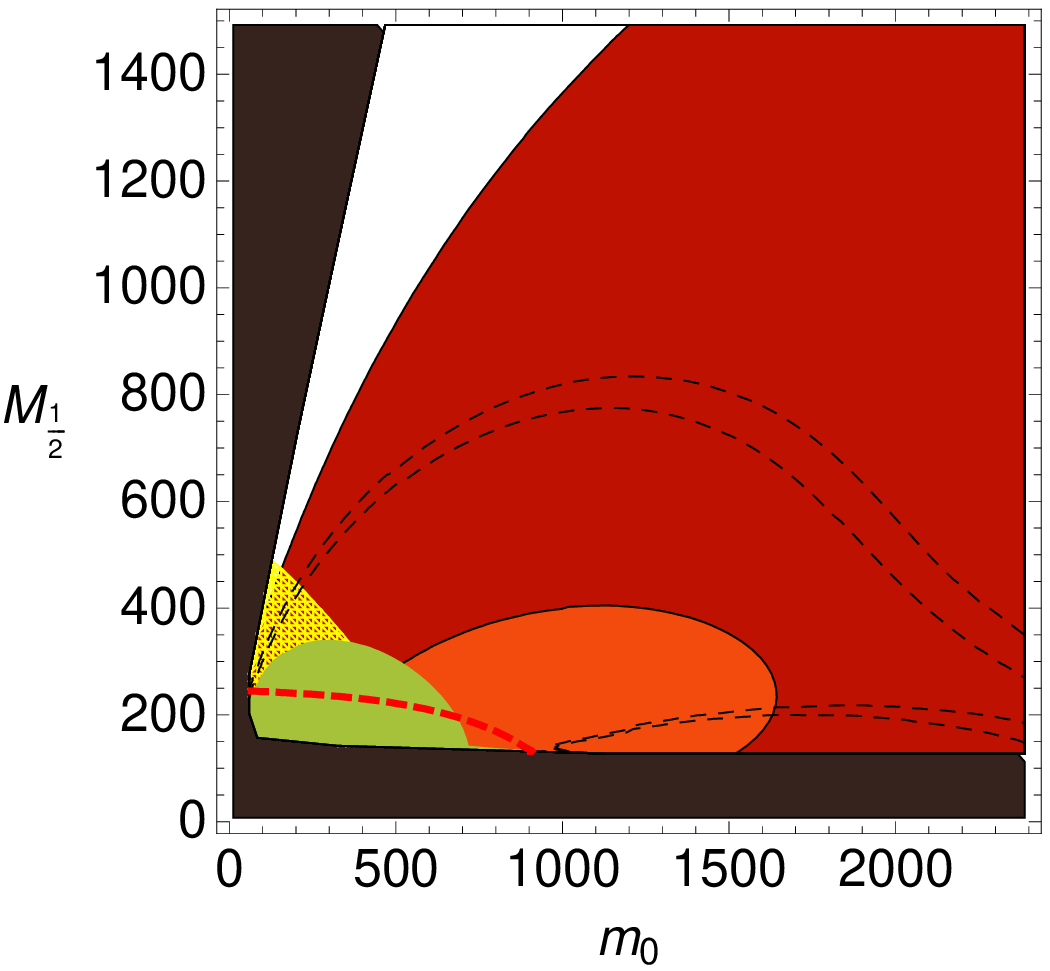} \qquad \includegraphics[scale=.5]{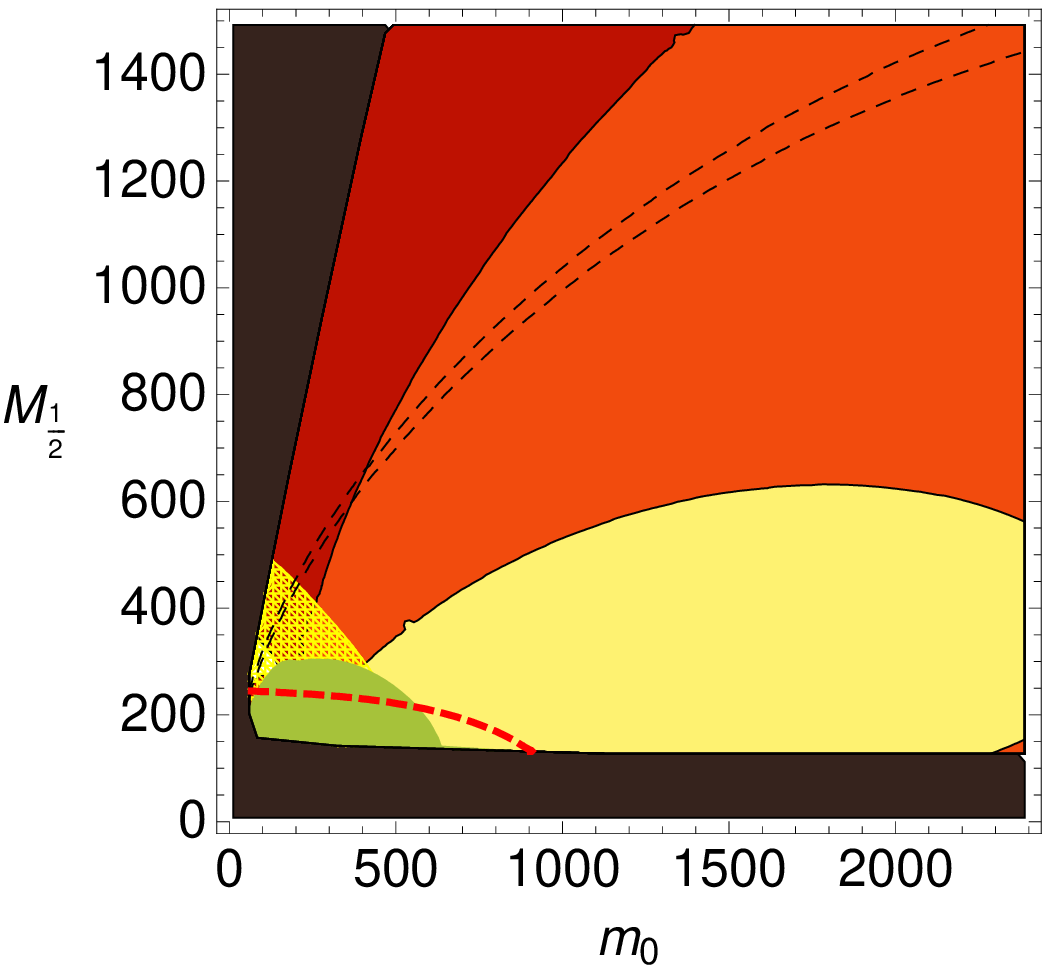} \\ 
\includegraphics[scale=.5]{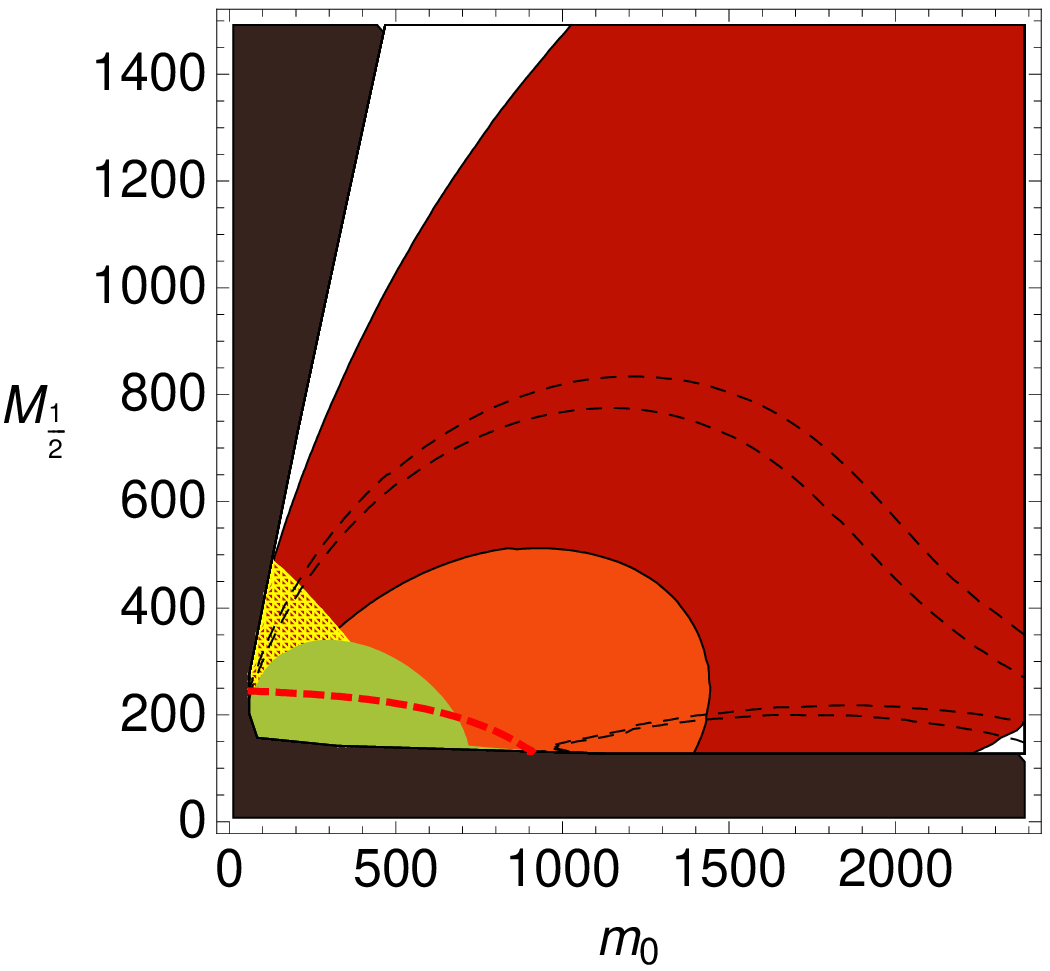} \qquad \includegraphics[scale=.5]{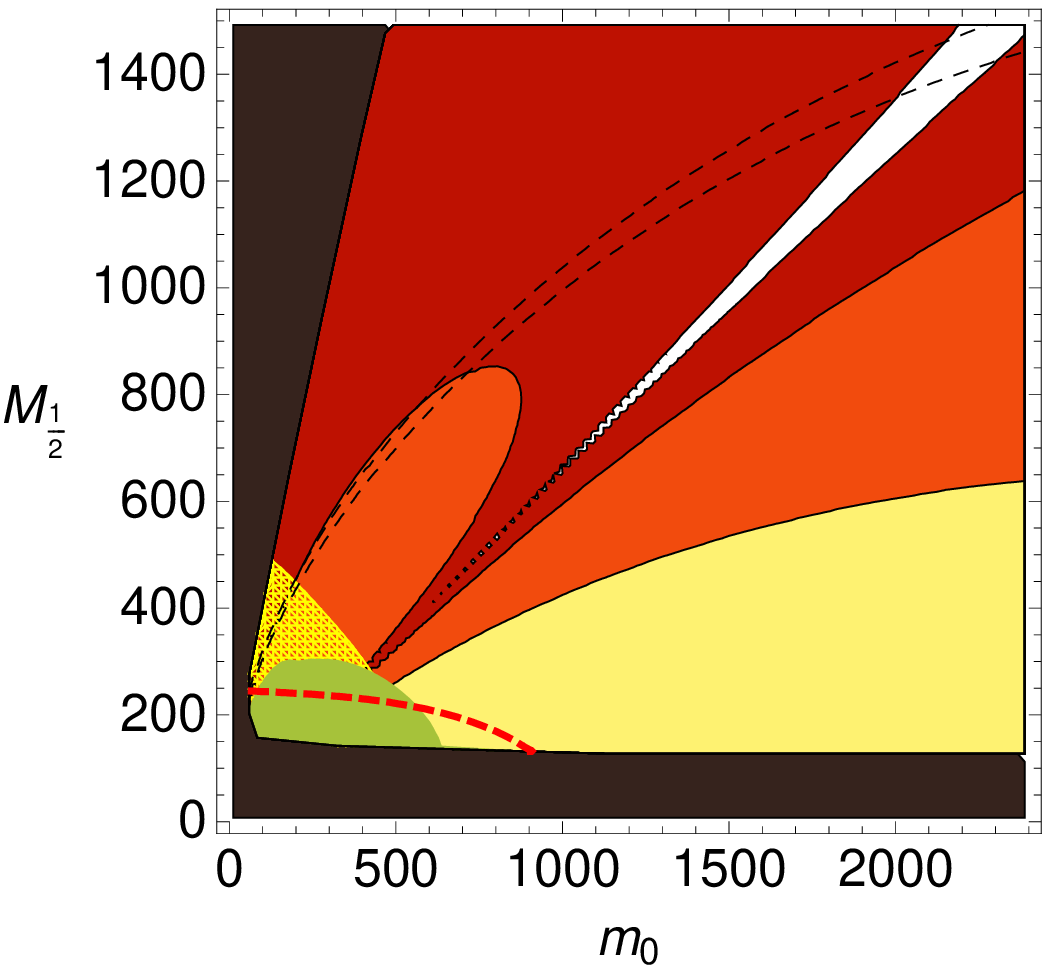} \\ 
\caption{Quark-parton (top) and chiral quark (bottom) contours of
  $|d_n|=1\times10^{-27}$ e cm (yellow) $|d_n|=1\times10^{-28}$ e cm (orange) and $|d_n|=1\times10^{-29}$ e cm
  (red) in the $m_0$-$M_{1/2}$ plane for $\tan\beta =10$ and $A_0=0$. We
  show predictions for RVV1 (left) and RVV2 (right). Current LFV bounds are also shown in green, and the $(g-2)_\mu$ favoured region is shown hatched in yellow.}
\label{fig:nedms}
\end{figure}

The $SU(3)$ predictions for $\tan\beta=10$ and $A_0=0$ are shown in
Figure~\ref{fig:nedms} for the quark-parton and chiral quark EDMs,
respectively. The EDMs receive a significant contribution from the $H^\pm$ loop in $d_d$. Furthermore, $d_n^{QP}$ is also dominated by gluino and $H^\pm$ contributions to $d_s$, while $d_n^{CQ}$ is influenced by their contribution to $d_d^c$, for both RVV1 and RVV2. The other contributions are smaller by at least half an order of magnitude.
We can see that current bounds are always weaker than the LEP and LFV
constraints and do not appear in the Figure.

The observation of $d_n$ in the near-future experiments is not always compatible with the solution of
the $\epsilon_K$ tension. In RVV1, which is mostly dominated by the $H^\pm$ loops, a $d_n$ of order $10^{-29}$
$e\textrm{ cm}^{-1}$ is usually favoured, one order of magnitude under the
reach of the next experiments~\cite{FutnEDM}. In RVV2, both gluino and $H^\pm$ effects are comparable, the former dominating in the chiral-quark and the latter in the quark-parton models. In both situations it is possible to obtain an observable $d_n$ which is compatible with the solution of the $\epsilon_K$ tension. 

It is interesting to see that in some regions of the parameter space we have a cancellation. A careful analysis proves that close to these regions the phase of the $(\delta^d_{13})_{LL}$ term vanishes, and that can only be due to a cancellation between the initial term at $M_{GUT}$ and the contributions from the running. This causes a change of sign in the gluino contribution, such that it can interfere destructively with the $H^\pm$ part. We can avoid the cancellations by changing the sign of the $O(1)$ terms at $M_{GUT}$, in which case the interference is constructive. This, of course, will turn the cancellation into a small enhancement, but in any case, $d_n$ never exceeds its current bounds.

\begin{figure}
\includegraphics[scale=.5]{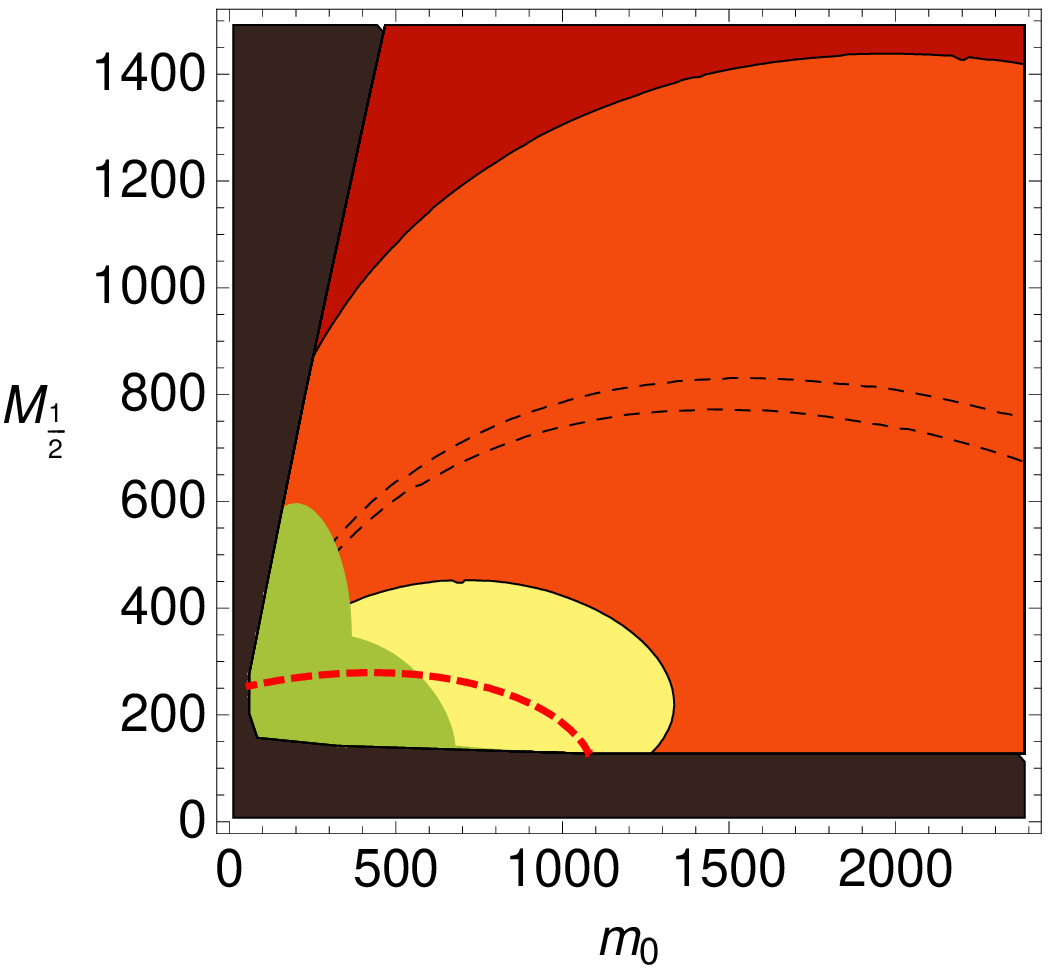} \qquad \includegraphics[scale=.5]{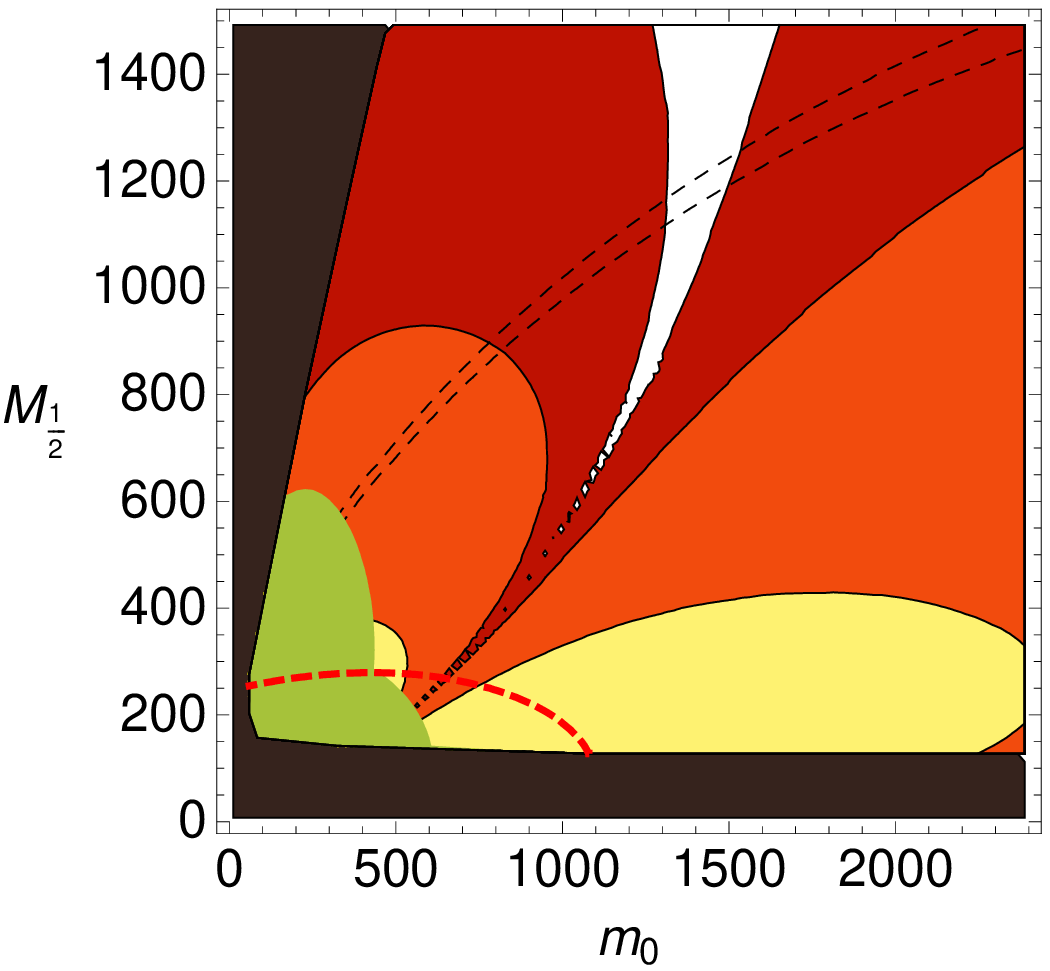} \\ 
\includegraphics[scale=.5]{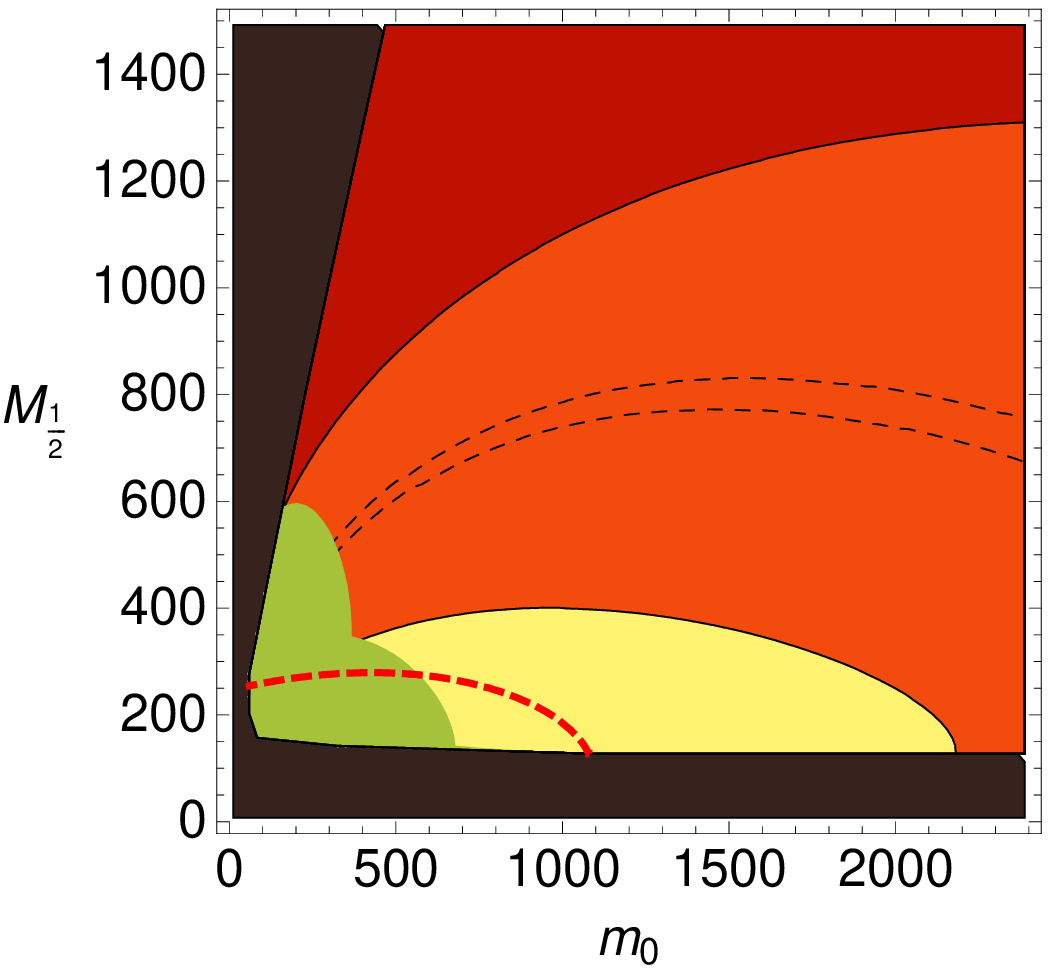} \qquad \includegraphics[scale=.5]{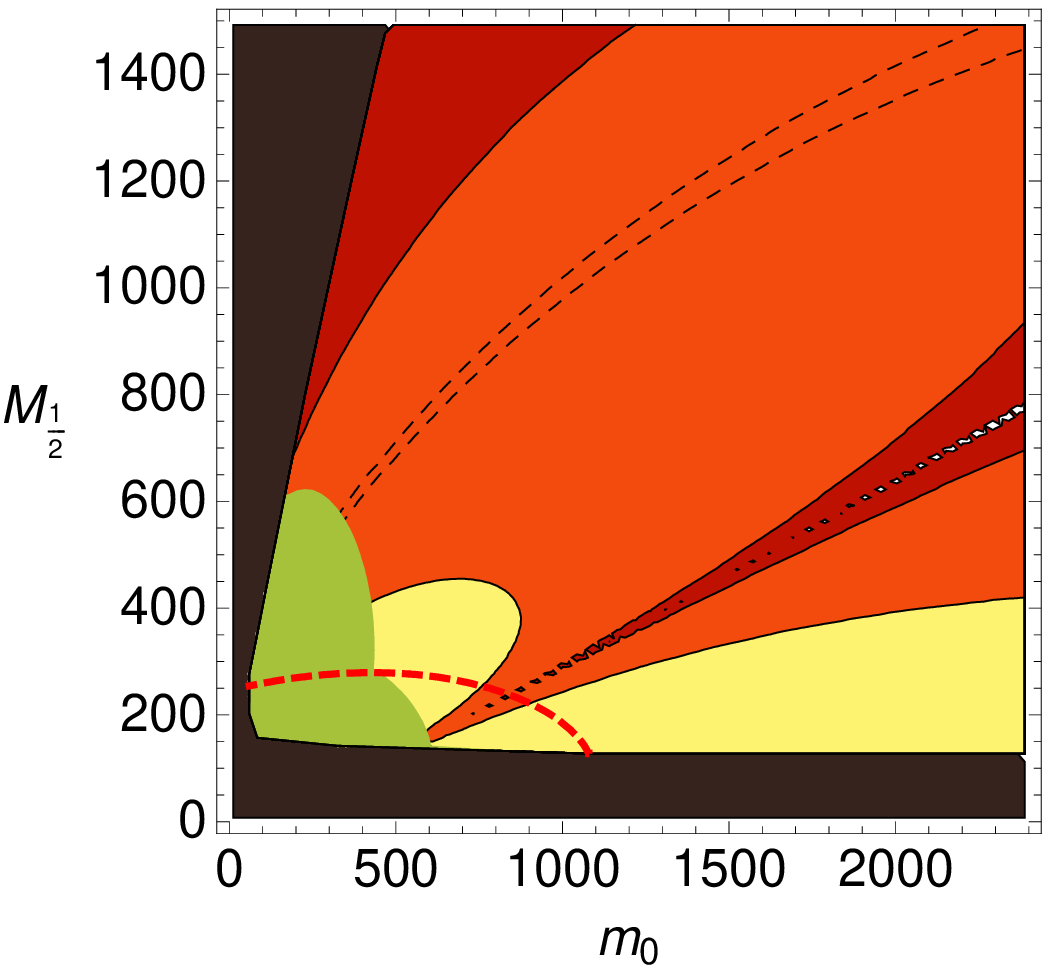} \\ 
\caption{Quark-parton (top) and chiral quark (bottom) contours of $|d_n|=1\times10^{-27}$ e cm (yellow) $|d_n|=1\times10^{-28}$ e cm (orange) and $|d_n|=1\times10^{-29}$ e cm (red) in the $m_0$-$M_{1/2}$ plane for $\tan\beta =10$ and $A_0=m_0$. We show predictions for RVV1 (left), RVV2 (right). Current LFV bounds are also shown in green, and the $(g-2)_\mu$ favoured region is shown hatched in yellow.}
\label{fig:nedms-a1}
\end{figure}

Finally, Figure~\ref{fig:nedms-a1} shows the same information for $A_0=m_0$. In this situation, the inclusion of the off-diagonal $(\delta^d_{ij})_{LR}$ terms allow $d_n$ contributions of the type $\Im m\left[(\delta^d_{ij})_{LL}(\delta^d_{ji})_{LR}\right]$ and $\Im m\left[(\delta^d_{ij})_{LR}(\delta^d_{ji})_{RR}\right]$. These are always proportional to $m_b$, although they do not receive a $\tan\beta$ enhancement. It was shown in~\cite{Calibbi:2008qt} that these terms could be important, but they at most remained within the order of magnitude of the $\Im m\left[(\delta^d_{i3})_{LL}(\delta^d_{33})_{LR}(\delta^d_{3i})_{RR}\right]$ terms. Notice that for $d_n$ the $(\delta^u_{ij})_{LR}$ insertions are of no interest: although RGE effects can make the off-diagonal terms as large as the $(\delta^d_{ij})_{LR}$ terms, they will usually enter the observables accompanied by a $(\delta^u_{ij})_{LL}$ or $(\delta^u_{ij})_{RR}$ insertion, which are proportional to powers of $\varepsilon$.

An important contribution that appears when $A_0\neq0$ comes from flavour-diagonal subleading phases in $(A_d)_{11}$. Although the rotation to the SCKM basis removes the leading phases in the flavour-diagonal elements of the trilinear couplings, the different $O(1)$ terms makes it impossible to remove the subleading phases in these terms. These phases are suppressed, but they still give important contributions, especially in RVV1, where the leading phases in the contributions from off-diagonal terms cancel. Such an effect is enhanced by the RGE evolution, and is particuarly relevant for the squark sector.

In Figure~\ref{fig:nedms-a1} we can see that, barring the cancellations of the mass-insertion phases, both quark-parton and chiral quark EDMs are similar in magnitude. Although both RVV1 and RVV2 are still not constrained by current bounds, the magnitude of the $d_n$ within the $\epsilon_K$ strip is now larger by an order of magnitude, enough to be observed in the near future. The main reason for the similarity in the order of magnitude between models lies in an enhancement of the gluino contribution, due to the $(A_d)_{11}$ subleading phase mentioned in the previous paragraph. For RVV2, this contribution mostly presents a correction to the off-diagonal contributions, but for RVV1 it becomes the main source for $d_n$.

\section{Combined Analysis in the SUSY Parameter Space}

\label{sec:combined}

In this Section, we are going to present the results of a combined analysis of
the observables studied in the previous sections. In particular, we will show
the predictions of the model for LFV decays and EDMs in the regions of the
parameter space where the SUSY contribution to $\epsilon_K$ can account for
the possible tension between the measured value and the SM prediction, as
discussed in \cite{Buras:2008nn}.  Such a requirement, and always up to
possible variations of $O(1)$ coefficients, is fulfilled in a restricted
portion of the parameter space and allows, as we will see, to do quite
definite predictions for the other flavour observables. If we require in
addition that the $(g-2)_\mu$ discrepancy between SM and data is explained by
SUSY, we are restricted in a region of rather light SUSY masses, where most of
the observables are expected to be close to the present experimental
bounds. Given both the presence of unknown $O(1)$ coefficients and the large
theoretical uncertainties in the calculation which don't allow us to speak of a real failure of the SM, we cannot take what outlined above too seriously. Nevertheless, we think this can be a useful exercise in order to show how the interplay of various flavour observables can be used for testing this kind of flavour models.

In order to understand the impact of the $O(1)$ coefficients, in the following plots we shall set all of the $O(1)$s in the soft mass matrices equal to unity. The $O(1)$s of the trilinears shall be kept random, but fixed, since setting them to one aligns them with the Yukawas. At the end of this section we analyze the impact of varying the $O(1)$ coefficients, in order to understand how much they affect our correlations.

\begin{figure}
\centering
\includegraphics[width=.8\textwidth]{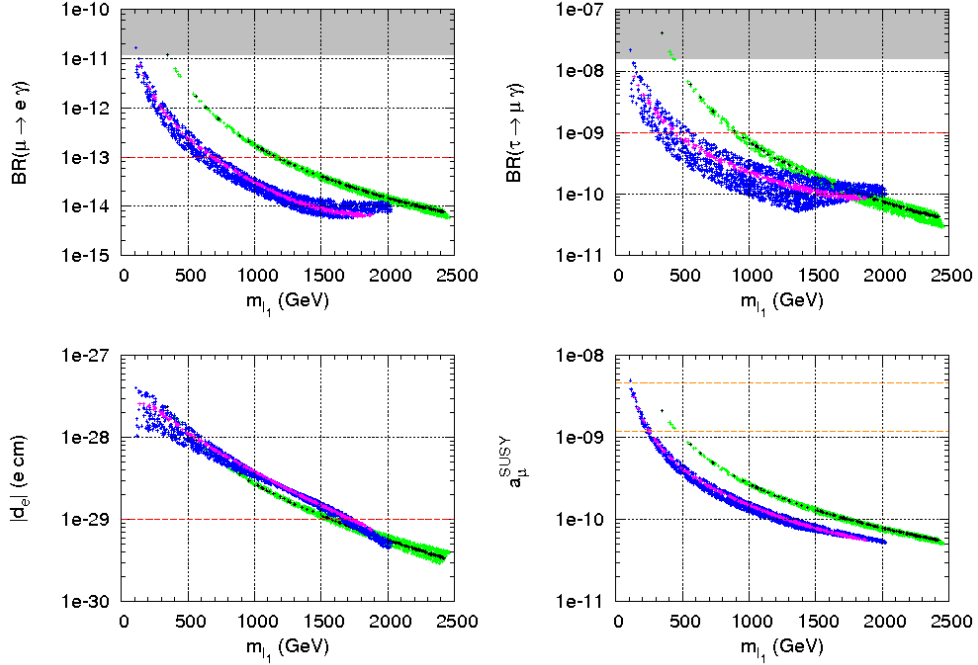}
\caption{Leptonic observables for model RVV2, $\tan\beta=10$, $A_0=0$, as functions of the lightest slepton mass. See the text for details.}
\label{fig:comb1}
\end{figure}

As explained in Section~\ref{epsK}, the SUSY contribution to $\epsilon_K$
depends on the sign of the $O(1)$ coefficient of the entry
${(\delta^d_{RR})_{12}}$, since a negative sign can induce a partial cancellation between the contributions proportional to $(\delta^d_{RR})^2_{12}$ and
$(\delta^d_{RR})_{12}(\delta^d_{LL})_{12}$. Therefore, we have to take into account both the possible signs for $(\delta^d_{RR})_{12}$. 
In Fig.~\ref{fig:comb1}, we show the leptonic observables as functions of the lightest slepton mass for the model RVV2 with  $\tan\beta=10$, $A_0=0$, $0 < m_0 < 2.5$ TeV, $0<M_{1/2}<1.5$ TeV. The green band corresponds to the points of the parameter space
for which $(|\epsilon^{\rm SM}_K + \epsilon^{\rm susy}_K| -
|\epsilon^{\rm exp}_K|) < \sigma^{\rm th}$, where we have checked that the correct sign has been obtained.
For the SM contribution
we took the central value computed in \cite{Buras:2009pj}, $|\epsilon^{\rm
  SM}_K|= 1.78\times10^{-3}$, and $\sigma^{\rm th}$ is the corresponding
theoretical error, $\sigma^{\rm th}=0.25\times10^{-3}$. The blue band
represents the same for the case of negative ${(\delta^d_{RR})_{12}}$. The
black and purple strips correspond to the more strict requirement
$(|\epsilon^{\rm SM}_K + \epsilon^{\rm susy}_K| - |\epsilon^{\rm
  exp}_K|) < 3 \sigma^{\rm exp}$, with $\sigma^{\rm exp}$ being the
experimental error, $0.012\times10^{-3}$ (in other words for the strips we
assume $|\epsilon^{\rm SM}_K|$ to take precisely to the central value in
\cite{Buras:2009pj}). The shaded regions are ruled out by experiment, and the red lines indicate the future experimental sensitivity. For $(g-2)_\mu$, the area between the yellow lines solve the tension below 2$\sigma$.

The leptonic predictions in these
``$\epsilon_K$-favoured'' regions are quite interesting. We see that ${\rm
  BR}(\mu\to e\gamma)$ (top-left panel) gives at present practically no
constraint, while the final MEG sensitivity ($\sim 10^{-13}$), will test the
model up to slepton masses around 0.5-1.2 TeV. Very interestingly, the
sensitivity of a Super Flavour Factory reaching ${\rm BR}(\tau\to
\mu\gamma)\simeq 10^{-9}$ can be rather similar (top-right panel). The reason
for this is that the ``$\epsilon_K$-favoured'' region selects rather low values of $M_{1/2}$, where the future bounds of MEG and the Super Flavour Factory are comparable (see Fig.~\ref{fig:meg} in Section~\ref{lfv}). Concerning eEDM (bottom-left panel), we see that reaching $d_e\sim 10^{-29}~{\rm e~cm^{-1}}$ would test this case up to $m_{\tilde{l}_1}\simeq 1.5$ TeV, well beyond the reach of the LFV experiments. Finally, requiring that the SUSY contribution to $(g-2)_\mu$ (bottom-right panel) lowers the tension with the experiments below 2$\sigma$, a rather light spectrum is selected, $m_{\tilde{l}_1}\lesssim 250-500$ GeV, so that all the other observables should be in the reach of running/future experiments, with, in particular, branching ratios of LFV decays being just below the present experimental limits. 
%
%
%
\begin{figure}
\centering
\includegraphics[width=0.8\textwidth]{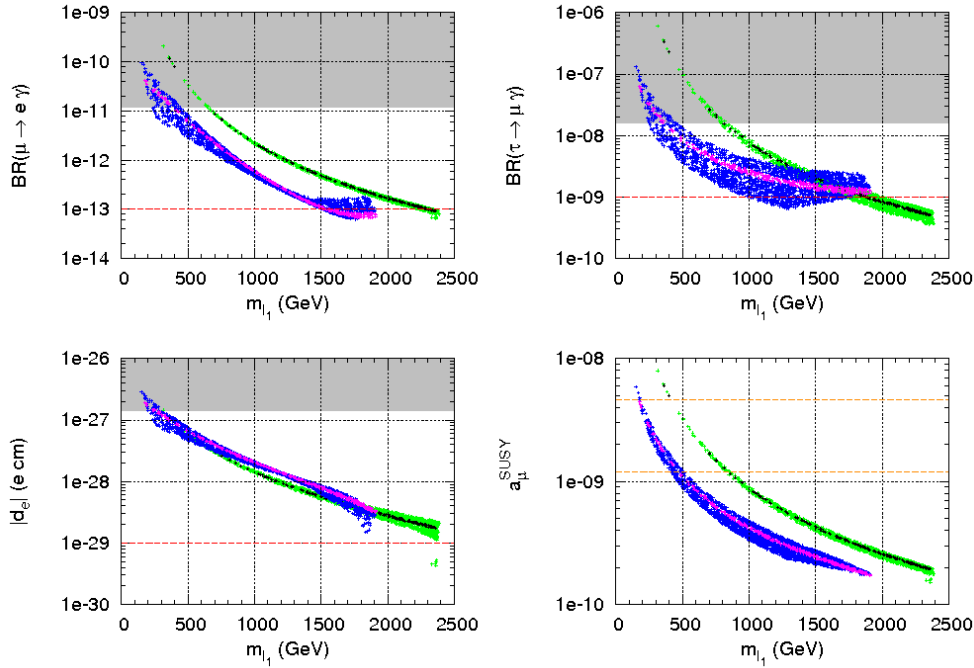}
\caption{Leptonic observables for model RVV2, $\tan\beta=30$, $A_0=0$, as functions of the lightest slepton mass.}
\label{fig:comb3}
\end{figure}
We checked that model RVV1 gives results for LFV decays and $(g-2)_\mu$ which are similar to the ones of RVV2, while $d_e$ is suppressed by approximately one order of magnitude: in this case, the future sensitivity on $d_e$ will test the parameter space up to slepton masses around 500-750 GeV.  

The predictions of Fig.~\ref{fig:comb1} are rather stable also for $a_0=1$. In this case, both for RVV1 and RVV2, only ${\rm BR}(\mu\to e\gamma)$ and $d_e$ gets slightly increased.
For larger values of $\tan\beta$ ($\sim 30$) we expect, both for RVV1 and RVV2, the future sensitivities for LFV decays and $d_e$ to completely test the parameter space in the range of parameters we considered, $0 < m_0 < 2.5$ TeV, $0<M_{1/2}<1.5$ TeV. This is shown in Fig.~\ref{fig:comb3} for RVV2. Notice that, also in this case, it is possible to account for the $(g-2)_\mu$ and the $\epsilon_K$ discrepancies at the same time, even if the leptonic observables are in general predicted to be very close to the present experimental bounds.
\begin{figure}
\centering
\includegraphics[width=0.3\textwidth]{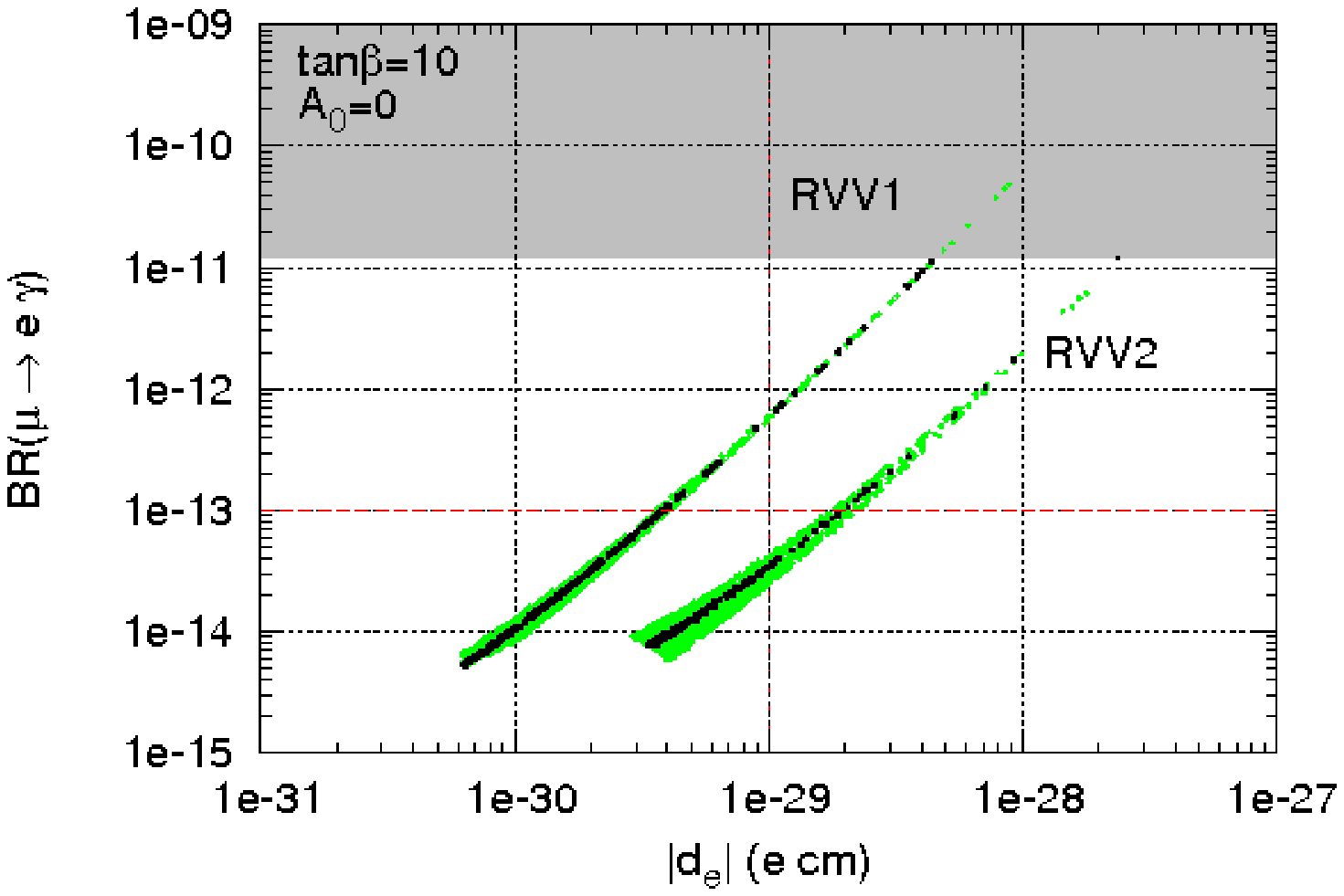}
\includegraphics[width=0.3\textwidth]{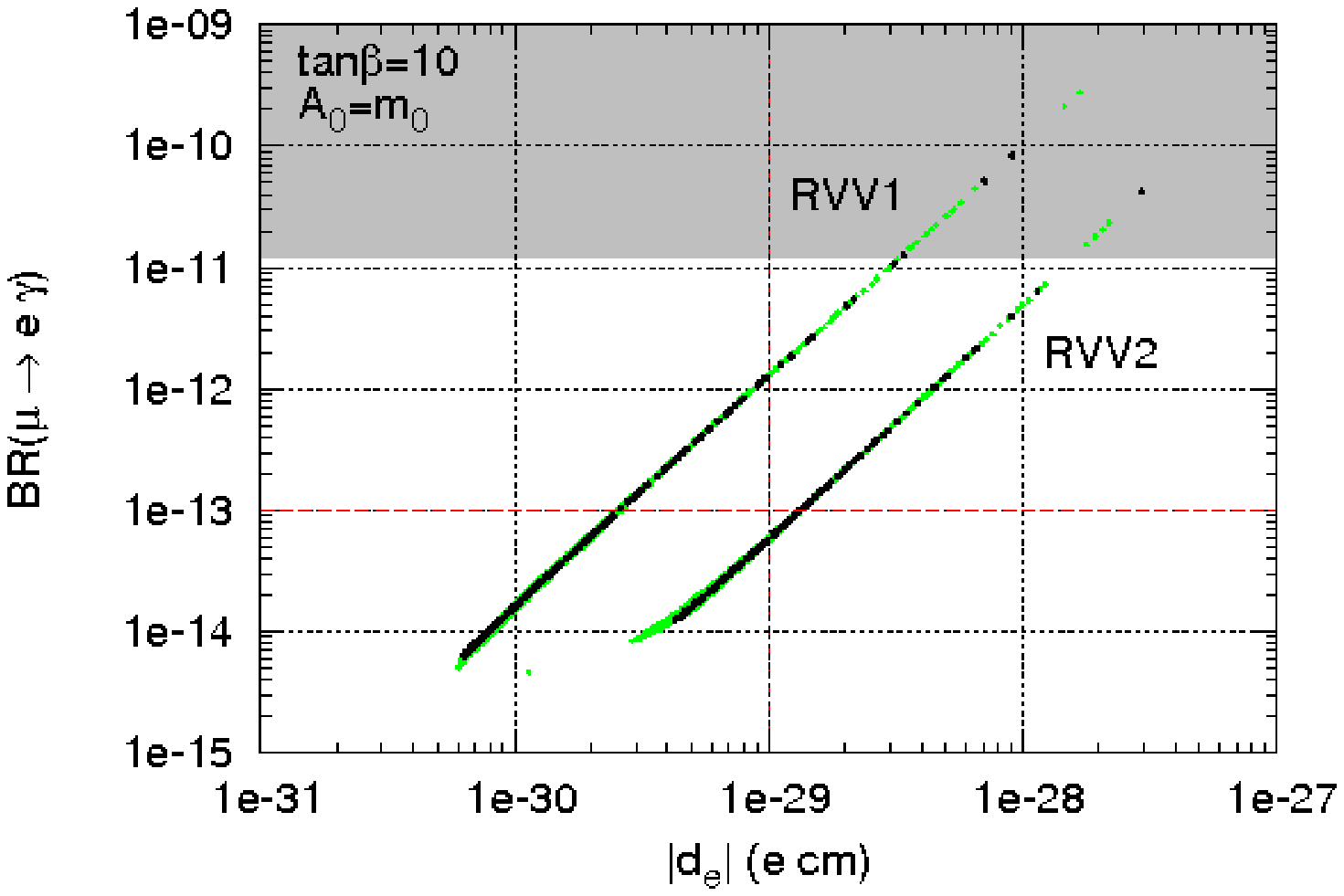}
\includegraphics[width=0.3\textwidth]{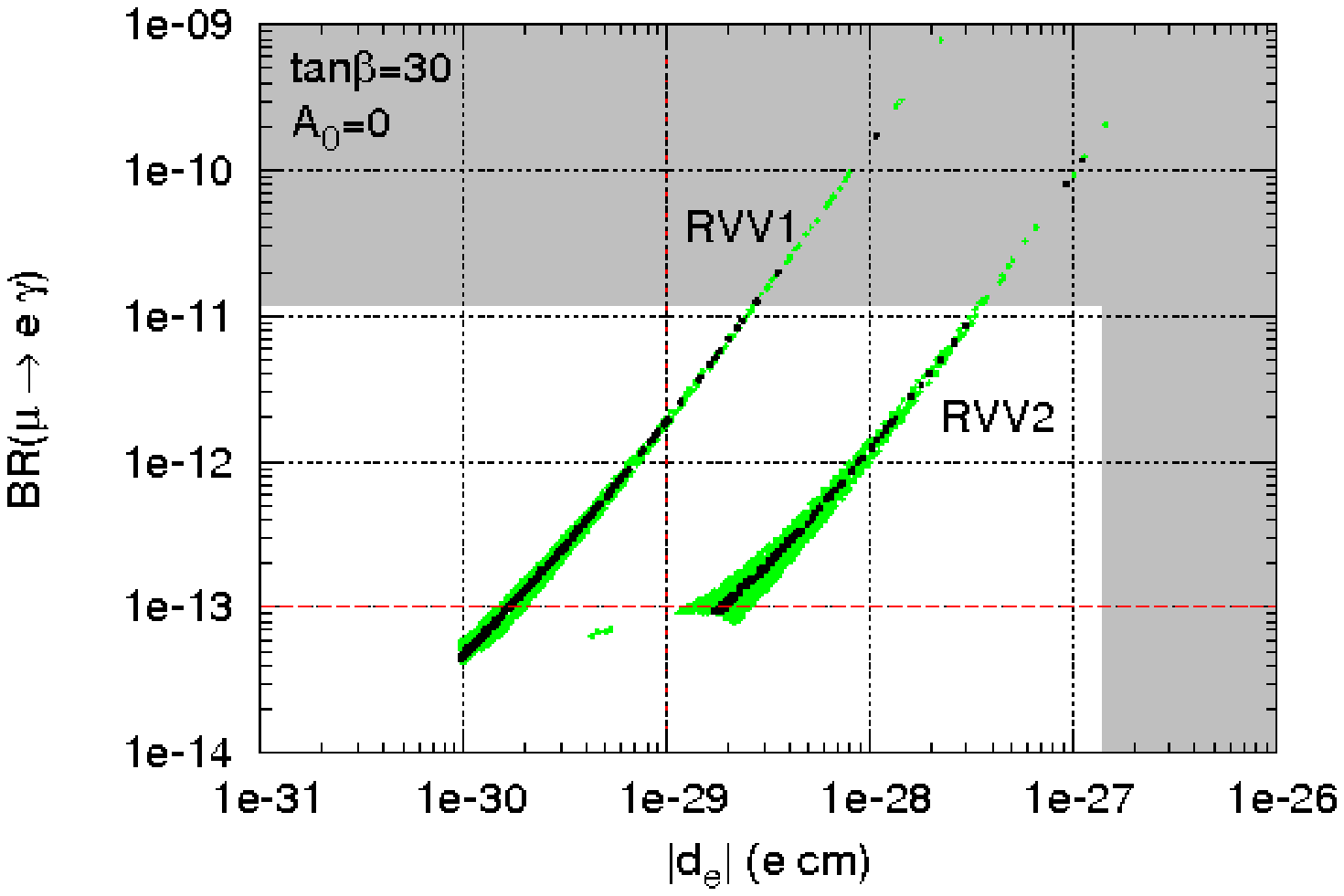}
\caption{${\rm BR}(\mu\to e\gamma)$ vs. $|d_e|$ for different scenarios. See the text for details.}
\label{fig:meg-edm}
\end{figure}

In Fig.~\ref{fig:meg-edm}, we compare the discovery potential of the two most promising leptonic observables, $\mu\to e\gamma$ and the electron EDM. The correlation of ${\rm BR}(\mu\to e\gamma)$ vs. $|d_e|$ is plotted for both RVV1 and RVV2, in the case 
$\tan\beta=10$, $a_0=0$ (left), $\tan\beta=10$, $a_0=1$ (center) and $\tan\beta=30$, $a_0=0$ (right). As before we studied the mass range: $0 < m_0 < 2.5$ TeV, $0<M_{1/2}<1.5$ TeV. In the figures, only the 
``$\epsilon_K$-favoured'' region with positive ${(\delta^d_{RR})_{12}}$
has been plotted. The horizontal line corresponds to the final sensitivity of
MEG, the vertical line to the sensitivity on $|d_e|$ of the running Yale-PdO
experiment. We see that, for RVV1, $\mu\to e\gamma$ should be able to constrain
the parameter space more strongly than eEDM, while for RVV2 it is $|d_e|$ the
most sensitive  observable (except for the large $\tan\beta$ case). These features could be useful in the future, in order to discriminate among different models and, more in general, shed light on the structure of mixings and phases in the slepton mass matrix. 
\begin{figure}
\centering
\includegraphics[width=0.8\textwidth]{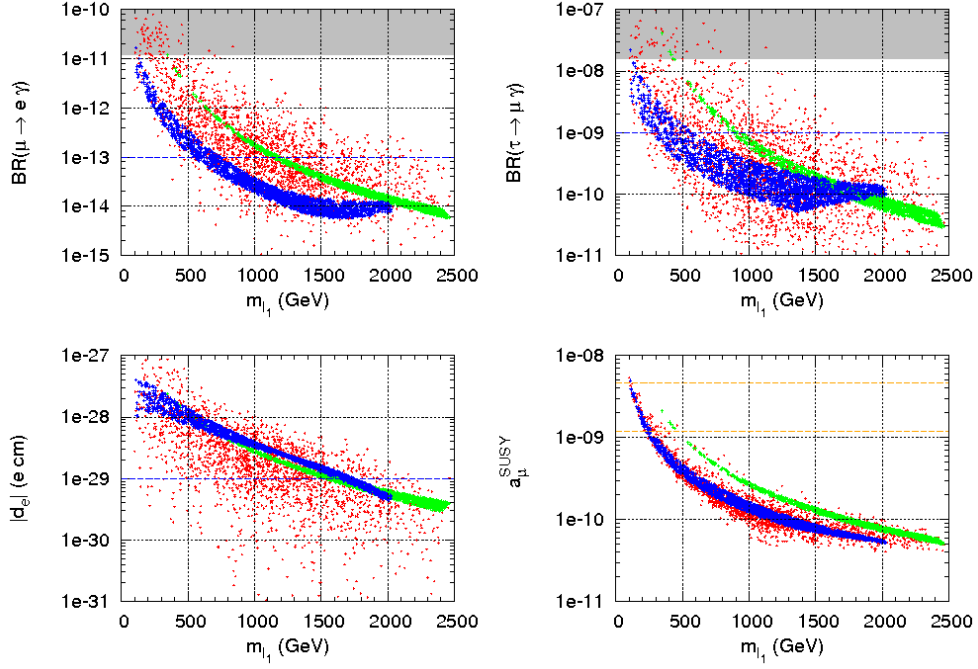}
\caption{Leptonic observables for model RVV2, $\tan\beta=10$, $A_0=0$, as functions of the lightest slepton mass, for a random variation of
the O(1) coefficients (red dots). See the text for details.}
\label{fig:O1var}
\end{figure}

\begin{figure}
\centering
\includegraphics[width=0.3\textwidth]{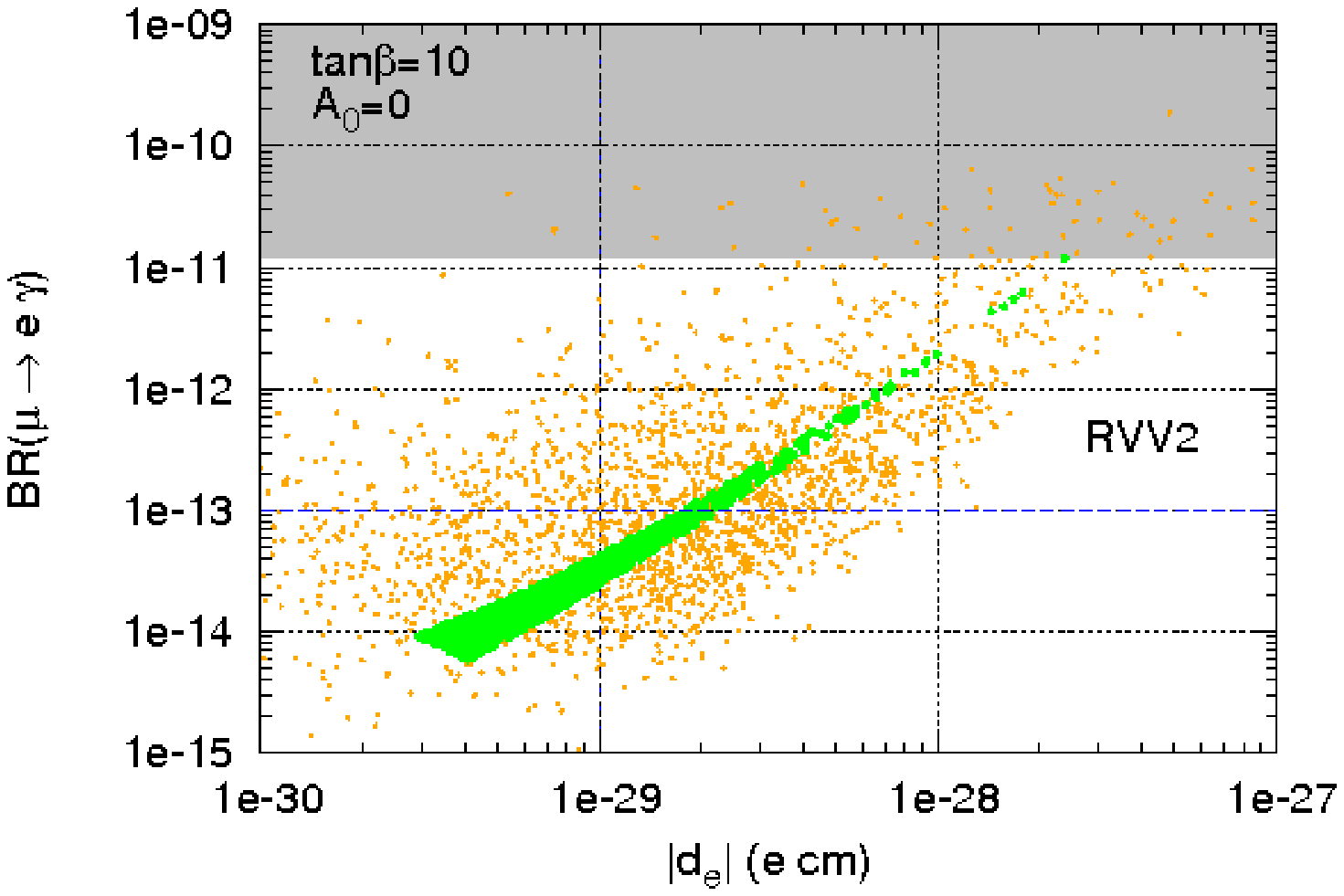}
\includegraphics[width=0.3\textwidth]{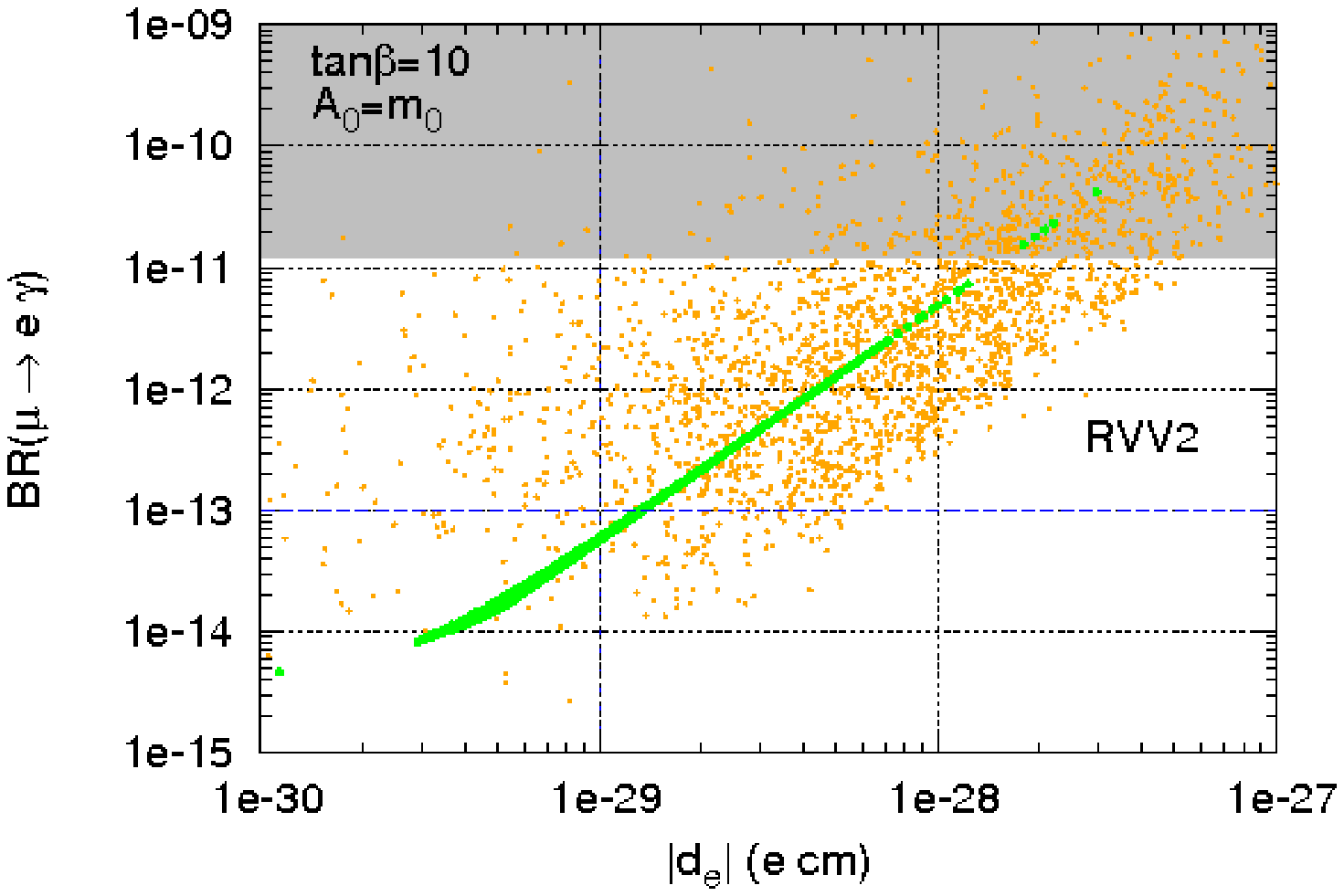}
\includegraphics[width=0.3\textwidth]{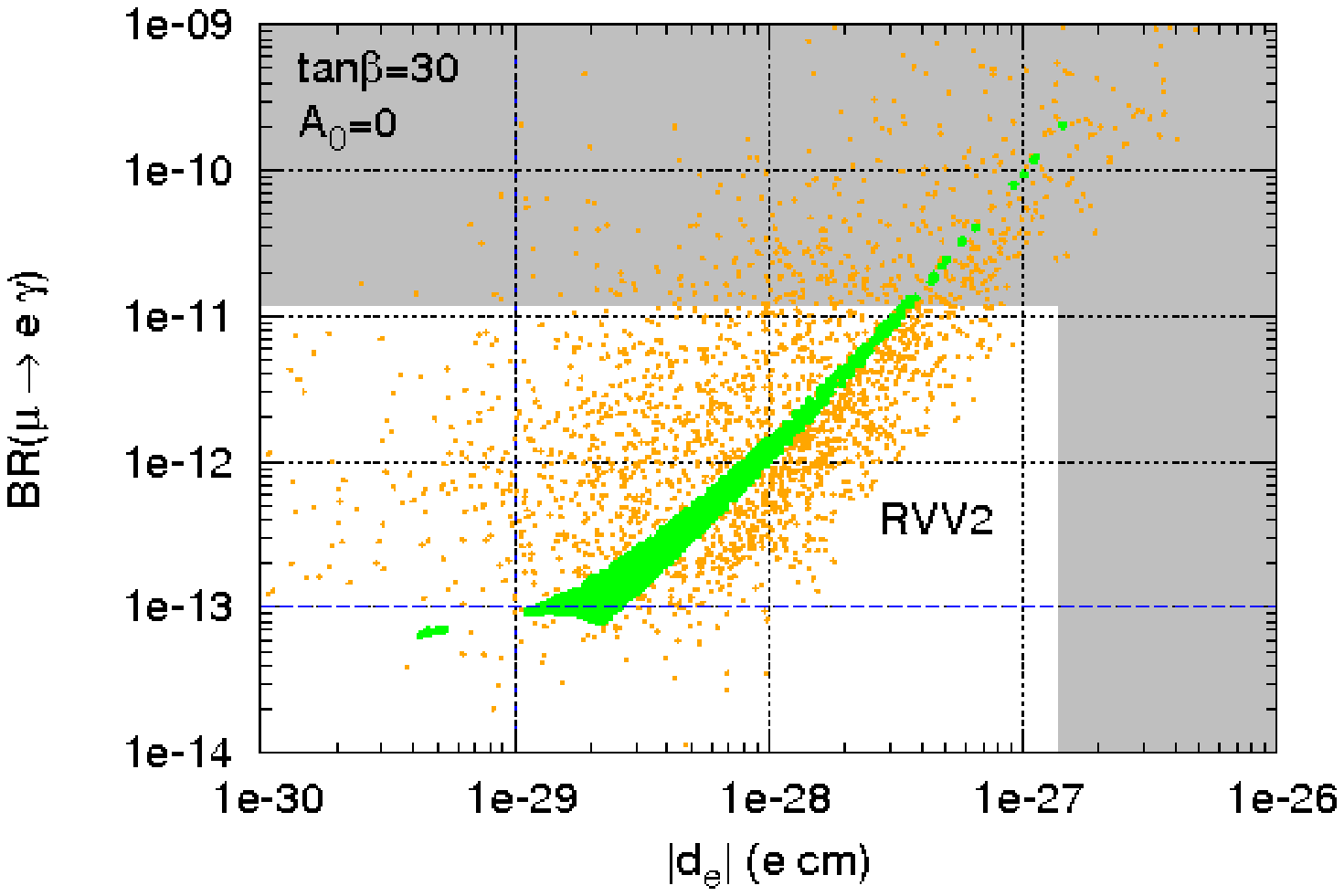}
\caption{Same as Fig.~\ref{fig:meg-edm}, but with a random variation of
the O(1) coefficients (red dots). See the text for details.}
\label{fig:O1var2}
\end{figure}

Finally, let us briefly comment about the impact of the unknown $O(1)$ coefficients on the analysis outlined above. In Fig.~\ref{fig:O1var}, we plot the same observables of Fig.~\ref{fig:comb1} (RVV2, $\tan\beta=10$, $a_0=0$) in red, performing a random variation of all the O(1) coefficients between 0.5 and 2 (in absolute value). For comparison, we superimpose to the scatter plot the green and blue bands of Fig.~\ref{fig:comb1}. We do the same for Fig.~\ref{fig:meg-edm} in Fig.~\ref{fig:O1var2}, only for RVV2, in yellow, superimposing the green bands.

From these Figures we see that, although the lines are broadened by the variation of $O(1)$s, as expected, the correlations are roughly mantained. In some cases we can see variations of even several orders of magnitude for the predictions. This is due to two different sources. On one hand, for fixed values of $m_0$ and $M_{1/2}$ we can have a variation of a factor 4 or $1/4$ in the product of two sleptonic $O(1)$ coefficients. On the other hand, if we allow the variation of $O(1)$ coefficients in $\epsilon_K$, the lines in the ($m_0$, $M_{1/2}$) plane of Figure~\ref{fig:epsK} become broad bands, increasing in turn the width of the bands of the sleptonic observables. Still, our predictions remain stable enough, especially for some observables such as ${\rm BR}(\mu\to e\gamma)$, to conclude that the main qualitative features discussed above are not affected too much by the unknown O(1) coefficients.

\section{Conclusions}
\label{sec:conclusions}

In this work, we have studied the phenomenology of a MSSM with a $SU(3)$
flavour symmetry and spontaneous CP violation. We have shown that in this
framework it is possible to fit the observed fermionic masses and mixings
simultaneously solving the so-called SUSY flavour and CP problems. As a proof
of existence, we have analyzed an explicit example based in the model of 
Ross, Velasco and Vives
\cite{Ross:2004qn} that successfully overcomes the present FCNC and
CP-violation constraints. At the same time, this model predicts a non-trivial
flavour structure in the SUSY soft-breaking terms that will show up in
near-future CP-violation, LFV and hadronic FCNC experiments.
We have analyzed a model that includes the minimal set of contributions to the
soft-breaking terms and we have presented two possible variations with
additional contributions to soft-breaking terms. 

In these models, we naturally expect to be able to observe the $\mu\to e\gamma$
decay at MEG for sfermion masses within LHC reach. Similarly, it is also
possible to measure the $\tau \to \mu \gamma$ branching ratio at the future 
Super Flavour Factory. In the hadronic sector the main effect of the model is
a sizeable contribution to $\epsilon_K$ that could explain the recently observed
discrepancy between the SM and the measured value taking the latest lattice
values for the $B_K$ factor \cite{Buras:2009pj}. The electric dipole moment 
of the electron is also probably to be within reach of the present experiments
while the neutron EDM  will be just beyond the planned sensitivity of the 
experiments. It is unfortunate that, at low $\tan\beta$ and without any conspiration of $O(1)$ parameters, the evaluated versions of this model cannot provide the required contribution to $\phi_{B_s}$ in order to solve the discrepancy reported in~\cite{Bona:2008jn}. This means that if the future data confirms the discrepancy, non-minimal versions of the model shall be required.

We have made a combined analysis of all the observables if we require that the
 $\epsilon_K$ discrepancy is solved by the SUSY contributions of the model. In
 this case we are able to relate the values of the different observables and
 again   $\mu\to e\gamma$, $\tau \to \mu \gamma$ and $d_e$ can be measured
 in the future experiments if SUSY is to be found at LHC. 

\acknowledgments
We would like to thank G.~G.~Ross, I.~de~Medeiros~Varzielas and M.~Passera 
for useful discussions.
L.~C. wishes to thank the University of Valencia for the kind hospitality 
and partial support during various stages of this work. J.~J.~P. thanks the
universities of Padova and W\"urzburg for hospitality during his visits to
complete this work. A.~M., J.-h.~P. and O.~V. would like to thank the Yukawa Institute, Kyoto and the research project YKIS
for providing the opportunity to develop an important part of this work in a stimulating and active environment. 
We acknowledge support from the MEC-INFN agreements FPA2008-04065-E and INFN2008-016.  
The work of O.~V. and J.~J.~P. was supported by the Spanish MICINN and FEDER
(EC) Grant No. FPA2008-02878. 
O.~V. was supported in part by European program MRTN-CT-2006-035482
``Flavianet'' and the  
Generalitat Valenciana under the grant PROMETEO/2008/004. 
A.~M. acknowledges partial support 
from the CARIPARO Project of Excellence (LHCosmology). 
W.~P. is partially supported by the German Ministry of Education and
Research (BMBF) under contract 05HT6WWA.
J.-h.~P. acknowledges Research Grants funded jointly by the Italian
Ministero dell'Istruzione, dell'Universit\`{a} e della Ricerca (MIUR),
by the University of Padova and
by INFN within the
\textit{Astroparticle Physics Project}. 
A.~M. and W.~P. are partially supported by MRTN-CT-2006-035505 (Hep-Tools).
A.~M. and J.-h.~P. acknowledge partial support from the INFN FA51 Grant
and from the RTN European Contracts MRTN-CT-2006-035863 (UniverseNet).

\appendix
\section{CKM Fit}
The $SU(3)$ model generates the following leading structure for the Yukawa matrices, in terms of $\varepsilon$ and
$\bar\varepsilon$:
\begin{eqnarray}  
\label{yukawas}
Y_d=\left( 
\begin{array}{ccc}
0 & x^d_{12}\,\bar\varepsilon^3\,e^{i\delta_d} & x^d_{13}\,\bar\varepsilon^3\,e^{i\delta_d+\beta_3} \\ 
x^d_{12}\,\bar\varepsilon^3\,e^{i\delta_d} & \Sigma_d\bar\varepsilon^2 & 
x^d_{23}\,\Sigma_d\,\bar\varepsilon^2\,e^{i\beta_3} \\ 
x^d_{13}\,\bar\varepsilon^3\,e^{i\delta_d+\beta_3} & x^d_{23}\,\Sigma_d\,\bar\varepsilon^{2}\,e^{i\beta_3} & e^{2i\chi}
\end{array}
\right)y_b \\
Y_u=\left(
\begin{array}{ccc}
0 & x^u_{12}\,\varepsilon^3\,e^{i\delta_u} & x^u_{13}\,\varepsilon^3\,e^{i\delta_u+\beta_3} \\ 
x^u_{12}\,\varepsilon^3\,e^{i\delta_u} & \Sigma_u\varepsilon^2 & 
x^u_{23}\,\Sigma_u\,\varepsilon^2\,e^{i\beta_3} \\ 
x^u_{13}\,\varepsilon^3\,e^{i\delta_u+\beta_3} & x^u_{23}\,\Sigma_u\,\varepsilon^{2}\,e^{i\beta_3} & 1
\end{array}
\right )y_t
\end{eqnarray}
where $\delta_f=2\alpha_f+\beta_3+\beta'_2$. It is necessary to fix the values for the $x^\alpha_{ij}$ parameters and the flavon phases. In the
quark sector, this means that we need to reproduce the quark masses and CKM
matrix and, in fact, we will see that the CKM matrix is the main source of constraints on $x^u_{ij}$ and $x^d_{ij}$.

After diagonalizing the Yukawas, keeping terms up to order $\bar\varepsilon^3$, $\varepsilon^2$ and $\varepsilon\bar\varepsilon^2$, and rephasing fields such that $V_{ud}$, $V_{us}$, $V_{cs}$, $V_{cb}$ and $V_{tb}$ are real, we get the following CKM matrix:
\begin{equation}
\label{CKM}
 V_{CKM}=\left(\begin{array}{ccc}
\left|1-\frac{1}{2}\left(\frac{x^d_{12}}{\Sigma_d}\right)^2\Lambda_{ud}\,\bar\varepsilon^2\right| &
\frac{x^d_{12}}{\Sigma_d}\,|\Lambda_{us}|\,\bar\varepsilon &
x^d_{13}\,|\Lambda_{ub}|\,\bar\varepsilon^3\,e^{i(\omega_{ub}-\omega_{us}-\omega_{cb}-\omega_{ud})} \\
-\frac{x^d_{12}}{\Sigma_d}\,|\Lambda_{us}|\,\bar\varepsilon &
\left|1-\frac{1}{2}\left(\frac{x^d_{12}}{\Sigma_d}\right)^2\Lambda_{ud}\,\bar\varepsilon^2\right| &
x^d_{23}\Sigma_d\,|\Lambda_{cb}|\,\bar\varepsilon^2 \\
(x^d_{12}x^d_{23}-x^d_{13})\,\bar\varepsilon^3\,e^{i(\omega_{cb}+\omega_{us})} &
-x^d_{23}\Sigma_d\,|\Lambda_{cb}|\,\bar\varepsilon^2\,e^{i\omega_{ud}} & 1
\end{array}\right)
\end{equation}
where $\Lambda_\alpha$ are complex corrections due to subleading terms, with
effective phase $\omega_\alpha$.  These subleading terms play a very important
role in the CKM phase $\delta_{CKM}$ which consists on a combination of 
subleading phases, the largest of these being $\omega_{ub}$:
\begin{eqnarray}
\omega_{ub} & = & \arg(\Lambda_{ub}) \nonumber \\
& = & \arg\left(1- x^d_{23} \left(\frac{x^u_{12}}{x^d_{13}}\right) \left(\frac{\Sigma_d}{\Sigma_u}\right) \frac{\varepsilon}{\bar\varepsilon}\, e^{-i(\delta_u-\delta_d)}\right),
\end{eqnarray}
followed by $\omega_{us}$:
\begin{equation}
\label{omega_us}
\omega_{us}=\arg\left(1-\left(\frac{x^u_{12}}{x^d_{12}}\right) \left(\frac{\Sigma_d}{\Sigma_u}\right) \frac{\varepsilon}{\bar\varepsilon}\,e^{-i(\delta_u-\delta_d)}
+(x^d_{12}x^d_{23}-x^d_{13})x^d_{23}\,\bar\varepsilon^2\,e^{2i(\beta_3-\chi)}+\ldots\right)
\end{equation}
Other $\Lambda_\alpha$ coefficients are $\Lambda_{ud},\Lambda_{cb} \simeq 1 +
O(\bar\varepsilon^2)$ and do not play an important role in the fit. However,
it is important to point out that the structure of these subleading terms is
highly model dependent, and that the values we get for them shall be valid
only in this model. Other $SU(3)$ models can have a completely different
subleading structure.

The phenomenological fit of~\cite{Roberts:2001zy}, based on structures similar
to those in Eq.~(\ref{yukawas}), was used in previous
works~\cite{Ross:2004qn,Calibbi:2008qt} to fix the $O(1)$ parameters of the
Yukawa matrices. In this work we make a new fit specifically for the $SU(3)$ model,
taking into account also flavon phases. Although the variation of the latter make very small changes in the CKM elements, such variations could make these elements exceed the $3\sigma$ bounds in~\cite{Amsler:2008zzb}.

The values found in~\cite{Amsler:2008zzb} for the CKM matrix are connected to
a large number of flavour observables. The analysis assumes that the
SM is the only source of flavour and CP violation. Nonetheless, the strong
consistency in the SM between all flavour observables, in particular those
participating in the construction of the unitarity triangle, suggests that new
physics effects are small, and that the main source of the registered flavour
and CP violation still lies in the CKM matrix. Thus, we shall adjust the
$O(1)$ parameters in $Y_u$ and $Y_d$ such that the CKM matrix
of~\cite{Amsler:2008zzb} is reproduced, assuming that SUSY contributions do
not affect the fit significantly.

To make a fit of the 11-dimensional parameter space of Eq.~(\ref{yukawas}), the Powell minimization method was used. The function to minimize was defined as a $\chi^2$ on the quark masses and on the four CKM parameters in the Wolfenstein parametrization: $\lambda$, $A$, $\bar\rho$ and $\bar\eta$, as in~\cite{Amsler:2008zzb}. We required the $O(1)$ parameters not to be larger than 2, and not smaller than $0.4$. We found that the following values give a very good fit on the masses and mixings:
\begin{subequations}
\begin{align}
\left(\begin{array}{c}
 x^d_{12} \\
 x^d_{13} \\
 x^d_{23}
\end{array}\right)=&
\left(\begin{array}{c}
 1.67 \\
 0.4 \\
 1.84
\end{array}\right); &
\left(\begin{array}{c}
 x^u_{12} \\
 x^u_{13} \\
 x^u_{23}
\end{array}\right)=&
\left(\begin{array}{c}
 1.42 \\
 2.0 \\
 2.0
\end{array}\right)
\end{align}
\begin{equation}
\alpha_u-\alpha_d= -0.69
\end{equation}
\end{subequations}
with $\beta'_2$, $\beta_3$ and $\chi$ non-uniquely determined. The results for $x^d_{ij}$ are similar to those in~\cite{Roberts:2001zy}, with a larger discrepancy in $x^d_{23}$. The value of $(\alpha_u-\alpha_d)$ is mainly fixed by the CKM phase and $\left|V_{us}\right|$, as reported in~\cite{Ross:2004qn}. To leading order, we can extract a $1\sigma$ uncertainty $\sigma^d_{ij}$ on the $x^d_{ij}$ constants from the errors on the CKM parameters~\cite{Amsler:2008zzb}. We roughly expect $\sigma^d_{12}\approx7\times10^{-3}$, $\sigma^d_{13}\approx5\times10^{-2}$ and $\sigma^d_{23}\approx4\times10^{-2}$. As the $x^u_{ij}$ parameters participate only in subleading terms, such a rough estimation is not possible. However, by fixing these constants at the above values, we can estimate the uncertainty on $\alpha_u-\alpha_d$, being $\sigma_{(\alpha_u-\alpha_d)}\sim0.04$.

\begin{figure}
\includegraphics[scale=.4]{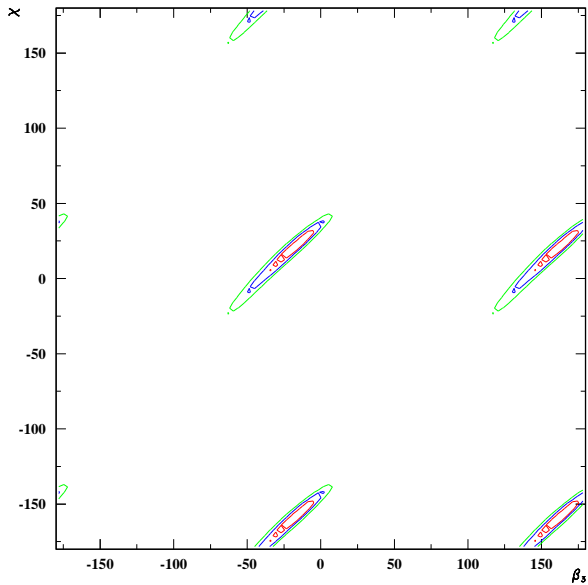}
\caption{Fit to Quark Masses and CKM matrix at one (red), two (blue) and three (green) sigma. There is a strong correlation between $\beta_3$ and $\chi$, but $\beta'_2$ does not affect CKM elements.} 
\label{fig:chifit}
\end{figure}

After fixing the determined $O(1)$s at these values, and taking $\alpha_d=0$, we make a grid-based $\chi^2$ analysis on the remaining three phases. We find out that the CKM matrix is completely independent of $\beta'_2$, as expected from Eq.~(\ref{CKM}). The result of the scan for $\beta_3$ and $\chi$ is shown in Figure~\ref{fig:chifit}. These two phases enter through subleading terms in the CKM matrix, and are severly constrained due to the precise determination of $\left|V_{us}\right|$, $\left|V_{cb}\right|$ and the diagonal elements. We are thus forced to place $(\beta_3,\chi)$ around $(-20^\circ,20^\circ)$, or on any of four degenerate spots obtained by adding $180^\circ$ to each. For definiteness, in the following we set $(\beta_3,\chi)=(-20^\circ,20^\circ)$, and $\beta'_2=0^\circ$.

Using these values, we compare the results of the fit with the measured values
for the CKM matrix and quark masses at $m_Z$~\cite{Dorsner:2006hw}. The results are shown in 
Tables~\ref{Table:MassFit} and~\ref{Table:CKMFit}. We
can see that the fit is very good and almost all of the SM parameters lie
within their $1\sigma$ values.

\begin{table}
\begin{center}
\begin{tabular}{|c||c|c|c||c|c|c|}
\hline
& $m_u$ & $m_c$ & $m_t$ & $m_d$ & $m_s$ & $m_b$ \\
\hline
$SU(3)$ (GeV) & $2.9\times10^{-3}$ & $0.57$ & $172.0$ & $4.1\times10^{-3}$ & $71\times10^{-3}$ & $2.85$ \\
Reference (GeV) & $(1.4\pm 0.5)\times10^{-3}$ & $0.63\pm.08$ & $170.3\pm2.4$ & $(3\pm1.2)\times10^{-3}$ & $(56\pm16)\times10^{-3}$ & $2.89\pm0.11$ \\
\hline
\end{tabular}
\end{center}
\caption{\label{Table:MassFit} $SU(3)$ predictions for quark masses at $m_Z$ and corresponding values obtained by running to the same scale~\cite{Dorsner:2006hw}.}
\end{table}

\begin{table}
\begin{center}
\begin{tabular}{|c||c|c|c|c|}
\hline
& $\lambda$ & $A$ & $\bar\rho$ & $\bar\eta$ \\
\hline
$SU(3)$ & $0.2250$ & $0.805$ & $0.156$ & $0.327$ \\
Reference & $0.2257^{+0.0009}_{-0.0010}$ & $0.814^{+0.021}_{-0.022}$ & $0.135^{+0.031}_{-0.016}$ & $0.349^{+0.015}_{-0.017}$ \\
\hline
\end{tabular}
\end{center}
\caption{\label{Table:CKMFit} $SU(3)$ predictions for CKM elements and corresponding measured values, taken from~\cite{Amsler:2008zzb}.}
\end{table}

Regarding the lepton sector, we shall unify the structure of quark and lepton flavoured matrices, so the Yukawas can be written as:
\begin{eqnarray}
Y_e=\left( 
\begin{array}{ccc}
0 & x^e_{12}\,\bar\varepsilon^3\,e^{i\delta_d} & x^e_{13}\,\bar\varepsilon^3\,e^{i\delta_d+\beta_3} \\ 
x^e_{12}\,\bar\varepsilon^3\,e^{i\delta_d} & \Sigma_e\bar\varepsilon^2 & 
x^e_{23}\,\Sigma_e\,\bar\varepsilon^2\,e^{i\beta_3} \\ 
x^e_{13}\,\bar\varepsilon^3\,e^{i\delta_d+\beta_3} & x^e_{23}\,\Sigma_e\,\bar\varepsilon^{2}\,e^{i\beta_3} & e^{2i\chi}
\end{array}
\right)y_b, \\
Y_\nu=\left(
\begin{array}{ccc}
0 & x^u_{12}\,\varepsilon^3\,e^{i\delta_u} & x^u_{13}\,\varepsilon^3\,e^{i\delta_u+\beta_3} \\ 
x^u_{12}\,\varepsilon^3\,e^{i\delta_u} & 0 & 0 \\ 
x^u_{13}\,\varepsilon^3\,e^{i\delta_u+\beta_3} & 0 & 1
\end{array}
\right )y_t
\end{eqnarray}
with and $\Sigma_e=3\Sigma_d$ and $\Sigma_\nu=0$. Notice that in $Y_\nu$ the zeroes are actually terms of order higher than $\varepsilon^3$. We use the same $O(1)$s in $Y_u$ for $Y_\nu$ and we find that the $O(1)$s that fit best the charged lepton masses are:
\begin{equation}
\left(\begin{array}{c}
 x^e_{12} \\
 x^e_{13} \\
 x^e_{23}
\end{array}\right)=
\left(\begin{array}{c}
 1.31 \\
 2.0 \\
 2.0
\end{array}\right)
\end{equation}
which give the values in Table~\ref{Table:LepMassFit}. Note that lepton masses
are known very precisely and thus subleading contributions will play an
important role in the fit. For this reason, we did not try to fit exactly the
reference values in Table~\ref{Table:LepMassFit}.

For more details on how the $SU(3)$ model accomodates neutrino mixing, we refer the interested reader to~\cite{Ross:2004qn}.

\begin{table}
\begin{center}
\begin{tabular}{|c||c|c|c|}
\hline
& $m_e$ & $m_\mu$ & $m_\tau$ \\
\hline
$SU(3)$ (GeV) &   $4.867808\times10^{-4}$     & $104.47907\times10^{-3}$     & $1.77634$ \\
Reference (GeV) & $4.866613(36)\times10^{-4}$ & $102.72899(44)\times10^{-3}$ & $1.74645^{+.00029}_{-.00026}$ \\
\hline
\end{tabular}
\end{center}
\caption{\label{Table:LepMassFit} $SU(3)$ predictions for lepton masses at
  $m_Z$ and corresponding values obtained by running to the same
  scale~\cite{Dorsner:2006hw}.} 
\end{table}

\section{$O(1)$ Coefficients in the SCKM Basis at the GUT Scale}





In general, flavour symmetry models contain unknown $O(1)$ coefficients at the flavour scale. In CPV and FCNC studies we are interested in the leading real and imaginary contributions. However, we find that in some particular observables, leading phases might cancel. In such a situation it is important to know the structure of subleading complex terms in order to make verifiable predictions. For this purpose, we present here the structure of the $O(1)$ coefficients for the leading terms after the SCKM rotation.

In the following, we shall define $x^f_{\delta}=(x^f_{12}x^f_{23}-x^f_{13})$, with $f=u,d,e$.

\subsection{RVV1}

Our soft matrices in the flavour basis have the following structure at the GUT scale:
\begin{equation}
M_{\tilde Q}^2=\left(\begin{array}{ccc}
1+Q_1\,\varepsilon^2\,y_t & 0 & 0 \\
0 & 1+Q_2\,\varepsilon^2 & Q_3\,\varepsilon^2\,e^{i\beta_3} \\
0 & Q_3\,\varepsilon^2\,e^{-i\beta_3} & 1+Q_4\,y_t
\end{array}\right)m_0^2, \nonumber
\end{equation} 
with exactly the same structure, but different $O(1)$ constants $U_i$ and $L_i$, for $M^2_{\tilde u_R^c}$ and $M^2_{\tilde L}$, respectively. For $M^2_{\tilde d_R^c}$ and $M^2_{\tilde e_R^c}$ we have an analogous structure, with $\varepsilon$ replaced by $\bar\varepsilon$, $y_t$ replaced by $y_b$, and different $O(1)$ constants $D_i$ and $E_i$.
This shall be true for all variations of the model, so in following subsections we shall only show $M_{\tilde Q}^2$ without repeating these specifications.

Rotation into the SCKM basis gives Eq~(\ref{sckm1}), with the following $O(1)$ coefficients:
\begin{subequations}
\begin{eqnarray}
M_{\tilde u_R^c}^2(2,1) & = &(U_2-U_1\,y_t)(x_{12}^u/\Sigma_u) \\
M_{\tilde u_R^c}^2(3,1) & = & U_3(x_{12}^u/\Sigma_u)-U_4\,x_{\delta}^u\,y_t\,e^{2i\beta_3} \\
M_{\tilde u_R^c}^2(3,2) & = & U_3 - U_4\,x_{23}^u\Sigma_u\,y_t\,e^{2i\beta_3}
\end{eqnarray}
\end{subequations}
\begin{subequations}
\begin{eqnarray}
M_{\tilde d_R^c}^2(2,1) & = &(D_2-D_1\,y_b)(x_{12}^d/\Sigma_d) \\
M_{\tilde d_R^c}^2(3,1) & = & D_3(x_{12}^d/\Sigma_d)-D_4\,x_{\delta}^d\,y_b\,e^{-2i(\chi-\beta_3)} \\
M_{\tilde d_R^c}^2(3,2) & = & D_3 - D_4\,x_{23}^d\Sigma_d\,y_b\,e^{-2i(\chi-\beta_3)}
\end{eqnarray}
\end{subequations}
\begin{subequations}
\begin{eqnarray}
M_{\tilde Q}^2(2,1) & = & (Q_2-Q_1\,y_t)(x_{12}^d/\Sigma_d) \\
M_{\tilde Q}^2(3,1) & = & -Q_4\,x_{\delta}^d + \frac{\varepsilon^2}{\bar\varepsilon^2}\frac{Q_3}{y_t}(x_{12}^d/\Sigma_d)\,e^{2i(\chi-\beta_3)} \\
M_{\tilde Q}^2(3,2) & = & Q_4\,x_{23}^d\,\Sigma_d - \frac{\varepsilon^2}{\bar\varepsilon^2}\,\frac{Q_3}{y_t}\,e^{2i(\chi-\beta_3)}
\end{eqnarray}
\end{subequations}
\begin{subequations}
\begin{eqnarray}
M_{\tilde e_R^c}^2(2,1) & = &(E_2-E_1\,y_b)(x_{12}^e/\Sigma_e) \\
M_{\tilde e_R^c}^2(3,1) & = & E_3(x_{12}^e/\Sigma_e)-E_4\,x_{\delta}^e\,y_b\,e^{-2i(\chi-\beta_3)} \\
M_{\tilde e_R^c}^2(3,2) & = & E_3 - E_4\,x_{23}^e\Sigma_e\,y_b\,e^{-2i(\chi-\beta_3)}
\end{eqnarray}
\end{subequations}
\begin{subequations}
\begin{eqnarray}
M_{\tilde L}^2(2,1) & = & (L_2-L_1\,y_t)(x_{12}^e/\Sigma_e) \\
M_{\tilde L}^2(3,1) & = & -L_4\,x_{\delta}^e + \frac{\varepsilon^2}{\bar\varepsilon^2}\frac{L_3}{y_t}(x_{12}^e/\Sigma_e)\,e^{2i(\chi-\beta_3)} \\
M_{\tilde L}^2(3,2) & = & L_4\,x_{23}^e\,\Sigma_e - \frac{\varepsilon^2}{\bar\varepsilon^2}\,\frac{L_3}{y_t}\,e^{2i(\chi-\beta_3)}
\end{eqnarray}
\end{subequations}

\subsection{RVV2}

The K\"ahler potential in RVV2 has got the minimal terms with the addition of the $\theta_3\bar\theta_{23}$ flavon vevs. This modifies the $(2,3)$ and $(3,2)$ elements of the soft mass matrices, giving them the following structure at the GUT scale:
\begin{equation}
M_{\tilde Q}^2=\left(\begin{array}{ccc}
1+Q_1\,\varepsilon^2\,y_t & 0 & 0 \\
0 & 1+Q_2\,\varepsilon^2 & Q_5\,\varepsilon\,y_t^{0.5}\,e^{i\beta'_2} \\
0 & Q_5\,\varepsilon\,y_t^{0.5}\,e^{-i\beta'_2} & 1+Q_4\,y_t
\end{array}\right)m_0^2. \nonumber
\end{equation} 

For simplicity, we have omitted the minimal $X_3$ ($X=U,D,Q,E,L$) $O(1)$s, replacing them by the corresponding $X_5$ that accompany the $\theta_3\bar\theta_{23}$ flavon vev. The $O(1)$ coefficients are:
\begin{subequations}
\begin{eqnarray}
M_{\tilde u_R^c}^2(2,1) & = &(U_2-U_1\,y_t)(x_{12}^u/\Sigma_u) \\
M_{\tilde u_R^c}^2(3,1) & = & U_5(x_{12}^u/\Sigma_u) - \varepsilon\,U_4\,x_{\delta}^u\,y_t^{0.5}\,e^{i(\beta_3+\beta'_2)} \\
M_{\tilde u_R^c}^2(3,2) & = & U_5 - \varepsilon\,U_4\,x_{23}^u\,\Sigma_u\,y_t^{0.5}\,e^{i(\beta_3+\beta'_2)}
\end{eqnarray}
\end{subequations}
\begin{subequations}
\begin{eqnarray}
M_{\tilde d_R^c}^2(2,1) & = &(D_2-D_1\,y_b)(x_{12}^d/\Sigma_d) \\
M_{\tilde d_R^c}^2(3,1) & = & D_5(x_{12}^d/\Sigma_d) - \bar\varepsilon\,D_4\,x_{\delta}^d\,y_b^{0.5}\,e^{-i(\chi-\beta_3-\beta'_2)} \\
M_{\tilde d_R^c}^2(3,2) & = & D_5 - \bar\varepsilon\,D_4\,x_{23}^d\,\Sigma_d\,y_b^{0.5}\,e^{-i(\chi-\beta_3-\beta'_2)}
\end{eqnarray}
\end{subequations}
\begin{subequations}
\begin{eqnarray}
M_{\tilde Q}^2(2,1) & = & (Q_2-Q_1\,y_t)(x_{12}^d/\Sigma_d)  \nonumber \\
& & - \frac{\bar\varepsilon^2}{\varepsilon}\,Q_5\,y_t^{0.5}\left(x_\delta^d\,e^{i(-2\chi+\beta_3+\beta'_2)} + x_{12}^dx_{23}^d\,e^{-i(-2\chi+\beta_3+\beta'_2)}\right) \\
M_{\tilde Q}^2(3,1) & = & Q_5(x_{12}^d/\Sigma_d) - \frac{\bar\varepsilon^2}{\varepsilon}Q_4\,x_\delta^d\,y_t^{0.5}\,e^{i(-2\chi+\beta_3+\beta_2')} \\
M_{\tilde Q}^2(3,2) & = & Q_5 - \frac{\bar\varepsilon^2}{\varepsilon}Q_4\,x_{23}^d\,\Sigma_d\,y_t^{0.5}\,e^{i(-2\chi+\beta_3+\beta_2')}
\end{eqnarray}
\end{subequations}
\begin{subequations}
\begin{eqnarray}
M_{\tilde e_R^c}^2(2,1) & = &(E_2-E_1\,y_b)(x_{12}^e/\Sigma_e) \\
M_{\tilde e_R^c}^2(3,1) & = & E_5(x_{12}^e/\Sigma_e) - \bar\varepsilon\,E_4\,x_{\delta}^e\,y_b^{0.5}\,e^{-i(\chi-\beta_3-\beta'_2)} \\
M_{\tilde e_R^c}^2(3,2) & = & E_5 - \bar\varepsilon\,E_4\,x_{23}^e\,\Sigma_e\,y_b^{0.5}\,e^{-i(\chi-\beta_3-\beta'_2)}
\end{eqnarray}
\end{subequations}
\begin{subequations}
\begin{eqnarray}
M_{\tilde L}^2(2,1) & = & (L_2-L_1\,y_t)(x_{12}^e/\Sigma_e) \nonumber \\
& & -\frac{\bar\varepsilon^2}{\varepsilon}\,L_5\,y_t^{0.5}\left(x_\delta^e\,e^{i(-2\chi+\beta_3+\beta'_2)} + x_{12}^ex_{23}^e\,e^{-i(-2\chi+\beta_3+\beta'_2)}\right) \\
M_{\tilde L}^2(3,1) & = & L_5(x_{12}^e/\Sigma_e) - \frac{\bar\varepsilon^2}{\varepsilon}L_4\,x_\delta^e\,y_t^{0.5}\,e^{i(-2\chi+\beta_3+\beta_2')} \\
M_{\tilde L}^2(3,2) & = & L_5 - \frac{\bar\varepsilon^2}{\varepsilon}L_4\,x_{23}^e\,\Sigma_e\,y_t^{0.5}\,e^{i(-2\chi+\beta_3+\beta_2')}
\end{eqnarray}
\end{subequations}

\subsection{RVV Model 3}

The addition of an effective term with the antisymmetric $(\theta_3\theta_{23})\theta_3$ flavon vevs gives the soft matrices the following structure, in the flavour basis, at the GUT scale:
\begin{equation}
M_{\tilde Q}^2=\left(\begin{array}{ccc}
1+Q_1\,\varepsilon^2\,y_t & 0 & Q_6\,\varepsilon\,y_t \\
0 & 1+Q_2\,\varepsilon^2 & Q_3\,\varepsilon^2\,e^{i\beta_3} \\
Q_6\,\varepsilon\,y_t & Q_3\,\varepsilon^2\,e^{-i\beta_3} & 1+Q_4\,y_t
\end{array}\right)m_0^2. \nonumber
\end{equation} 

The additional term in the $(1,3)$ and $(3,1)$ sectors of the mass matrices induce large terms in the $(1,2)$ and $(2,1)$ sectors of the mass matrices, and have a moderate influence on the $(2,3)$ and $(3,2)$ sectors. The structure of the $O(1)$ coefficients follow:
\begin{subequations}
\begin{eqnarray}
M_{\tilde u_R^c}^2(2,1) & = &(U_2-U_1\,y_t)(x_{12}^u/\Sigma_u) + U_6\,x_{23}^u\,\Sigma_u\,y_t\,e^{-i(\beta_3+\delta_u)} \\
M_{\tilde u_R^c}^2(3,1) & = & U_6 \\
M_{\tilde u_R^c}^2(3,2) & = & U_3 - U_4\,x_{23}^u\Sigma_u\,y_t\,e^{2i\beta_3} + U_6\,y_t\,(x_{12}^u/\Sigma_u)\,e^{i(\beta_3-\delta_u)}
\end{eqnarray}
\end{subequations}
\begin{subequations}
\begin{eqnarray}
M_{\tilde d_R^c}^2(2,1) & = &(D_2-D_1\,y_b)(x_{12}^d/\Sigma_d) + D_6\,x^d_{23}\,\Sigma_d\,y_b\,e^{i(2\chi-\beta_3-\delta_d)} \\
M_{\tilde d_R^c}^2(3,1) & = & D_6 \\
M_{\tilde d_R^c}^2(3,2) & = & D_3 - D_4\,x_{23}^d\,\Sigma_d\,y_b\,e^{-2i(\chi-\beta_3)} + D_6\,y_b\,(x_{12}^d/\Sigma_d)\,e^{i(\beta_3-\delta_d)}
\end{eqnarray}
\end{subequations}
\begin{subequations}
\begin{eqnarray}
M_{\tilde Q}^2(2,1) & = & Q_6\,x_{23}^d\,\Sigma_d + \frac{\varepsilon}{\bar\varepsilon}\,\frac{(Q_2-Q_1\,y_t)}{y_t}(x_{12}^d/\Sigma_d)\,e^{i(-2\chi+\beta_3+\delta_d)} \\
M_{\tilde Q}^2(3,1) & = & Q_6 \\
M_{\tilde Q}^2(3,2) & = & Q_4\,x_{23}^d\,\Sigma_d - \frac{\varepsilon}{\bar\varepsilon}\left[Q_6(x_{12}^d/\Sigma_d)e^{-i\delta_d} + \frac{\varepsilon}{\bar\varepsilon}\frac{Q_3}{y_t}e^{-i\beta_3}\right]\,e^{i(2\chi-\beta_3)}
\end{eqnarray}
\end{subequations}
\begin{subequations}
\begin{eqnarray}
M_{\tilde e_R^c}^2(2,1) & = &(E_2-E_1\,y_b)(x_{12}^e/\Sigma_e) + E_6\,x^e_{23}\,\Sigma_e\,y_b\,e^{i(2\chi-\beta_3-\delta_d)} \\
M_{\tilde e_R^c}^2(3,1) & = & E_6 \\
M_{\tilde e_R^c}^2(3,2) & = & E_3 - E_4\,x_{23}^e\,\Sigma_e\,y_b\,e^{-2i(\chi-\beta_3)} + E_6\,y_b\,(x_{12}^e/\Sigma_e)\,e^{i(\beta_3-\delta_d)}
\end{eqnarray}
\end{subequations}
\begin{subequations}
\begin{eqnarray}
M_{\tilde L}^2(2,1) & = & L_6\,x_{23}^e\,\Sigma_e + \frac{\varepsilon}{\bar\varepsilon}\,\frac{(L_2-L_1\,y_t)}{y_t}(x_{12}^e/\Sigma_e)\,e^{i(-2\chi+\beta_3+\delta_d)} \\
M_{\tilde L}^2(3,1) & = & L_6 \\
M_{\tilde L}^2(3,2) & = & L_4\,x_{23}^e\,\Sigma_e - \frac{\varepsilon}{\bar\varepsilon}\left[L_6(x_{12}^e/\Sigma_e)e^{-i\delta_d} + \frac{\varepsilon}{\bar\varepsilon}\frac{L_3}{y_t}e^{-i\beta_3}\right]\,e^{i(2\chi-\beta_3)}
\end{eqnarray}
\end{subequations}

\subsection{A-Terms}

For all of the evaluated models, the A-Terms will have the same structure the Yukawas have, but with different $O(1)$ constants. We parametrize them in the following way:
\begin{equation}
A_d=\left(\begin{array}{ccc}
0 & A_1^d\,x_{12}^d\,\bar\varepsilon^3\,e^{i\delta_d} & A_2^d\,x_{13}^d\,\bar\varepsilon^3\,e^{i(\beta_3+\delta_d)} \\
A_1^d\,x_{12}^d\,\bar\varepsilon^3\,e^{i\delta_d} & A_3^d\,\Sigma_d\,\bar\varepsilon^2 & A_4^d\,x_{23}^d\,\Sigma_d\,\bar\varepsilon^2\,e^{i\beta_3} \\
A_2^d\,x_{13}^d\,\bar\varepsilon^3\,e^{i(\beta_3+\delta_d)} & A_4^d\,x_{23}^d\,\Sigma_d\,\bar\varepsilon^2\,e^{i\beta_3} & A_5^d\,e^{2i\chi}
\end{array}\right)A_0\,y_b
\end{equation}
and similarly for $A_u$ and $A_e$.

The A-Terms are rotated into the SCKM basis using the same unitary matrices that diagonalize the Yukawas. However, as the $O(1)$ constants in each entry are different from the ones in the Yukawas, the A-Terms are not diagonalized, and mantain the same structure. For squarks, the remaining off-diagonal terms are then affected by the rephasings, and we obtain:
\begin{eqnarray}
\label{aterms-sckm}
A_u=\left(\begin{array}{ccc}
0 & A_2^{\prime u}\,x_{12}^u\,\varepsilon^3\,e^{-i\omega'} & A_2^{\prime u}\,x_{13}^u\,\varepsilon^3\,e^{i(2\chi-\omega')} \\
A_1^{\prime u}\,x_{12}^u\,\varepsilon^3\,e^{i\omega'} & A_3^u\,\Sigma_u\,\varepsilon^2 & A_4^{\prime u} \,x_{23}^u
\,\Sigma_u\,\varepsilon^2\,e^{2i\chi} \\
A_2^{\prime u}\,x_{13}^u\,\varepsilon^3\,e^{i(\omega'+2\beta_3-2\chi)} & A_4^{\prime u} \,x_{23}^u
\,\Sigma_u\,\varepsilon^2\,e^{2i(\beta_3-\chi)} & A_5^u
\end{array}\right)A_0\,y_t \\
A_d=\left(\begin{array}{ccc}
0 & A_2^{\prime d}\,x_{12}^d\,\bar\varepsilon^3\,e^{-i\omega_{us}} & A_2^{\prime d}\,x_{13}^d\,\bar\varepsilon^3\,e^{-i\omega_{us}} \\
A_1^{\prime d}\,x_{12}^d\,\bar\varepsilon^3\,e^{-i\omega_{us}} & A_3^d\,\Sigma_d\,\bar\varepsilon^2 & A_4^{\prime d} \,x_{23}^d\,\Sigma_d\,\bar\varepsilon^2 \\
A_2^{\prime d}\,x_{13}^d\,\bar\varepsilon^3\,e^{i(\omega_{us}+2\beta_3-2\chi)} & A_4^{\prime d}\,x_{23}^d
\,\Sigma_d\,\bar\varepsilon^2\,e^{2i(\beta_3-\chi)} & A_5^d
\end{array}\right)A_0\,y_b \\
A_e=\left(\begin{array}{ccc}
0 & A_2^{\prime e}\,x_{12}^e\,\bar\varepsilon^3 & A_2^{\prime e}\,x_{13}^e\,\bar\varepsilon^3\,e^{i(\beta_3-\chi)} \\
A_1^{\prime e}\,x_{12}^e\,\bar\varepsilon^3 & A_3^e\,\Sigma_e\,\bar\varepsilon^2 & A_4^{\prime e} \,x_{23}^e\,\Sigma_e\,\bar\varepsilon^2\,e^{i(\beta_3-\chi)} \\
A_2^{\prime e}\,x_{13}^e\,\bar\varepsilon^3\,e^{i(\beta_3-\chi)} & A_4^{\prime e}\,x_{23}^e
\,\Sigma_e\,\bar\varepsilon^2\,e^{i(\beta_3-\chi)} & A_5^e
\end{array}\right)A_0\,y_b
\end{eqnarray}
with:
\begin{subequations}
 \begin{eqnarray}
A_1^{\prime f} & = & (A_1^f-A_3^f) \\
A_2^{\prime f} & = & (A_2^f-A_5^f)-(A_4^f-A_5^f)(x_{12}^f x_{23}^f/x_{13}^f) \\
A_4^{\prime f} & = & (A_4^f-A_5^f)
 \end{eqnarray}
\end{subequations}
with no subleading phases generated by the SCKM rotation. Notice that if the initial $A_i^f=1$, the A-terms are aligned with the Yukawas. Thus, the new $A_i^{\prime f}$ go to zero, and the A-terms are diagonal at the SCKM basis.


\section{O(1) Constants due to RGEs}

Using the leading log approximation, we can estimate the effect of the running on the flavoured matrices. For this we define the negative parameter $\Delta$, such that:
\begin{equation}
 |\Delta| = \frac{1}{16\pi}\left|\log{\frac{M_{SUSY}}{M_{GUT}}}\right|\gtrsim\mathcal{O}(\bar\varepsilon) \nonumber
\end{equation}
The rotation to the SCKM basis must be done at $M_W$. However, since the off-diagonal RGE contributions to the Yukawa matrices are negligible, we can perform the rotation at the GUT scale, and then apply the corrections from the running.

For $M_{\tilde L}$, one must take into account that only the $(3,3)$ term of $Y_\nu$ gives a significant contribution to the running. This parameter only participates until the heavy $(\nu_R^c)_3$ decouples, at the scale $M_3$. To take into account this effect, we introduce the parameter:
\begin{equation}
 \Delta_\nu = \frac{1}{\Delta}\frac{1}{16\pi}\log{\frac{M_3}{M_{GUT}}} \nonumber
\end{equation}

With the exception of the A-Terms, the general result for all models is that the structure in terms of $\varepsilon$ and $\bar\varepsilon$ remains unchanged. Even though the generated parameters are not necessarily of $O(1)$, the $\Delta$ parameter provides an additional suppression, which approximately keeps the matrix structure very similar to the one at the GUT scale.

These induced coefficients can have two parts: one independent of the flavour structure of the soft masses (the MFV contribution, generated by the misalignment of the Yukawa matrices), and one dependent on the flavour structure. In the following equations, those parts independent of the soft mass flavour structure can be distinguished because they do not vanish when the $X_i$ terms ($X=U,D,Q,E,L$) are set to zero and when the $A_i$ terms are set to one (this also sets all primed variables to zero). They are only present in the $M_{\tilde Q}^2$, $M_{\tilde L}^2$ and $A_i$ matrices.

In the following, we shall denote $M_{\tilde X}^2(i,j)$, at the scale $\mu$, as $X_{ij}^\mu$, with $X=U,D,Q,E,L$. For all soft mass terms, we shall factorize the $\bar\varepsilon$ and $\varepsilon$ factors that appear in the mass matrices presented in Eq.~(\ref{sckm1}), as well as the leading phase. Since we shall also factorize $m_0^2$, we define $a_0=A_0/m_0$. 

\subsection{RVV Model 1}

\begin{subequations}
 \begin{eqnarray}
U_{21}^{EW} & = & U_{21}^{GUT} \\
U_{31}^{EW} & = & U_{31}^{GUT} + 2\left\{ U^{GUT}_{31} - 2a_0^2\,A_2^{\prime u}\,A^u_5\,x_{13}^u\,e^{2i\beta_3}\right\}y_t^2\,\Delta \\
U_{32}^{EW} & = & U_{32}^{GUT} + 2\left\{ U^{GUT}_{32} + 2a_0^2\,A_4^{\prime u}\,A^u_5\,x_{23}^u\,\Sigma_u\,e^{2i\beta_3}\right\}y_t^2\,\Delta
 \end{eqnarray}
\end{subequations}
\begin{subequations}
 \begin{eqnarray}
D_{21}^{EW} & = & D_{21}^{GUT} \\
D_{31}^{EW} & = & D_{31}^{GUT} + 2\left\{ D^{GUT}_{31} - 2a_0^2\,A_2^{\prime d} \,A^d_5\,x_{13}^d\,e^{-2i(\chi-\beta_3)}\right\}y_b^2\,\Delta \\
D_{32}^{EW} & = & D_{32}^{GUT} + 2\left\{ D^{GUT}_{32} + 2a_0^2\,A_4^{\prime d} \,A^d_5\,x_{23}^d\,\Sigma_d\,e^{-2i(\chi-\beta_3)}\right\}y_b^2\,\Delta
 \end{eqnarray}
\end{subequations}
\begin{subequations}
 \begin{eqnarray}
Q_{21}^{EW} & = & Q_{21}^{GUT} 
- 2\frac{\bar\varepsilon^4}{\varepsilon^2}(3+a_0^2(A_5^u)^2)x_\delta^d\,x_{23}^d\,\Sigma_d\,y_t^2\,\Delta \nonumber \\
& & + \frac{\bar\varepsilon^4}{\varepsilon^2}
\left\{2(Q_4\,y_t+U_4\,y_t)x_\delta^d\,x_{23}^d\,y_t^2
+2a_0^2(A_2^{\prime d}A_4^{\prime d}x^d_{13}x^d_{23}+A_1^{\prime d}A^d_3x^d_{12})y_b^2\right\}\Sigma_d\,\Delta     \\
Q_{31}^{EW} & = & Q_{31}^{GUT} - 2(3+a_0^2\,(A_5^u)^2)x_{\delta}^d\,y_t\,\Delta
+2\frac{\varepsilon^2}{\bar\varepsilon^2}(3+a_0^2A_4^uA_5^u)(x_{12}^d/\Sigma_d)x_{23}^u\,\Sigma_u\,y_t\,e^{2i\chi}\,\Delta \nonumber \\
& & -\left\{Q_{31}^{GUT}\,(y_b^2+y_t^2) + (Q_4+2U_4)x_{\delta}^dy_t^2 + 2a_0^2\,A_2^{\prime d} \,A^d_5\,x_{13}^d\,\frac{y_b^2}{y_t}\right\}\,\Delta \nonumber \\
& & +\frac{\varepsilon^2}{\bar\varepsilon^2}(Q_4+2U_4)(x_{12}/\Sigma_d)x_{23}^u\Sigma_u\,y_t^2\,e^{2i\chi}\,\Delta \\
Q_{32}^{EW} & = & Q_{32}^{GUT} + 2(3+a_0^2\,(A_5^u)^2)x_{23}^d\,\Sigma_d\,\,y_t\,\Delta
-2\frac{\varepsilon^2}{\bar\varepsilon^2}(3+a_0^2A_4^uA_5^u)x_{23}^u\,\Sigma_u\,y_t e^{2i\chi}\Delta \nonumber \\
& + & \left\{\frac{Q_{32}^{GUT}}{x_{23}^d\Sigma_d}(y_b^2+y_t^2) + (Q_4+2U_4)\left(1-\frac{\varepsilon^2}{\bar\varepsilon^2}\frac{x_{23}^u\Sigma_u}{x_{23}^d\Sigma_d}e^{2i\chi}\right)y_t^2 - 2a_0^2A_4^{\prime d}A^d_5\frac{y_b^2}{y_t}\right\}x_{23}^d\Sigma_d\Delta 
 \end{eqnarray}
\end{subequations}
\begin{subequations}
 \begin{eqnarray}
E_{21}^{EW} & = & E_{21}^{GUT} \\
E_{31}^{EW} & = & E_{31}^{GUT} + 2\left\{ E^{GUT}_{31} - 2a_0^2\,A_2^{\prime e} \,A^e_5\,x_{13}^d\,e^{-2i(\chi-\beta_3)}\right\}y_b^2\,\Delta \\
E_{32}^{EW} & = & E_{32}^{GUT} + 2\left\{ E^{GUT}_{32} + 2a_0^2\,A_4^{\prime e} \,A^e_5\,x_{23}^e\,\Sigma_e\,e^{-2i(\chi-\beta_3)}\right\}y_b^2\,\Delta
 \end{eqnarray}
\end{subequations}
\begin{subequations}
 \begin{eqnarray}
L_{21}^{EW} & = & L_{21}^{GUT} 
- 4\frac{\bar\varepsilon^4}{\varepsilon^2}x_\delta^e\,x_{23}^e\,\Sigma_e\,y_t^2\,\Delta_\nu\,\Delta \nonumber \\
& & + \frac{\bar\varepsilon^4}{\varepsilon^2}
\left\{2(L_4\,y_t)x_\delta^e\,x_{23}^e\,y_t^2\,\Delta_\nu
+2a_0^2(A_2^{\prime e}A_4^{\prime e}x^e_{13}x^e_{23}+A_1^{\prime e}A^e_3x^e_{12})y_b^2\right\}\Sigma_e\,\Delta     \\
L_{31}^{EW} & = & Q_{31}^{GUT} - 4x_{\delta}^e\,y_t\,\Delta_\nu\,\Delta \nonumber \\
& & -\left\{Q_{31}^{GUT}\,(y_b^2+y_t^2\,\Delta_\nu) + L_4x_{\delta}^e y_t^2\,\Delta_\nu + 2a_0^2\,A_2^{\prime e} \,A^e_5\,x_{13}^e\,\frac{y_b^2}{y_t}\right\}\,\Delta \\
L_{32}^{EW} & = & Q_{32}^{GUT} + 4x_{23}^e\,\Sigma_e\,\,y_t\,\Delta_\nu\,\Delta \nonumber \\
& + & \left\{\frac{Q_{32}^{GUT}}{x_{23}^e\Sigma_e}(y_b^2+y_t^2\,\Delta_\nu) + L_4y_t^2\,\Delta_\nu - 2a_0^2A_4^{\prime e}A^e_5\frac{y_b^2}{y_t}\right\}x_{23}^e\Sigma_e\Delta 
 \end{eqnarray}
\end{subequations}

\subsection{RVV Model 2}

\begin{subequations}
\begin{eqnarray}
U^{EW}_{21} & = & U^{GUT}_{21} \\
U^{EW}_{31} & = & U^{GUT}_{31} + 2\left\{U^{GUT}_{31} - 2a_0^2\,\frac{\varepsilon}{y_t^{0.5}}A_2^{\prime u}\,A^u_5\,x_{13}^u\,e^{i(\beta_3+\beta'_2)}\right\}y_t^2\,\Delta \\
U^{EW}_{32} & = & U^{GUT}_{32} + 2\left\{ U^{GUT}_{32} + 2a_0^2\,\frac{\varepsilon}{y_t^{0.5}}A_4^{\prime u}\,A^u_5\,x_{23}^u\,\Sigma_u\,e^{i(\beta_3+\beta'_2)}\right\}y_t^2\,\Delta
\end{eqnarray}
\end{subequations}
\begin{subequations}
\begin{eqnarray}
D^{EW}_{21} & = & D^{GUT}_{21} \\
D^{EW}_{31} & = & D^{GUT}_{31} + \left\{2D^{GUT}_{31} - 2a_0^2\,\frac{\bar\varepsilon}{y_b^{0.5}}\,A_2^{\prime d} \,A^d_5\,x_{13}^d\,e^{-i(\chi-\beta_3-\beta'_2)}\right\}y_b^2\Delta \\
D^{EW}_{32} & = & D^{GUT}_{32} + 2\left\{D^{GUT}_{32} + 2a_0^2\,\frac{\bar\varepsilon}{y_b^{0.5}}\,A_4^{\prime d} \,A^d_5\,x_{23}^d\,\Sigma_d\,e^{-i(\chi-\beta_3-\beta'_2)}\right\}y_b^2\,\Delta
\end{eqnarray}
\end{subequations}
\begin{subequations}
\begin{eqnarray}
Q^{EW}_{21} & = & Q_{21}^{GUT} 
+ 2\frac{\bar\varepsilon^4}{\varepsilon^2}(3+a_0^2(A_5^u)^2)x_\delta^d\,x_{23}^d\,\Sigma_d\,y_t^2\,\Delta \nonumber \\
& & + \frac{\bar\varepsilon^4}{\varepsilon^2}
\left\{2(Q_4\,y_t+U_4\,y_t)x_\delta^d\,x_{23}^d\,y_t^2
+2a_0^2(A_2^{\prime d}A_4^{\prime d}x_{13}^dx_{23}^d+A_1^{\prime d}A^d_3x_{12}^d)y_b^2\right\}\Sigma_d\,\Delta  \nonumber \\
& & - \frac{\bar\varepsilon^2}{\varepsilon}\,Q_5\,y_t^{0.5}\left(x_\delta^d\,e^{i(-2\chi+\beta_3+\beta'_2)} + x_{12}^dx_{23}^d\,e^{i(2\chi-\beta_3-\beta'_2)}\right)y_t^2\Delta \\
Q^{EW}_{31} & = & Q^{GUT}_{31} + 2\frac{\bar\varepsilon^2}{\varepsilon}(3+a_0^2(A_5^u)^2)\frac{x_\delta^d}{y_t^{0.5}}y_t^2e^{i(-2\chi+\beta_3+\beta'_2)}\Delta \nonumber \\ 
& - & \left\{Q^{GUT}_{31}(y_t^2+y_b^2)
- \frac{\bar\varepsilon^2}{\varepsilon}\left((Q_4+2U_4)x_\delta^dy_t^{0.5}y_t^2+2a_0^2A_2^{\prime d}A_5^dx_{13}^d\frac{y_b^2}{y_t^{0.5}} \right)e^{i(-2\chi+\beta_3+\beta_2')}\right\}\Delta \nonumber \\ \\
Q^{EW}_{32} & = & Q^{GUT}_{32} - 2\frac{\bar\varepsilon^2}{\varepsilon}(3+a_0^2\,(A_5^u)^2)\frac{x_{23}^d\,\Sigma_d}{y_t^{0.5}}\,y_t^2\,\Delta \nonumber \\
& + & \left\{Q^{GUT}_{32}(y_t^2+y_b^2) 
-\frac{\bar\varepsilon^2}{\varepsilon}\left((Q_4+2U_4)y_t^{0.5}y_t^2+2a_0^2A_4^{\prime d}A_5^d\frac{y_b^2}{y_t^{0.5}} \right)x_{23}^d\Sigma_de^{i(-2\chi+\beta_3+\beta_2')}\right\}\Delta \nonumber \\
\end{eqnarray}
\end{subequations}
\begin{subequations}
\begin{eqnarray}
E^{EW}_{21} & = & E^{GUT}_{21} \\
E^{EW}_{31} & = & E^{GUT}_{31} + \left\{2E^{GUT}_{31} - 2a_0^2\,\frac{\bar\varepsilon}{y_b^{0.5}}\,A_2^{\prime e} \,A^e_5\,x_{13}^e\,e^{-i(\chi-\beta_3-\beta'_2)}\right\}y_b^2\Delta \\
E^{EW}_{32} & = & E^{GUT}_{32} + 2\left\{E^{GUT}_{32} + 2a_0^2\,\frac{\bar\varepsilon}{y_b^{0.5}}\,A_4^{\prime e} \,A^e_5\,x_{23}^e\,\Sigma_e\,e^{-i(\chi-\beta_3-\beta'_2)}\right\}y_b^2\,\Delta
\end{eqnarray}
\end{subequations}
\begin{subequations}
\begin{eqnarray}
L^{EW}_{21} & = & L_{21}^{GUT} 
+ 4\frac{\bar\varepsilon^4}{\varepsilon^2}x_\delta^e\,x_{23}^e\,\Sigma_e\,y_t^2\,\Delta_\nu\,\Delta \nonumber \\
& & + \frac{\bar\varepsilon^4}{\varepsilon^2}
\left\{2(L_4\,y_t)x_\delta^e\,x_{23}^e\,y_t^2\,\Delta_\nu
+2a_0^2(A_2^{\prime e}A_4^{\prime e}x^e_{13}x^e_{23}+A_1^{\prime e}A^e_3x_{12}^e)y_b^2\right\}\Sigma_e\,\Delta  \nonumber \\
& & - \frac{\bar\varepsilon^2}{\varepsilon}\,L_5\,y_t^{0.5}\left(x_\delta^e\,e^{i(-2\chi+\beta_3+\beta'_2)} + x_{12}^ex_{23}^e\,e^{i(2\chi-\beta_3-\beta'_2)}\right)y_t^2\,\Delta_\nu\Delta \\
L^{EW}_{31} & = & L^{GUT}_{31} + 4\frac{\bar\varepsilon^2}{\varepsilon}\frac{x_\delta^d}{y_t^{0.5}}y_t^2e^{i(-2\chi+\beta_3+\beta'_2)}\,\Delta_\nu\,\Delta \nonumber \\
& - & \left\{L^{GUT}_{31}(y_t^2\,\Delta_\nu+y_b^2)
- \frac{\bar\varepsilon^2}{\varepsilon}\left(L_4x_\delta^dy_t^{0.5}y_t^2\,\Delta_\nu+2a_0^2A_2^{\prime d}A_5^dx_{13}^d\frac{y_b^2}{y_t^{0.5}} \right)e^{i(-2\chi+\beta_3+\beta_2')}\right\}\Delta \nonumber \\ \\
L^{EW}_{32} & = & L^{GUT}_{32} - 4\frac{\bar\varepsilon^2}{\varepsilon}\frac{x_{23}^e\,\Sigma_e}{y_t^{0.5}}\,y_t^2\,\Delta_\nu\,\Delta \nonumber \\
& + & \left\{L^{GUT}_{32}(y_t^2\,\Delta_\nu+y_b^2) 
-\frac{\bar\varepsilon^2}{\varepsilon}\left(L_4y_t^{0.5}y_t^2\,\Delta_\nu+2a_0^2A_4^{\prime e}A_5^e\frac{y_b^2}{y_t^{0.5}} \right)x_{23}^e\Sigma_ee^{i(-2\chi+\beta_3+\beta_2')}\right\}\Delta \nonumber \\
\end{eqnarray}
\end{subequations}

\subsection{RVV Model 3}

\begin{subequations}
\begin{eqnarray}
U^{EW}_{21} & = & U^{GUT}_{21} \\
U^{EW}_{31} & = & U^{GUT}_{31} + 2U^{GUT}_{31}\,y_t^2\Delta \\
U^{EW}_{32} & = & U^{GUT}_{32} + 2\left\{U^{GUT}_{32}+ 2a_0^2\,A_4^{\prime u}A_5^u\,\Sigma_u\,e^{2i\beta_3}\right\}y_t^2\Delta
\end{eqnarray}
\end{subequations}
\begin{subequations}
\begin{eqnarray}
D^{EW}_{21} & = & D^{GUT}_{21} \\
D^{EW}_{31} & = & D^{GUT}_{31} + 2D^{GUT}_{31}\,y_b^2\Delta \\
D^{EW}_{32} & = & D^{GUT}_{32} + 2\left\{D^{GUT}_{32}+ 2a_0^2\,A_4^{\prime d}A_5^d\,\Sigma_d\,e^{-2i(\chi-\beta_3)}\right\}y_b^2\Delta
\end{eqnarray}
\end{subequations}
\begin{subequations}
\begin{eqnarray}
Q^{EW}_{21} & = & Q^{GUT}_{21} + \left(1+\frac{\varepsilon^2}{\bar\varepsilon^2} \frac{x_{23}^u\Sigma_u}{x_{23}^d\Sigma_d}e^{-2i\chi}\right)(Q_6\,x_{23}^d\Sigma_d)y_t^2\Delta \\
Q^{EW}_{31} & = & Q^{GUT}_{31} + Q_6(y_t^2+y_b^2)\Delta \\
Q^{EW}_{32} & = & Q^{GUT}_{32} + 2(3+a_0^2\,(A_5^u)^2)x_{23}^d\Sigma_d\,y_t\,\Delta
-2\frac{\varepsilon^2}{\bar\varepsilon^2}(3+a_0^2A_4^uA_5^u)x_{23}^u\,\Sigma_u\,y_t e^{2i\chi}\Delta \nonumber \\
& + & \left\{\frac{Q_{32}^{GUT}}{x_{23}^d\Sigma_d}\,(y_b^2+y_t^2) + (Q_4+2U_4)\left(1-\frac{\varepsilon^2}{\bar\varepsilon^2}\frac{x_{23}^u\Sigma_u}{x_{23}^d\Sigma_d}e^{2i\chi}\right)y_t^2 - 2a_0^2\,A^d_5\,A_4^{\prime d}\,\frac{y_b^2}{y_t}\right\}x_{23}^d\Sigma_d\,\Delta \nonumber \\
\end{eqnarray}
\end{subequations}
\begin{subequations}
\begin{eqnarray}
E^{EW}_{21} & = & E^{GUT}_{21} \\
E^{EW}_{31} & = & E^{GUT}_{31} + 2E^{GUT}_{31}\,y_b^2\Delta \\
E^{EW}_{32} & = & E^{GUT}_{32} + 2\left\{E^{GUT}_{32}+ 2a_0^2\,A_4^{\prime e}A_5^e\,\Sigma_e\,e^{-2i(\chi-\beta_3)}\right\}y_b^2\Delta
\end{eqnarray}
\end{subequations}
\begin{subequations}
\begin{eqnarray}
L^{EW}_{21} & = & L^{GUT}_{21} + (L_6\,x_{23}^e\Sigma_e)y_t^2\,\Delta_\nu\,\Delta \\
L^{EW}_{31} & = & L^{GUT}_{31} + L_6(y_t^2\,\Delta_\nu+y_b^2)\Delta \\
L^{EW}_{32} & = & L^{GUT}_{32} + 4x_{23}^e\Sigma_e\,y_t\,\Delta_\nu\,\Delta \nonumber \\
& + & \left\{\frac{L_{32}^{GUT}}{x_{23}^e\Sigma_e}\,(y_b^2+y_t^2\,\Delta_\nu) + L_4y_t^2\,\Delta_\nu - 2a_0^2\,A^e_5\,A_4^{\prime e}\,\frac{y_b^2}{y_t}\right\}x_{23}^e\Sigma_e\,\Delta
\end{eqnarray}
\end{subequations}

\subsection{A-Terms}
To write the RGE contribution for the A-Terms, we shall first denote:
\begin{subequations}
\begin{eqnarray}
\kappa_d^A & = & -\frac{7}{15}g_1^2-3g_2^2-\frac{16}{3}g_3^2+3y_b^2+y_\tau^2 \\
\kappa_d^Y & = & \frac{7}{15}g_1^2 M_1 +3g_2^2 M_2 +\frac{16}{3}g_3^2 M_3 + (3y_b^2 A_5^d+y_\tau^2 A_5^e)A_0 \\
\kappa_u^A & = & -\frac{13}{15}g_1^2-3g_2^2-\frac{16}{3}g_3^2+3y_t^2+\sum_i y_{\nu,i}^2 \\
\kappa_u^Y & = & \frac{13}{15}g_1^2 M_1 +3g_2^2 M_2 +\frac{16}{3}g_3^2 M_3 +3y_t^2 A_5^u A_0 \\
\kappa_e^A & = & -\frac{9}{5}g_1^2-3g_2^2+3y_b^2+y_\tau^2 \\
\kappa_e^Y & = & \frac{9}{5}g_1^2 M_1 +3g_2^2 M_2 + (3y_b^2 A_5^d+y_\tau^2 A_5^e)A_0
\end{eqnarray}
\end{subequations}

As the A-Terms are not hermitian matrices, the RGE evolution is different for each off-diagonal term. For each element, we shall factorize in Eq.~(\ref{aterms-sckm}) the leading order of magnitude in terms of $\varepsilon$ or $\bar\varepsilon$, the phase, and the global term $A_0\,y_i$. Then, after the running, we get:
\begin{subequations}
\begin{eqnarray}
A_d(1,1) & = & \mathcal{O}(\bar\varepsilon^4) \\
A_d(1,2) & = & (1+\kappa_d^A\Delta)A_1^{\prime d}x_{12}^d \\
A_d(1,3) & = & \left(1+(\kappa_d^A+A_{13}^{d,run})\Delta\right) A_2^{\prime d}x_{13}^d \\
A_d(2,1) & = & A_d(1,2) \\
A_d(2,2) & = & \left(1+\left(\kappa_d^A +2\frac{\kappa_d^Y}{A^d_3A_0}\right)\Delta\right)A^d_3\,\Sigma_d \\
A_d(2,3) & = & \left(1+(\kappa_d^A+A_{23}^{d,run})\Delta\right) A_4^{\prime d}\,x_{23}^d\,\Sigma_d \\
A_d(3,1) & = & \left(1+(\kappa_d^A+5y_b^2+y_t^2)\Delta\right) A_2^{\prime d} \,x_{13}^d \\
A_d(3,2) & = & \left(1+(\kappa_d^A+5y_b^2+y_t^2)\Delta\right) A_4^{\prime d} \,x_{23}^d\,\Sigma_d \\
A_d(3,3) & = & \left(1+\left(9y_b^2+\frac{(A_5^d+2 A_5^u)}{A_5^d}y_t^2+\kappa_d^A +2\frac{\kappa_d^Y}{A_5^dA_0}\right)\Delta\right)A_5^d \\
\end{eqnarray}
\end{subequations}
\begin{subequations}
\begin{eqnarray}
A_u(1,1) & = & \mathcal{O}(\varepsilon^4) \\
A_u(1,2) & = & (1+\kappa_u^A\Delta)A_1^{\prime u}x_{12}^u \\
A_u(1,3) & = & \left(1+(\kappa_u^A+A_{13}^{u,run})\Delta\right) A_2^{\prime u}x_{13}^u \\
A_u(2,1) & = & A_u(1,2) \\
A_u(2,2) & = & \left(1+\left(\kappa_u^A +2\frac{\kappa_u^Y}{A^u_3A_0}\right)\Delta\right)A^u_3\,\Sigma_u \\
A_u(2,3) & = & \left(1+\left(\kappa_u^A+A^{u,run}_{23}\right)\Delta\right)A_4^{\prime u} x_{23}^u\,\Sigma_u \\
A_u(3,1) & = & \left(1+(\kappa_u^A+5y_t^2+y_b^2)\Delta\right) A_2^{\prime u}\,x_{13}^u \\
A_u(3,2) & = & \left(1+(\kappa_u^A+5y_t^2+y_b^2)\Delta\right) A_4^{\prime u}\,x_{23}^u\,\Sigma_u \\
A_u(3,3) & = & \left(1+\left(9y_t^2+\frac{(A_5^u+2 A_5^d)}{A_5^u}y_b^2+\kappa_u^A +2\frac{\kappa_u^Y}{A^u_5A_0}\right)\Delta\right)A^u_5
\end{eqnarray}
\end{subequations}
with:
\begin{subequations}
 \begin{eqnarray}
A^{d,run}_{13} & = & 4y_b^2+\left(\frac{2A_5^u+A_5^d}{A_2^{\prime d}}\right)\frac{x_{\delta}^d}{x_{13}^d}y_t^2
-\frac{\varepsilon^2}{\bar\varepsilon^2}\left(\frac{2A_4^u+A_5^d}{A_2^{\prime d}}\right)
\left(\frac{x_{12}^dx_{23}^u\Sigma_u}{x_{13}^d\Sigma_d}\right)y_t^2\,e^{2i\chi} \\
A^{d,run}_{23} & = & 4y_b^2-3\frac{A_5^d}{A_4^{\prime d}}y_t^2
+\frac{\varepsilon^2}{\bar\varepsilon^2}\left(\frac{2A_4^u+A_5^d}{A_4^{\prime d}}\right) \left(\frac{x^u_{23}\Sigma_u}{x^d_{23}\Sigma_d}\right)y_t^2\,e^{2i\chi} \\
A^{u,run}_{13} & = & 4y_t^2+
\frac{\bar\varepsilon^3}{\varepsilon^3}\left[\left(\frac{2A_2^d+A_5^u}{A_2^{\prime u}}\right)\frac{x_{13}^d}{x_{13}^u} \,e^{i(\omega'-\omega_{us})}
-\frac{\varepsilon}{\bar\varepsilon}\left(\frac{2A_4^d+A_5^u}{A_2^{\prime u}}\right)\frac{x_{12}^u x_{23}^d\Sigma_d}{x_{13}^u\Sigma_u}\right]y_b^2\,e^{-2i\chi} \\
A^{u,run}_{23} & = & 4y_t^2-\left(\frac{2A_5^d+A_5^u}{A_4^{\prime u}}\right)y_b^2
+\frac{\bar\varepsilon^2}{\varepsilon^2}\left(\frac{2A_4^d+A_5^u}{A_4^{\prime u}}\right) \left(\frac{x^d_{23}\Sigma_d}{x^u_{23}\Sigma_u}\right)y_b^2\,e^{-2i\chi} \\
 \end{eqnarray}
\end{subequations}

We see that the running can generate subleading phases, which are contained within the $A^{f,run}_{ij}$ terms. In particular, it is noticeable that these phases do not vanish when aligning the A-Terms with the Yukawas (setting all $A_i^{\prime f}\to0$), which means they belong to a MFV contribution.

It is also important to remark the fact that $A^{u,run}_{13}$ and $A^{u,run}_{23}$ contain terms of order $(\bar\varepsilon/\varepsilon)^3$ and $(\bar\varepsilon/\varepsilon)^2$, which are enhancement factors. Again, these are due to MFV contributions. Thus, the RGE evolution has the potential of changing the structure of $A_u$ noticeably.

We do not list the $A_e$ matrices, since their structure is identical to $A_d$, with the replacements $y_t^2\to y_t^2\,\Delta_\nu$, $\kappa^\alpha_d\to\kappa^\alpha_e$, $x_i^d\to x_i^e$, $A_i^d\to A_i^e$ and $A_i^u\to0$. One must also consider $\Sigma_e=3\Sigma_d$ and $\Sigma_\nu=0$.

\end{document}